\documentclass[12pt,preprint]{aastex}
\begin{document}
\title{Polarimetric Observations of 15 Active Galactic Nuclei at High Frequencies: 
Jet Kinematics from Bimonthly Monitoring with the Very Long Baseline Array}
\author{Svetlana G. Jorstad\altaffilmark{1,2}, Alan P. Marscher\altaffilmark{1},
Matthew L. Lister\altaffilmark{3}, Alastair M. Stirling\altaffilmark{4,5},
Timothy V. Cawthorne\altaffilmark{4}, Walter K. Gear\altaffilmark{6}, 
Jos\'e L. G\'omez\altaffilmark{7}, Jason A. Stevens\altaffilmark{8},
Paul S. Smith\altaffilmark{9}, James R. Forster\altaffilmark{10}, 
and E. Ian Robson\altaffilmark{8}}
\altaffiltext{1}{Institute for Astrophysical Research, Boston University,
725 Commonwealth Ave., Boston, MA 02215-1401; jorstad@bu.edu, marscher@bu.edu}
\altaffiltext{2}{Sobolev Astronomical Institute, St. Petersburg State University,
Universitetskij pr. 28, 198504 St. Petersburg, Russia}
\altaffiltext{3}{Department of Physics, Purdue University, 525 Northwestern Ave.,
West Lafayette, IN 47907-2036; mlister@physics.purdue.edu}
\altaffiltext{4}{Center for Astrophysics, University of Central Lancashire, Preston,
PR1 2HE, UK; tvcawthorne@uclan.ac.uk}
\altaffiltext{5}{University of Manchester, Jodrell Bank Observatory, Macclesfield,
Cheshire, SK11 9DL, UK (current address); ams@jb.man.ac.uk}
\altaffiltext{6}{School of Physics and Astronomy, Cardiff University, 5, The Parade
 Cardiff CF2 3YB, Wales, UK; Walter.Gear@astro.cf.ac.uk}
\altaffiltext{7}{Insituto de Astrof\'{\i}sica de Andaluc\'{\i}a (CSIC), Apartado 3004,
Granada 18080, Spain; jlgomez@iaa.es}
\altaffiltext{8}{Astronomy Technology Centre, Royal Observatory, Blackford Hill, 
Edinburgh EH9 3HJ, UK; jas@roe.ac.uk, eir@roe.ac.uk}
\altaffiltext{9}{Steward Observatory, University of Arizona, Tucson, AZ 85721;psmith@as.arizona.edu}
\altaffiltext{10}{Hat Creek Observatory, University of California, Berkeley, 42231 
Bidwell Rd. Hatcreek, CA 96040; rforster@astro.berkeley.edu}

\shorttitle{7~mm VLBA Polarization Monitoring}
\shortauthors{Jorstad et al.}
\begin{abstract}
We present total and polarized intensity images 
of 15 active galactic nuclei obtained with the Very Long Baseline Array at 
7~mm wavelength at 17 epochs from 1998 March to 2001 April. At some epochs
the images are accompanied by nearly simultaneous polarization measurements 
at 3~mm, 1.35/0.85~mm, and optical wavelengths. Here we analyze the 7~mm images 
to define the properties of the jets of two radio galaxies, five BL~Lac objects, 
and eight quasars on angular scales $\gtrsim 0.1$ milliarcseconds.
We determine the apparent velocities of 106 features in the jets. 
For many of the features we derive Doppler factors using a new method based 
on comparison of timescale of decline in flux density with the light-travel 
time across the emitting region. 
This allows us to estimate the Lorentz factors, $\Gamma$, intrinsic
brightness temperatures, and viewing angles of 73 superluminal
knots, as well as the opening angle of the jet for each source. 
The Lorentz factors of the jet flows in the different blazars range 
from $\Gamma\sim 5$ to 40 with the majority of the quasar components
having $\Gamma\sim 16-18$ while the values in the BL~Lac objects are more
uniformly distributed. The brightest knots in the quasars have the highest 
apparent speeds, while the more slowly moving components are pronounced 
in the BL~Lac objects. The quasars in our sample have similar opening angles 
and marginally smaller viewing angles than the BL Lacs. The two radio galaxies 
have lower Lorentz factors and wider viewing angles than the blazars.  
Opening angle and Lorentz factor are 
inversely proportional, as predicted by gas-dynamic models.
The brightness temperature drops more abruptly with distance from the core in
the BL~Lac objects than in the quasars and radio galaxies, perhaps owing to stronger
magnetic fields in the former resulting in more severe synchrotron losses 
of the highest energy electrons. 
In nine sources we detect statistically meaningful deviations from
ballistic motion, with the majority of components
accelerating with distance from the core. In six sources we identify jet features with
characteristics of trailing shocks that form behind the
primary strong perturbations in jet simulations. The apparent speeds of these 
components increase with distance from the core, suggestive of 
acceleration of the underlying jet.
\end{abstract}
 
\section{Introduction}
The jets observed emanating from the nuclei of quasars and other
active galactic nuclei (AGN) represent the most energetic long-lived phenomenon 
in the universe. Although jet imaging is no longer the sole prerogative of radio
astronomy, the exquisite resolution of Very Long Baseline Interferometry (VLBI) 
at radio wavelengths remains an unreached goal for sub-millimeter (sub-mm)
and shorter wavelengths. It is thought that accretion onto a black hole drives the 
jet outward via magnetic forces \citep[e.g.][]{Meier00}. Currently, the most 
direct way to provide observational evidence for such a model is to study the 
linear polarization of jets at different 
frequencies (from radio to optical) along with changes in the innermost jet 
structure. The most promising objects in such investigations 
are blazars, flat-spectrum radio-loud quasars and BL~Lac objects characterized 
by high optical polarization up to 46\% \citep{Mead90,IT90}, pronounced and 
rapid variability of flux, and one-sided jets with knots that move at superluminal 
apparent velocities, as fast as $\gtrsim$30$c$ \citep{J01,KL04}. According to \citet
{Beverley92}, there is a highly significant correlation between optical polarization and
dominance of a compact radio core-jet structure, which suggests beaming of the optical 
flux along with the radio emission. Comparison of radio-loud quasars with low and
high optical polarization shows a higher fractional 
polarization of the radio core for the latter \citep{LS00}, which implies a co-spatial
origin of the emission at these wavelengths. In support of this, the optical
polarization in blazars is affected by the formation and emergence of new VLBI 
knots \citep{G94,G96}. 
These results indicate that simultaneous multifrequency polarization 
monitoring, together with high resolution polarimetric imaging of the radio jets, 
provides a unique tool for identifying the location of the regions responsible for
variability at different wavelengths and for relating the magnetic field geometry to 
the structure of the jet.  

We have obtained total and polarized intensity images of 15 AGNs with the Very Long 
Baseline Array (VLBA) at 7~mm (43~GHz) at 17 epochs over three years. The VLBA observations 
are accompanied at many epochs by nearly simultaneous (within two weeks) measurements 
of polarization at 1.35/0.85 mm (230/350~GHz) and at optical wavelengths. In the 
second half of the program simultaneous polarization observations at 3~mm were
performed at several epochs. The main goals of the project are to relate emission
regions at high frequencies to the parsec-scale jet structure and to investigate
the strength, direction, and variability of the magnetic field close to the central
engine. These can be achieved only after detailed study of the jet kinematics. 
This paper is devoted to an analysis of the jet structure and its variability 
associated with ejection and propagation of disturbances down the jet, which
appear on radio maps as knots of enhanced brightness. 
The paper is the first in a series based on the entire data set collected during 
the project. Some results on individual sources have been published already 
by \citet[3C~120]{MAR02}, \citet[BL~Lac]{AS03}, and \citet[3C~279]{J04}.
Other papers will include (1) comparison of the polarization parameters at different
frequencies with the jet structure and disturbances in the jet; (2) analysis of the
available radio, sub-mm, optical, and X-ray light curves to relate the flux variability
to activity in the jet; (3) results regarding stability
of the VLBI core position based on phase-referencing observations obtained for 
five sources in the sample; and (4) structure of the core and intraday total and polarized
intensity variability in the parsec-scale jets.

Although much progress has been made in studying the kinematics of jets since 
the VLBA started to operate fully in 1995 \citep{Pear98,KL98,HO01,J01,G43,KL04}, 
our program is unique regarding the number of sources observed in such detail 
over a rather long
period of time. This allows us to separate the fast and slow, ballistic and curved 
motion in the jet flow, define physical parameters for both individual jet features 
and entire parsec-scale jets, and compare the results across different classes of AGNs.

\section{Sample Selection} 
Our program is designed for comparison of the linear polarization at high frequencies
(mm, sub-mm, and optical wavelengths) with the parsec scale jet structure of AGNs. 
This has defined the main criteria for selecting the sample and frequency of our VLBA
observations: 
\begin{enumerate}
\item{The sources should be bright, $\geq$0.5~Jy, and polarized, $\geq$3\%, 
at sub-mm wavelengths.}
\item{The size of the sample and brightness of sources should allow us to perform  VLBA 
observations at a single epoch during 24 hours with sufficient uv-coverage to produce 
total and polarized intensity images at 43~GHz with high dynamic range.}
\item{The sample should contain sources with resolved radio structure from
different sub-classes of AGNs for which variability in the jet flow can 
be expected on timescales of months.}
\item{The sources should be convenient for monitoring in the northern hemisphere
and their coordinates should cover the whole range of right ascensions.}
\end{enumerate}

Following these constraints, we have formed a sample of 15 AGNs that consists of 8 
quasars, 5 BL~Lac objects, and 2 radio galaxies (3C~120 of Fanaroff-Riley type 1
and 3C~111 of type 2) after obtaining information on the 
linear polarization at sub-mm wavelengths from \citet{N98}. The sources are listed 
in Table \ref{Sample}.

\section{Observations and Data Reduction}
We have observed the objects in our sample in four different regions of the 
electromagnetic spectrum: at 43~GHz (7 mm) with the VLBA, at 350/230~GHz (0.85/1.3~mm)
with the {\it James Clerk Maxwell Telescope}  (JCMT, Mauna Kea, Hawaii) using SCUBA 
\citep{SCUBA} and its polarimeter \citep{POL03}, at the Steward Observatory 1.5~m 
telescope (Mt. Lemmon, Arizona) with the Two-Holer Polarimeter/Photometer 
\citep{SSS85} over an effective wavelength range of $\sim$6000--7000${\AA}$, 
and at 86~GHz (3~mm) with the {\it Berkeley-Illinois-Maryland Array} 
(BIMA, Hat Creek, California). The observations were performed from 1998 March to 
2001 April. The VLBA monitoring was carried out roughly bimonthly (17 epochs). 
For the JCMT and optical observations the number of epochs depends on the source, 
with the maximum being 11 and seven epochs and the minimum five and three epochs, respectively,
except for the BL~Lac object 1803+784, which was not observed in the optical
region owing an inaccessibly high declination. 
The majority of the JCMT polarization observations were accompanied 
by total flux measurements. Differential $V$-band photometry
($R$-band photometry in the case of 3C~273) was carried out 
when conditions were photometric. The BIMA polarization observations started
in 2000 April and for many sources in the sample were performed simultaneously
with the VLBA observations at 3-4 epochs. However, each of these epochs
usually was accompanied by a series of BIMA observations separated by several 
days so that for some sources polarization measurements at 3~mm were obtained at 
20 epochs. The polarization data at 3~mm include all four Stokes parameters. 
In this paper the optical, JCMT, and BIMA polarization results are shown only graphically
to illustrate the data collected. Detailed descriptions of the data reduction
and tables of measurements will be presented in a later paper.

\subsection{Radio Observations}
The 43~GHz observations were carried out with the VLBA recording
system using eight 8 MHz wide channels, each in right and left circular 
polarization, with 15-20 scans of 3-5 minute duration for each object. 
All 10 antennas were used for each source except at epochs affected by weather or
receiver failure.
Table \ref{Antenna} lists the antennas operated at each epoch and typical parameters of
the synthesized beam (those of the OJ 287 observations) for uniform weighting
(used for imaging all sources) to illustrate the consistency of the $uv$-coverage 
over epochs. Initial correlation was carried out at the National Radio 
Astronomy Observatory (NRAO) Array Operations Center in Socorro, NM. 
Subsequent calibration was performed with the Astronomical Image Processing 
System (AIPS) software supplied by NRAO, while images were made with 
the Caltech software Difmap \citep{Difmap}. The calibration included 
application of 
the nominal antenna-based gain curves and system temperatures, and correction 
for sky opacity, followed by iterative imaging plus phase and amplitude
self-calibration. For each epoch we calculated
the total flux density in the images of sources that are known to have very 
weak emission outside the angular size range of the VLBA images
(0420-014, 0528+134, OJ 287, and BL~Lac). These values were compared
with total flux densities obtained by interpolating in time the 
measurements of the monitoring program at 37 GHz at the Mets\"ahovi Research
Station, Finland \citep{T04}. The comparison produced the flux density correction
factors, $f_{\rm amp}$, given in Table \ref{Antenna}. The factors were
applied for the final adjustment of the flux density scale in the images. 
We have constructed  
light curves of the VLBI core for each source to confirm the absence of correlation
between their behavior, indicating no evidence for residual amplitude calibration
errors. The light curves of the 15 sources peak at 9 different epochs
and the correlation coefficients are spread between $-$0.4 and +0.3, consistent
with no systematic calibration errors in the flux density scaling.

We performed a cross-hand fringe fit using a scan of 3C~279 averaged over
all baselines (the data were corrected for parallactic angle rotation 
of phases in advance). 
The resulting right-left phase and rate delay corrections
were processed with the AIPS task POLSN and applied to the full data set.
After preliminary images were produced, the images were used in AIPS 
task CALIB to self-calibrate the phases of all sources; this includes
the removal of residual R-L phase differences owing to calibration errors.
The instrumental polarization ``D-terms'' were determined via the
method of \citet{RWB94} and \citet{LZD95}. A final set of D-terms at each epoch
was obtained by averaging of the solutions  for those sources in the sample
for which there is the best agreement between the D-terms 
(usually, 3C~111, OJ~287, 3C~345, and BL~Lac). The electric vector position angle
(EVPA) calibration was obtained by different methods: comparison between the 
Very Large Array (VLA) and VLBA integrated EVPAs at quasi-simultaneous epochs,  
D-terms method (see below), and using EVPA-stable features in the images 
of the jets of 3C~279, OJ 287, and CTA102. Over the period 1998 March -- 1999 February 
we used results of the VLA observations performed for 0420$-$014, 3C~120, 
OJ 287, 3C~279, BL~Lac, and 3C~454.3 by \citet{Dterm}. Later we obtained polarization
measurements for 0420$-$014, 0528+134, OJ 287, and 1803+784 at two epochs, 2000 April 
and 2000 July, with the VLA $C$ and $CnD$ configuration, respectively. At epochs 
where VLA data were not available we used the D-terms method, which 
is based on the assumption that the instrumental polarization parameters change 
slowly with time \citep{LZD95}. A detailed description of the method is given in 
\citet{Dterm}. In addition, later we checked our calibration for these epochs using 
the NRAO data base at {\bf http://www.vla.nrao.edu/astro/calib/polar/} for sources 
0420$-$014, 0528+134, 
OJ287, and BL~Lac when the NRAO data were available at epochs close to the VLBA dates. 
The results are completely consistent with our calibration. 
The EVPA-stable feature in 3C~279 is the well-known superluminal component 
$C4$ \citep[e.g.,][]{H03}, which underwent a change in EVPA in the beginning of 
1998 \citep{J04} but maintained this EVPA until the end of our program. 
The EVPA-stable features in OJ~287 and CTA~102
are quasi-stationary components at $\sim$1~mas and $\sim$2~mas from the core,
respectively. However, these both became very weak in the second half of 2000 in the
43~GHz images. The final EVPA calibration is a result of the best agreement among
the different methods. The accuracy of the calibrated EVPA measurements, 
$\sigma$(EVPA), is indicated in Table \ref{Antenna}.

We have combined the images obtained for every source in a sequence  
of total intensity maps convolved with the same beam, corresponding
to the average beam during the epochs when all 10 antennas were in operation. 
The contours are in terms
of the global peak of the maximum total intensity observed over all epochs.  
The sequences are presented in Figures \ref{Alla}-\ref{Allo}, which also show the JCMT, BIMA,
and optical polarization measurements. All images are oriented with north toward the
top and east toward the left. The scale in mas is indicated along one of the axes,
while the epoch of each image is given along the other axis. The peaks of the total, 
$I_{\rm peak}$, and polarized, $I^p_{\rm peak}$, intensity, parameters of the average beam,
the total intensity of the lowest contour, $I_{\rm min}$, and the lowest level 
of the polarized intensity, $I^p_{\rm min}$, of the combined maps are indicated 
in Table \ref{Sample}. The polarization at 7~mm is shown inside the
total intensity images by line segments, which are plotted if the local polarized 
intensity exceeds $I^p_{\rm min}$. The segments are oriented in the local direction 
of the polarization and their length is proportional to the local polarized intensity,
with the maximum length corresponding to the global peak of the polarized intensity 
over all epochs, $I^p_{\rm peak}$. Table \ref{Antenna} marks epochs when the BIMA, JCMT, 
and optical observations were performed within two weeks of the corresponding VLBA epoch.
 
\subsection{Model Fitting of the VLBA Images}
We employed the task MODELFIT in Difmap to represent each total intensity image 
(Stokes parameter $I$) as a sequence of circular Gaussian components that are 
characterized by flux density, size, and position relative to the map center.
Initially, point-like components are used to obtain an image with 100:1 dynamic
range. When a set of point-like components was found, a final group of 100 iterations 
was executed with all parameters of all components allowed to vary to define their
properties. In addition, because we have roughly bimonthly observations, we repeated the
model fitting with the components from the previous epoch used as the initial model 
(except for the first epoch, 1998 March). This improves identification of components 
over epochs and provides an estimate of the accuracy of the parameters by comparison 
of the outcomes obtained with different initial models. For several sources (3C~66A, 
OJ 287, 3C~279, PKS~1510$-$089, and 1803+784) at four epochs (1998 March 1998, 1999
October, 2000 July 2000, and 2001 January) we derived estimates of the 
1$\sigma$ uncertainties of the best-fit models using the method described 
by \citet{DIFER} and realized in the package Difwrap 
\citep{Difwrap}. The uncertainties depend significantly on the brightness and size 
of components:
1) for {\it bright} (flux of knot $\geq 100~\rm{rms}$ noise level) and {\it compact} 
(size $\leq$0.1~mas) features, 
uncertainties in flux density $\sim$1\%, in position $\sim$0.01~mas, and in size
$<$ 1\%; 
2) for the majority of components with size 0.1-0.3~mas and flux $\geq$50~mJy, 
uncertainties in the flux density $\sim$3\%, in position  
$\sim$1/5 of the beam size, and in size $\sim$5\%; 
3) for diffuse components,
uncertainties in flux density $\sim$10\%, in position are
comparable with half the size of the knot, and in size $\sim$10\%;
4) for components weaker than 50~mJy, uncertainties in flux density $\sim$50\%
and positional uncertainties correspond to the size of the average beam 
(see Table \ref{Sample}); however, for the majority of 
weak components these uncertainties were later modified as discussed in \S 4 
(Errors of Polynomial Parameters).

We determined the parameters of polarized components by applying the task MODELFIT 
to the $uv$-data in Stokes parameters $Q$ and $U$ separately. For this purpose only 
point-like components were used in the models. Because the Q and U components can 
be either positive or negative, the procedure needs to be carried out carefully by 
checking with the polarized intensity image to avoid the inclusion of false features
generated by the imaging procedure. The derived positional parameters of the $Q$ and $U$ 
components were compared with the parameters of $I$ components. We consider a $Q$ and/or 
$U$ component to be associated with an $I$ component if they are co-spatial to within  
the uncertainties. If both polarized components are present, 
the position of the polarized component is defined by the average location, weighted by 
the absolute values of the $Q$ and $U$ fluxes. There are few cases
when a significant polarized component does not have an $I$ counterpart.
Uncertainties in the values of the parameters of polarized components are 
difficult to define. Estimates of the accuracy of polarization parameters in jet
features at 15 and 22~GHz are discussed by \citet{HO02}. For components that are bright
in the total intensity and highly polarized (percent polarization $\ge$5\%), we 
find that the uncertainties in fractional polarization $\sim 1$\% and in polarization angle 
$\sim 5^\circ$. 

A table of parameters of jet features for each source over all epochs can be found at 
the web-site {\bf www.bu.edu/blazars/multi.html}. The columns of these tables are as follows:
1 - epoch, 2 - flux density in Jy, 3 - relative right ascension in mas, 
4 - relative declination 
in mas, 5 - distance from the core in mas, 6 - position angle relative to the core in
degrees, 7 - angular size in mas, 8 - polarized flux density of a polarized component
associated with a total intensity component, in Jy, 9 - distance of polarized component 
from the core in mas, 10 - position angle of polarized component relative to the core
in degrees, 11 - EVPA in degrees.
  
\section{Technique to Measure Motion in the Jets}
The data allow us to follow the evolution of major features of the jets 
that are observed in the total and polarized intensity VLBA images 
of the 15 AGNs at 17 epochs over 3 years. For each image we identify
a component, $A0$, as the VLBI core. For all sources $A0$ is located at one 
end of the jet, although it is not always the brightest feature. The position of the core
in right ascension $x$ and declination $y$ is defined as $x=0$ and $y=0$. In our
analysis we assume that the core is stationary over all epochs. Locations of other
features are determined relative to the core. Components are categorized as follows: 
knots $A$ (other than $A0$) are features that either are stationary
(the proper motion is less than or equal to its uncertainty), undergo reverse motion
(toward the core), or move at subluminal apparent speeds. 
Knots $B$, $C$, and $D$ are superluminal features, where $B$ components are the 
fastest knots in the jet, $D$ components are the farthest knots
from the core, ejected before our monitoring period started, and the remainder are
labeled as $C$. The designation
of components is different for 3C~120, 3C~279, 3C~345, and BL~Lac, where the naming 
corresponds to \citet{G43}, \citet{W01}, \citet{RZL00}, and \citet{AS03}, respectively.
Each component is characterized by the following parameters: total $S$ and polarized 
$S_p$ flux density; positions $x$ (RA), $y$ (Dec), and $R$, where $R=\sqrt{x^2+y^2}$; 
position angle $\Theta=\tan^{-1}(x/y)$; size (FWHM) of component $a$; and electric 
vector position angle, EVPA. To define the temporal evolution of the jet 
and determine the apparent velocities of the jet flow we perform the following 
steps.

1. {\it Identification of components at different epochs.} 
The identification of a component is based on comparison of the parameters during 
the epochs when it is visible on the images. We assume that the 
same component has similar total, $S$, and polarized, $S_{\rm p}$, fluxes, 
position angles $\Theta$ and EVPA, and size, $a$, at successive epochs. 
However, some components evolve dramatically even over a two-month interval
between epochs, splitting into one or more subcomponents or merging with 
other features of the jet. In this paper we analyze those jet features
whose identification is supported by the similarity of a number of 
parameters at different epochs.

2. {\it Fitting Various Polynomials.}
The thorough sequences of images allow us to search for acceleration/deceleration 
of the jet flow and non-ballistic projected trajectories.
We fit the $x$,$y$ positions of a component over $N$ epochs by different 
polynomials of order $l$:
\begin{equation}
x(t_i)=a_o+a_1\times (t_i-t_{mid})+a_2\times (t_i-t_{mid})^2+...
a_l\times (t_i-t_{mid})^l,\label{e1}
\end{equation}
\begin{equation}
y(t_i)=b_o+b_1\times (t_i-t_{mid})+b_2\times (t_i-t_{mid})^2+...
b_l\times (t_i-t_{mid})^l, \label{e2}
\end{equation}
where $t_i$ is the epoch of observation, $i$=1,...,N, and $t_{mid}$=$(t_1+t_N)/2$.
We use the program LSQPOL of the FORTRAN version of the package DATAN \citep{DATAN} 
to find the optimal polynomials of order $l$, where $l$ runs from $0$ to $4$. 
The upper limit, $l=4$, is a consequence of the maximum number of epochs of observation 
of a knot, equal to 17. The program provides the value $M_l$, which is the 
goodness-of-fit by a polynomial of order $l$. For each order, we perform a $\chi^2$ 
test to determine the polynomial that best fits the data. We choose a polynomial of the 
lowest order for which $M_l< M_{\chi^2}$, where $M_{\chi^2}$ is the
value of the $\chi^2$ distribution corresponding to significance level 
$\zeta$=0.05 for $f=N-l-1$ degrees of freedom \citep{BL72}.
 
Examples of selection of polynomials that fit the data are given in Tables 
\ref{Poly1} and \ref{Poly2} for component $C1$ identified in the jet of 
3C~66A at 16 epochs and component $B1$, also seen at 16 epochs, in 3C~273.
These tables show parameters of the best-fit polynomials of order from 0 to 4,
the corresponding values of the goodness of fit to the data $M$, and the  
goodness required by the $\chi^2$-test.
Table \ref{Poly1} shows that $l$=1 satisfies the $\chi^2$ test for coordinates
$x$ and $y$ in the case of $C1$ for 3C~66A. 
For $B1$, $l$=4 is required to match motion in RA, while a second-order
polynomial adequately describes the data for declination.

In a few cases the $\chi^2$ test is not satisfied even by polynomials of order
4. An increase in the order of the polynomial does not improve 
the situation since, given the limited number of observations, the number of degrees 
of freedom becomes too small. Such cases are approximated by a straight line 
with special consideration for uncertainties in the $x$,$y$ values 
(see below). These cases are marked in Table \ref{Speed}
by ``1*'' under the polynomial order.

3. {\it Uncertainties in Polynomial Parameters.}
After the best-fit polynomial is found, we run the subroutine LSQASN of the package DATAN
to estimate errors in the derived parameters. This method is valid for
a normal distribution of the unknowns and when the true value lies with probability 
$W$ within the confidence region around the estimated value ($W$=0.95 is used).
However, the errors depend significantly on the uncertainties in the individual
observations that are obtained from the model fitting (see \S 3). This allows us 
to lessen the weight of data at epochs with bad weather, a failure of one or more antennas, 
or when a given component was very weak or diffuse. For component 
motion that cannot be represented well by polynomials $l\leq$4 (see above), 
we approximate the motion by a first-order polynomial and determine uncertainties 
in the $x$,$y$ position 
using the program LSQPOL and the method suggested by \citet{HO01}. We set the uncertainty 
for each data point equal to the average beam size and
calculate the optimal polynomial of first order corresponding to a preliminary 
$\chi^2$ value. Taking this preliminary $\chi^2$ value, we uniformly re-scale the
uncertainties in the data points such that $\chi^2$ would correspond to the $\chi^2$
value for $\zeta$=0.05 and $f=N-2$. 

4. {\it Calculation of Proper Motion, Acceleration, and Ejection Time.}
We define the average proper motion as a vector ($<\mu>$,$<\Phi>$), where
$<\mu>$ represents the mean angular speed of motion and $<\Phi>$ gives the average
direction of motion. The average values are derived as follows:
$<\mu>=\sqrt{<\mu_x>^2+<\mu_y>^2}$ and $<\Phi>=\tan^{-1}(\frac{<\mu_x>}{<\mu_y>})$,
where $<\mu_x>=\int_{t_1}^{t_N}{\dot{x}dt}/(t_N-t_1)$ and 
$<\mu_y>=\int_{t_1}^{t_N}{\dot{y}dt}/(t_N-t_1)$.
In the case of first or second-order polynomials, $<\mu_x>$=$a_1$ and $<\mu_y>$=$b_1$. 
For a polynomial of order higher than unity we compute an average vector of
acceleration, ($\dot{\mu}_\parallel$,$\dot{\mu}_\perp$), 
where $\dot{\mu}_\parallel$ is along the direction of the average velocity, $<\Phi>$. 
In the case of a second-order polynomial
$\dot{\mu}_x=2\;a_2$ and $\dot{\mu}_y=2\;b_2$. For a higher order 
polynomial, $\dot{\mu}_x=\int_{t_1}^{t_N}{\ddot{x}dt}/(t_N-t_1)$ and 
$\dot{\mu}_y=\int_{t_1}^{t_N}{\ddot{y}dt}/(t_N-t_1)$.

\section {Jet Velocities}
We find superluminal apparent speeds for 19 of 22 components in the radio galaxies, 19 of
31 knots in the BL Lac objects, and 46 of 53 knots in the quasars. 
Table \ref{Speed} lists the average apparent speed, $<\beta_{app}>$, calculated in units 
of the speed of light, $c$, using the average proper motion, $<\mu>$.
An inhomogeneous Friedmann-Lema$\hat{\rm i}$tre-Robertson-Walker cosmology, 
with $\Omega_m=0.3$, $\Omega_\Lambda=0.7$, and Hubble constant 
$H_\circ$=70~km~s$^{-1}$~Mpc$^{-1}$ \citep{KKT00}, is adopted for the calculations.
We derive the uncertainties in 
the average proper motion, acceleration, and apparent speed from the uncertainties in the
polynomial coefficients. Table \ref{Speed} contains also the number of epochs, $N$,
at which the component has been observed; the average total flux, $<S>$, in Jy;
the average distance from the core, $<R>$, in mas; the average position angle
of the component, $<\Theta>$, in degrees, with uncertainty corresponding to the scatter 
across the epochs; the average direction of the velocity vector, $<\Phi>$, in degrees;
and the ejection time (epoch of zero separation), $T_\circ$.  
The ejection time is the extrapolated time of coincidence of the position of 
a moving knot with the core in the VLBA images. $T_\circ$ is the average of 
$t_{x\circ}$ and $t_{y\circ}$ weighted 
by their uncertainties, where $t_{x\circ}$ and $t_{y\circ}$ are roots of the best-fit
polynomials. In the case of a polynomial of order 3 and higher, the roots are computed
by successive iterations with an accuracy of $10^{-5}$.   
 
Table \ref{Accel} gives the parameters of acceleration for
components in Table \ref{Speed} having best-fit polynomials of order $\ge$2
(some columns repeat information from Table \ref{Speed} for convenience).
Figure \ref{Ident} shows the positions of the jet components relative 
to the core for both the total (open circles) and polarized (filled
circles) intensity images of each source. The solid lines/curves indicate the  
best polynomial fit to the data. Components presented in Figure \ref{Ident}
are marked at the first epoch in Figures \ref{Alla}-\ref{Allo} if they are detected
then or at the later epoch at which they appear in the jet.  
Figure \ref{Map} displays an image from one particular 
epoch for each source when the most prominent jet features are seen. Table \ref{Tmap} 
lists the parameters of the maps shown in Figure \ref{Map} (rms(I) and rms(Ip) are
the root mean squares of the residual total and polarized intensity on the images, 
respectively). The trajectories of all components with non-ballistic motion are plotted
in Figure \ref{traj}. Appendix A contains a description of our results for each 
individual object in the sample. 

\section{Apparent Speed as a Probe of Parsec-Scale Jets}
Our intensive, prolonged high-frequency VLBI monitoring reveals complex, 
changing structure of parsec scale jets, including significant variations
in the apparent speed of knots in the jet, $\beta_{app}$, within individual 
objects. The apparent speed is determined by the intrinsic velocity, 
$\beta$, and the angle between the trajectory and the line of sight, $\Theta_\circ$:
\begin{equation} 
\beta_{app}=\beta\;\sin\;\Theta_\circ(1-\beta\;\cos\;\Theta_\circ)^{-1}. \label{e3}
\end{equation} 
The intrinsic velocity defines the Lorentz factor of the knot, $\Gamma=1/\sqrt{1-\beta^2},$
where $\beta$ is in units of the speed of light. 
Large scatter in the apparent speed within a source can be caused by
(1) different patterns of jet components, such as ``blobs'' of energetic
plasma or forward, reverse, or stationary shocks; (2) variable Lorentz factor of 
the jet flow; or (3) different trajectories of components in the jet. 
The latter can result from a change in the direction of the jet 
or in the paths of components if each given knot does not fill 
the entire cross-section of the jet. 
 
The observed flux density of a superluminal component is boosted in the 
direction of the observer by a factor $\delta^{3+\alpha}$, where
\begin{equation}  
\delta=[\Gamma(1-\beta\;\cos\;\Theta_\circ)]^{-1}\label{e4}
\end{equation}
 is the Doppler factor and 
$\alpha$ is the spectral index ($S_\nu\propto\nu^{-\alpha}$) of the knot. 
The flux of a component decreases with distance from the core as the result 
of radiative energy losses and expansion (``adiabatic'' cooling). The emergence
of a new component is associated with one or more flares in the radio light 
curves \citep[e.g.,][]{SAV02}. The flare usually has a sharp peak and 
nearly symmetric exponential rise and decay at high frequencies \citep{TV94}. 
In the case of the shock-in-jet model for the appearance of superluminal knots,
variations in flux, spectral energy distribution, and polarization \citep{MG85,HAA85}, 
the symmetric light curves of flares suggest that the flux variability is 
controlled by light travel delays across the shocked region \citep[see][]{SOK04}. 
This assumption allows us to calculate the 
Doppler factor for each superluminal component using its flux density variability 
and size measured from the VLBA data.
We can then use the derived Doppler factors to study the physical parameters 
of the jets in our sample.

\subsection{Physical Parameters of the Jet Components} 
\begin{figure}
\epsscale{1.0}
\plotone{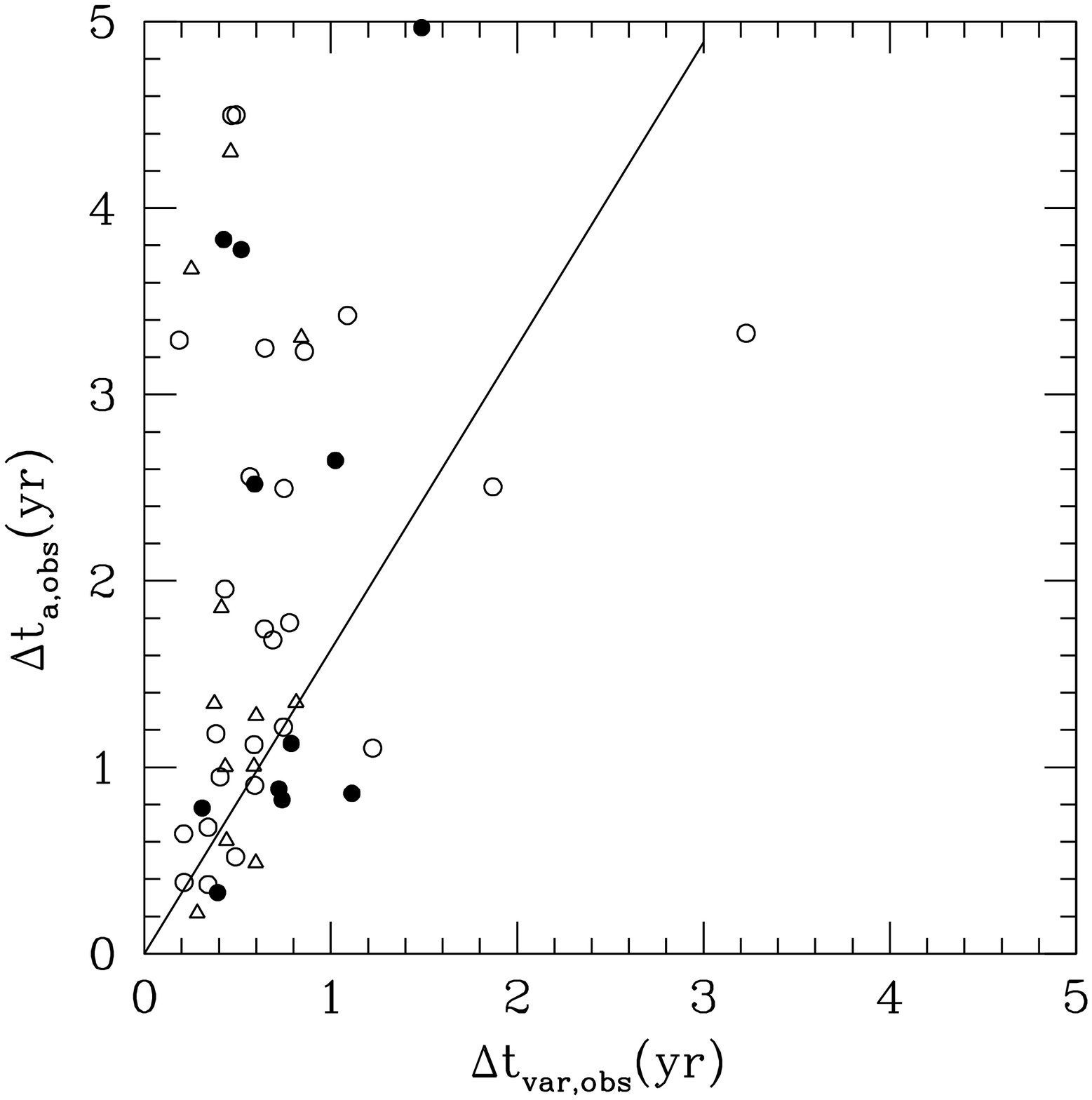}
\caption{Size variability timescale $\Delta t_{\rm a,obs}$ vs. flux variability
timescale $\Delta t_{\rm var,obs}$ for superluminal components in the quasars (open
circles), BL Lac objects (filled circles), and radio galaxies (triangle). These
symbols are applied for all subsequent figures where radio galaxies, BL~Lacs,
and quasars are plotted.
The solid line shows the expected relationship during the ``adiabatic'' stage
defined by \citet{MG85}. \label{tau_tau}}
\end{figure}

Figure \ref{flux} shows light curves of all superluminal components listed in 
Table \ref{Speed} (except knots classified as trailing features, see \S 6.6). 
Each light curve is normalized by the corresponding average flux indicated 
in Table \ref{Speed}. We define the timescale of the variability for each
superluminal component as $\Delta t_{\rm var}=dt/ln(S_{\rm max}/S_{\rm min})$ 
\citep{BJO74}, where $S_{\rm max}$ and $S_{\rm min}$ are the measured maximum 
and minimum flux densities, respectively, and $dt$ is the time in years between
$S_{\rm max}$ and $S_{\rm min}$. The variability Doppler
factors are derived as
\begin{equation} 
\delta_{\rm var}=\frac{s\;D}{c\;\Delta t_{\rm var}\;(1+z)}, \label{e5}
\end{equation} 
where $D$ is the luminosity distance, $s$ is the angular size of the component, 
equal to $1.6a$ for a Gaussian with FWHM=$a$ measured at the epoch of maximum flux
if the true geometry is similar to a uniform face-on disk. 
In the case of a point-like knot, we adopt $a$=0.1~mas, which
yields an upper limit to $\delta_{\rm var}$. This definition of the Doppler 
factor assumes that the variability timescale corresponds to the light-travel time
across the knot. This will be true if the radiative cooling time is shorter
than the light crossing time, which in turn is shorter than the timescale for
cooling by adiabatic expansion. We can roughly verify this assumption by using
the relation between flux density and size of a shocked region derived in 
\citet{MG85} during the ``adiabatic'' stage. We calculate the timescale, 
$\Delta t_{\rm a}$, of the variability in size for each non-point-like component 
as $\Delta t_{\rm a}=dt/ln(a_{\rm max}/a_{\rm min})$
(we use $a_{\rm max}=a_{\rm S_{min}}$ and $a_{\rm min}=a_{\rm S_{max}}$ 
if $a_{\rm S_{min}}>a_{\rm S_{max}}$, which is valid for the majority of 
components). Figure \ref{tau_tau} plots the size variability timescale  
versus the flux variability timescale. The straight line indicates the expected 
relation between $\Delta t_{\rm a}$ and $\Delta t_{\rm var}$ for adiabatic
losses for optically thin shocked gas with $\alpha=0.7$ \citep{MG85}. 
Figure \ref{tau_tau} shows that the majority of components have shorter flux
variability timescales than those predicted for adiabatic expansion.
This implies that at high radio frequencies the decay in flux is 
driven by radiative losses and, therefore, equation \ref{e5} should apply to our data.
 
\begin{figure}
\epsscale{1.0}
\plotone{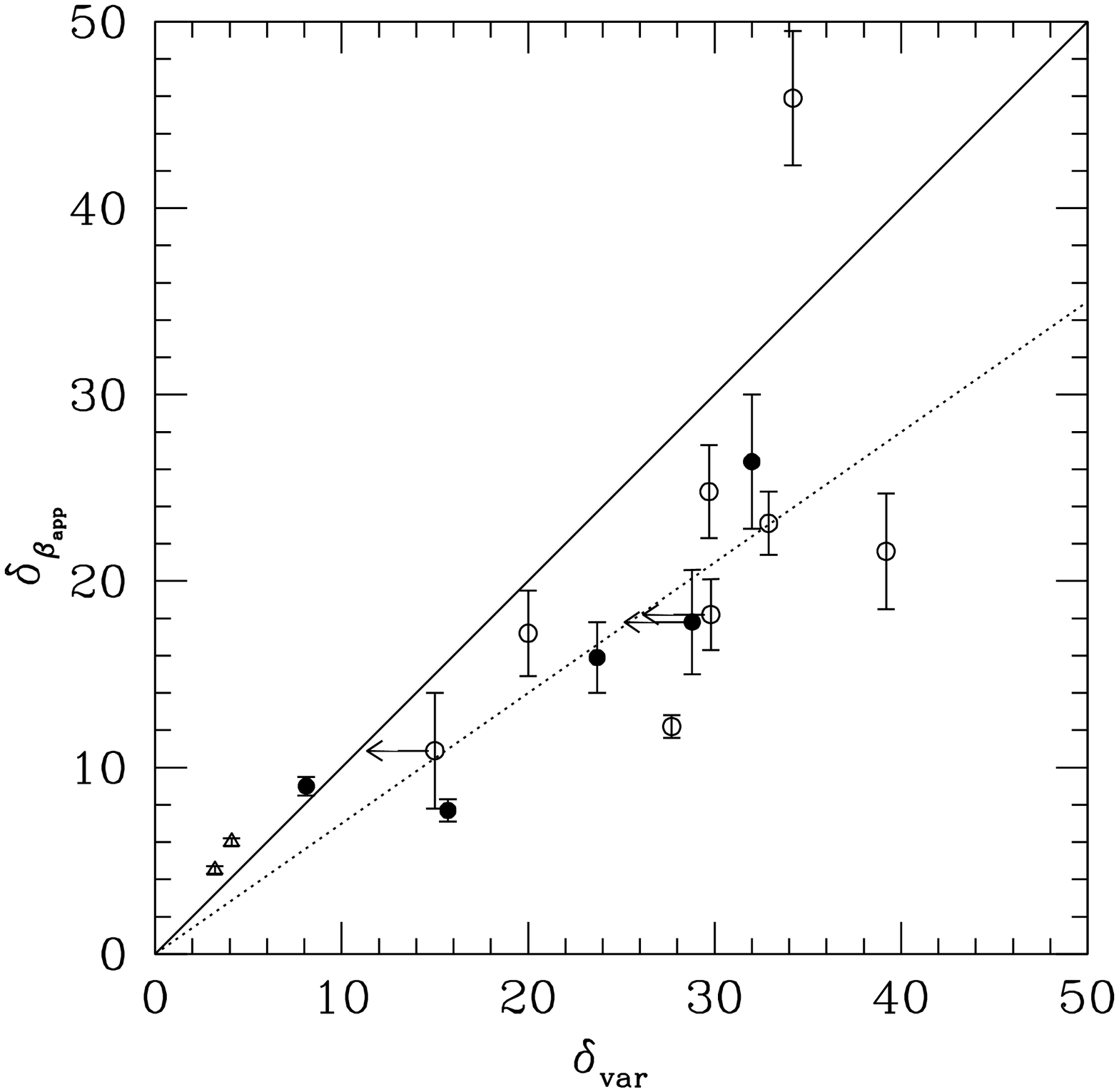}
\caption{Doppler factor $\delta_{\beta_{\rm app}}$ estimated from the
apparent speed for components with the highest apparent speed in the jets
vs. Doppler factor $\delta_{\rm var}$ computed from the flux variability and size 
of the components. The solid line shows the relation $\delta_{\beta_{\rm app}}=\delta_{\rm var}$,
the dotted line shows the least-square fit of the {\it observed} dependence. \label{delta_delta}}
\end{figure}

A widely used method to 
estimate the Doppler factor from VLBI data assumes
that the highest apparent speed detected in a source defines the lowest possible 
Lorentz factor of the jet, $\Gamma\ge\sqrt{1+\beta_{\rm app}^2}$. 
The viewing angle is then taken to be $\Theta_\circ\leq\sin^{-1}(1/\beta_{\rm app})$;  
this leads to $\delta_{\beta_{\rm app}}\sim\beta_{\rm app}$.
In Figure \ref{delta_delta} Doppler factors thus derived are
plotted versus the Doppler factor computed from the 
flux variability and size of components having the highest apparent speed in the jets.  
Figure \ref{delta_delta} demonstrates
that there is reasonable agreement between the Doppler factors estimated
by these two different methods. The best least-square linear fit to the dependence 
is $\delta_{\beta_{\rm app}}=(0.72\pm 0.15)\delta_{\rm var}$,
consistent with $\delta_{\rm var}$ being the true value and $\delta_{\beta_{\rm app}}$
being a lower limit taking in account that $\delta_{\beta_{\rm app}}$ is estimated
for the lowest possible Lorentz factor. We conclude that the combination of
variability of flux and measurement of angular sizes provides a new, robust
method for deriving Doppler factors from well-sampled sequences of VLBI data.

Tables \ref{Q_Param}, \ref{B_Param}, and \ref{G_Param} give estimates of 
the Lorentz factor, viewing angle, Doppler factor, observed brightness 
temperature, $T_{\rm b,obs}$, and intrinsic brightness temperature, 
$T_{\rm b,int}$, for 43 knots in the quasars, 19 knots in the BL Lacs, and 
15 knots in the radio galaxies. The Lorentz factor and viewing
angle are solutions of the system that combines equations (\ref{e3}) and (\ref{e4})
for a knot having measured apparent speed with the Doppler factor derived via equation 
(\ref{e5}). The observed brightness
temperature is computed based on the VLBI measurements as 
$T_{\rm b,obs}=7.5\times 10^8\;S_{\rm max}/s^2$~K, where $S_{\rm max}$ is 
the maximum flux of the component in Jy and $s$ is its measured angular size 
in mas at the epoch of maximum flux (see above). We apply the Doppler factors derived with 
equation (\ref{e5}) to estimate the intrinsic brightness temperature, 
$T_{\rm b,int}= T_{\rm b,obs}\;(1+z)^{1.7}/\delta^{1.7}$, where we adopt 
$\alpha=0.7$.

\subsection{Opening Angle of the Jet} 
\begin{figure}
\epsscale{1.1}
\vspace{-3cm}
\plotone{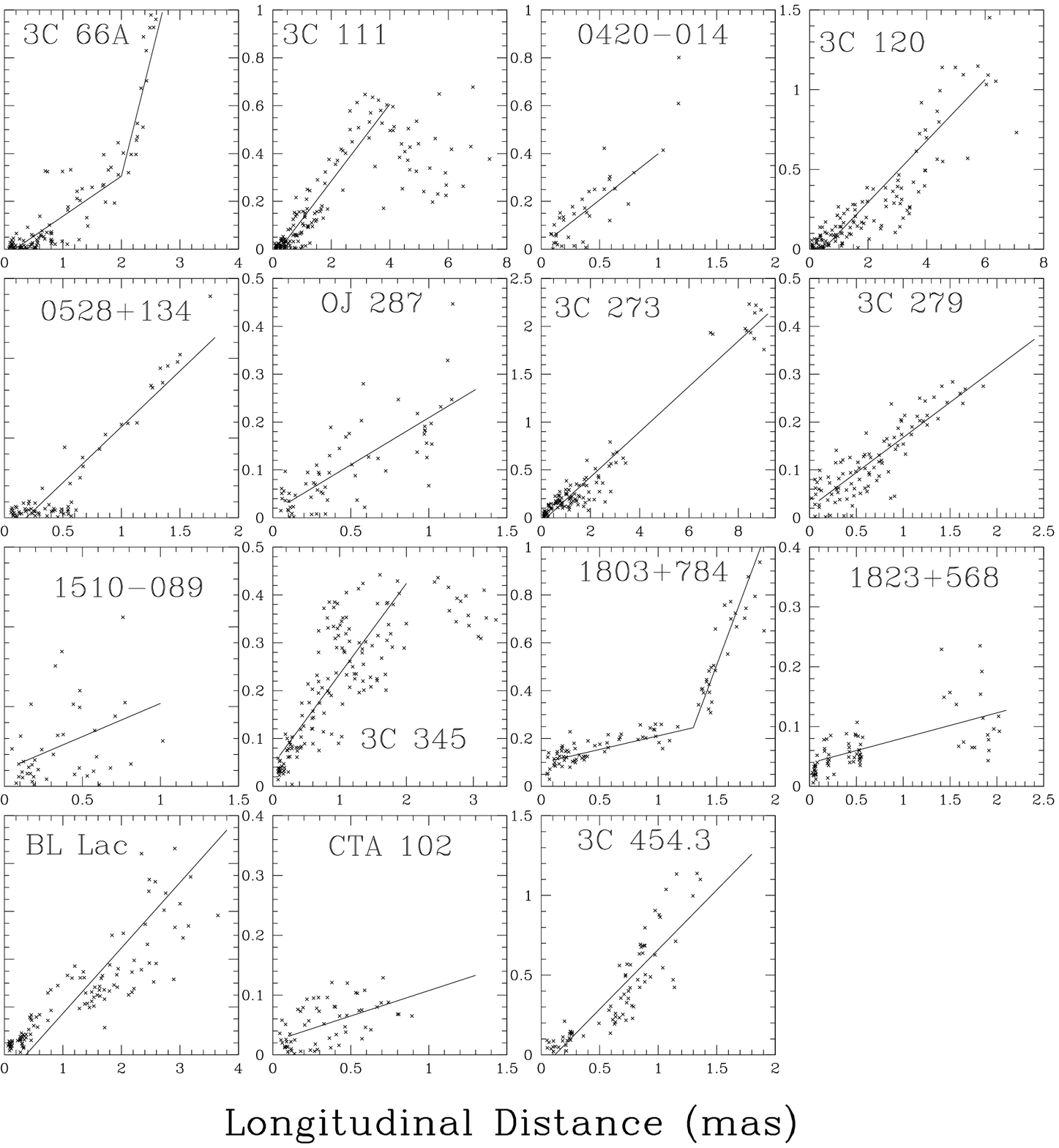}
\vspace{-5cm}
\caption{Projected transverse size of the jet vs. longitudinal distance from the core.
The solid lines repesent the best linear fit to the data. \label{Opan}}
\end{figure}

We assume that jets have conical structure and constant angle, $\theta$,
between the jet axis and surface of the cone that contains the entire 
region of emission. Therefore, $\theta$ is the actual half opening angle of the jet.
This assumption is likely correct for the section of the 
jet within 1-2~mas from the core, while at larger distances ($\ge$10~mas) 
the jet could trace out a helical structure covering a wide range of position angles
even if the body of the jet remains narrow \citep[e.g.,][]{LK03}.  
We estimate the projected half opening angle, $\theta_{\rm p}$, for each source
using the ratio between apparent transverse size, $s_{\rm t}$, of the jet and 
apparent longitudinal distance, $s_{\rm l}$, of components:
$\theta_{\rm p}=\tan^{-1}\;\psi$, where $\psi$ is the slope 
of the best linear fit to the relation between $s_{\rm t}$ and $s_{\rm l}$ as defined 
at the position of each component that is brighter than 1\% of the peak intensity. 
The values of $s_{\rm t}$ and $s_{\rm l}$ are calculated as follows: $s_{\rm l}=R$, 
where $R$ 
is the observed separation of the component from the core; and 
$s_{\rm t}=R\;\sin\;(|\Theta_{\rm jet}-\Theta|)+a/2$, where $\Theta_{\rm jet}$ is the 
projected direction of the jet as defined by the mean position angle of all
sufficiently bright components over all epochs, $\Theta$ is the position
angle of the component, and $a$ is the size of the component. We first plot 
the values of $s_{\rm t}$ against $s_{\rm l}$ to estimate a preliminary linear dependence. 
The preliminary relationship is used to remove 
points that deviate from the dependence by more than 3$\sigma$  
of the linear model. Then the value of $\psi$ is obtained by minimizing
$\chi^2$ of the adjusted data. In Figure \ref{Opan} all pairs of $(s_{\rm t},s_{\rm l})$ are 
plotted and the best linear fit of the adjusted data is presented for each source. 
Figure \ref{Opan} shows that a jet model with constant opening angle provides
a good approximation to the inner jet, although the plots of two BL~Lac objects
(3C~66A and 1803+784) display an increase in $\theta_{\rm p}$ beyond 2 and 1~mas,
respectively, which might be a general feature of many BL~Lac jets. In the radio 
galaxy 3C~111 and quasar 3C~345 a decrease in $\theta_{\rm p}$ is observed, 
and is most likely caused by the weakness and diffuse nature of components at large 
distance from the 
core, such that our images only contain portions of these features.
Table \ref{Source_par} gives the parameters $\Theta_{\rm jet}$, $\theta_{\rm p}$, 
$<\Theta_\circ>$,
$\theta$, $<\Gamma>$, and $<\delta>$ for each source. Parameters $<\Gamma>$, 
$<\delta>$, and $<\Theta_\circ>$ are weighted averages of the values for individual
components listed in Tables \ref{Q_Param}, \ref{B_Param}, and \ref{G_Param}, with 
weights inversely proportional to the uncertainty in apparent speed. The intrinsic
half opening angle, $\theta$, is estimated as 
$\theta=\theta_{\rm p}\;\sin<\Theta_\circ>$, where $<\Theta_\circ>$ is the angle 
between the jet axis and the line of sight.

\subsection{Jet Parameters}
\begin{figure}
\epsscale{1.0}
\plotone{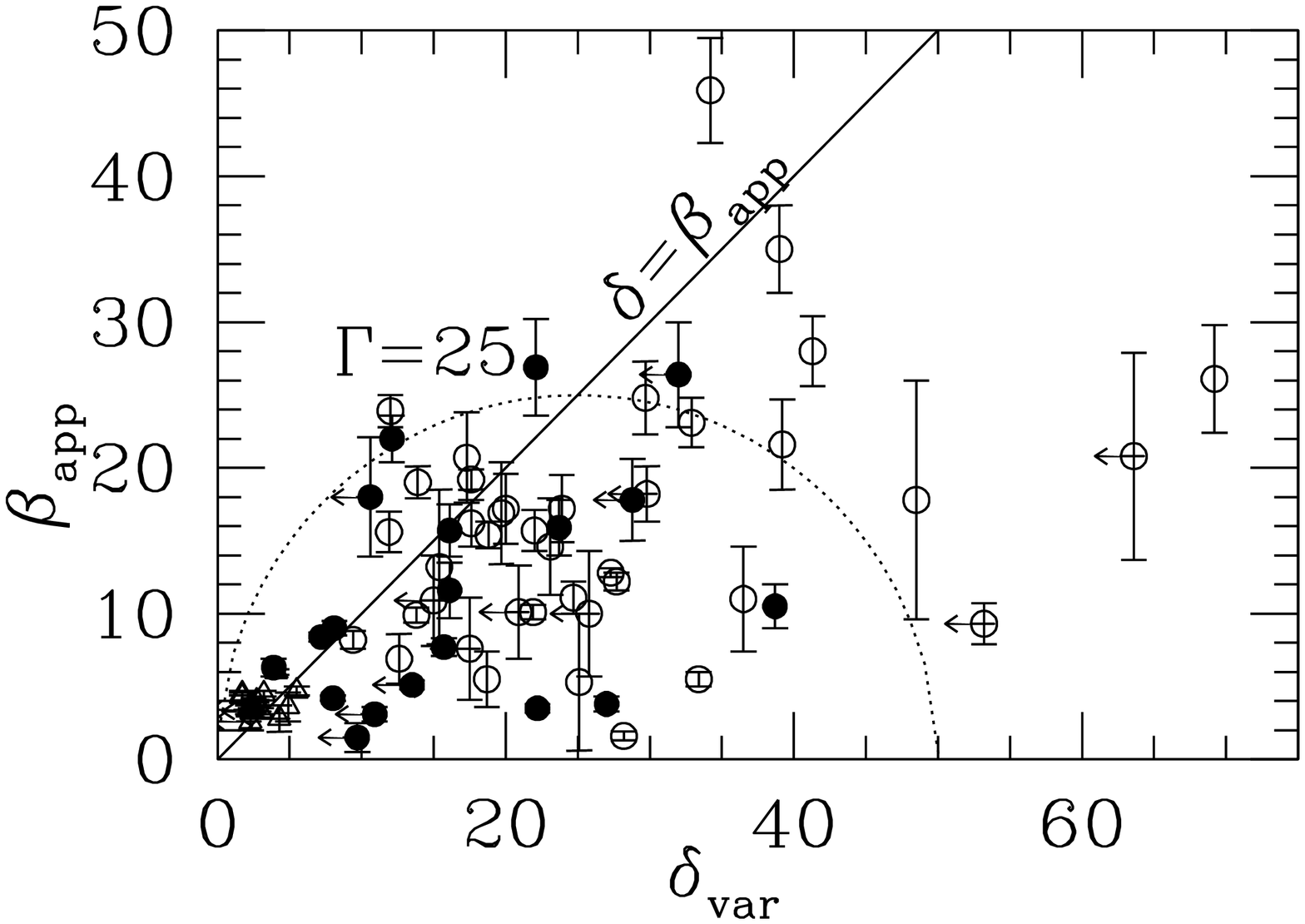}
\caption{Apparent speed vs. variability Doppler factor for superluminal
components. The solid line denotes $\delta_{\rm var}=\beta_{\rm app}$.
The dotted curve encloses the expected location of points corresponding to 
$\Gamma\leq 25$.\label{V_Doppler}}
\end{figure}

\begin{figure}
\epsscale{1.0}
\plotone{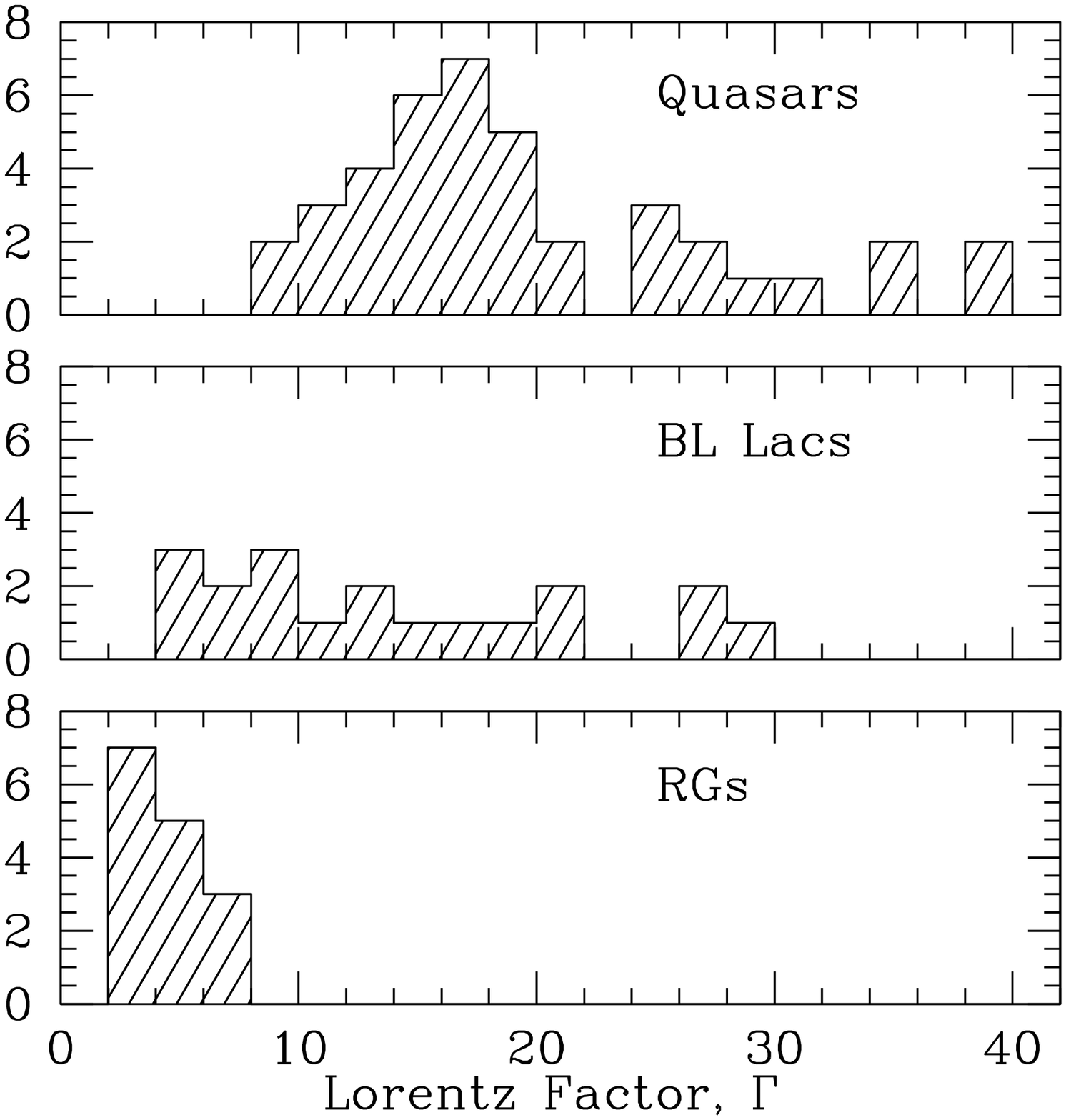}
\caption{Distribution of Lorentz factors of superluminal jet components in the 8 quasars,
5 BL~Lac objects, and 2 radio galaxies.\label{h_Gamma}}
\end{figure}

We have constructed the dependence between the Doppler beaming factor, $\delta_{\rm var}$,
and apparent speed (Fig. \ref{V_Doppler}) to compare derived parameters
with results from the 2-cm VLBA survey \citep{KL04}. These authors determined
Doppler factors using the variability method of \citet{LV99}, with an intrinsic
brightness temperature of $2\times 10^{10}$~K that they assume corresponds to 
the region of the VLBI core where the main variability at radio frequencies 
occurs \citep{SAV02}. In Figure \ref{V_Doppler}
the majority of points follow the expectation that $\beta_{app}\lesssim\delta_{\rm var}$ and 
lie inside the ``$1/ \beta_{app}$ cone''. Moreover, despite our sample consisting 
of blazars or blazar-like sources for which a higher Lorentz factor is expected,
the upper limit to $\beta_{app}$ corresponding to $\Gamma=25$ \citep{KL04}
applies to our sample as well (dotted curve in Fig. \ref{V_Doppler}).
Figure \ref{h_Gamma} shows the distribution of Lorentz factors of superluminal
components. There is a significant scatter in $\Gamma$ for the quasars and BL~Lacs, 
and the Lorentz factors of the radio galaxies are lower than for the other classes. 
The distributions, in general, agree with the distribution of Lorentz factors
found in the 2-cm survey. 

\begin{figure}
\epsscale{1.0}
\plotone{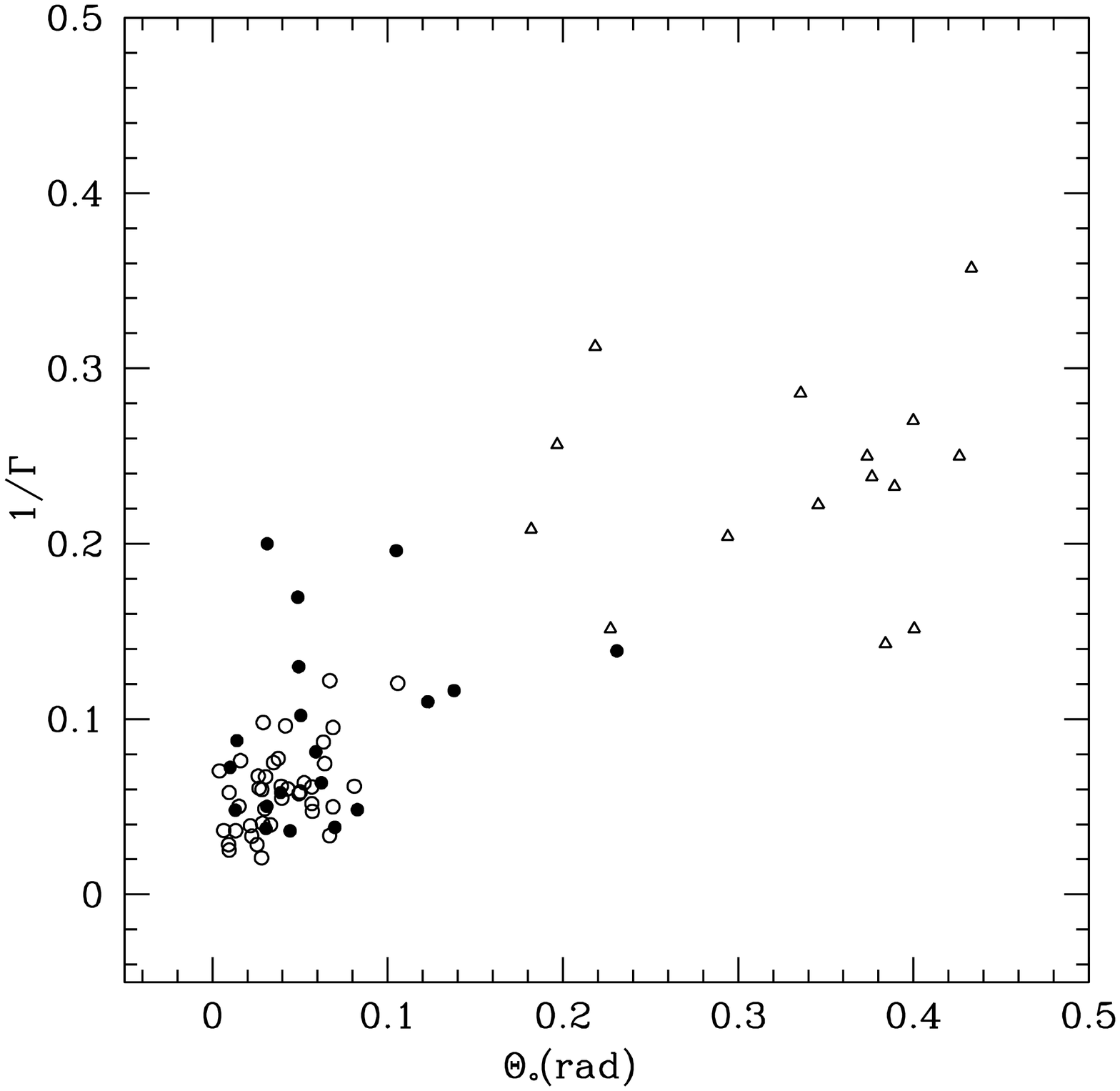}
\caption{Inverse Lorentz factor vs. viewing angle for superluminal jet
components. \label{GT}}
\end{figure}

\begin{figure}
\epsscale{1.0}
\plottwo{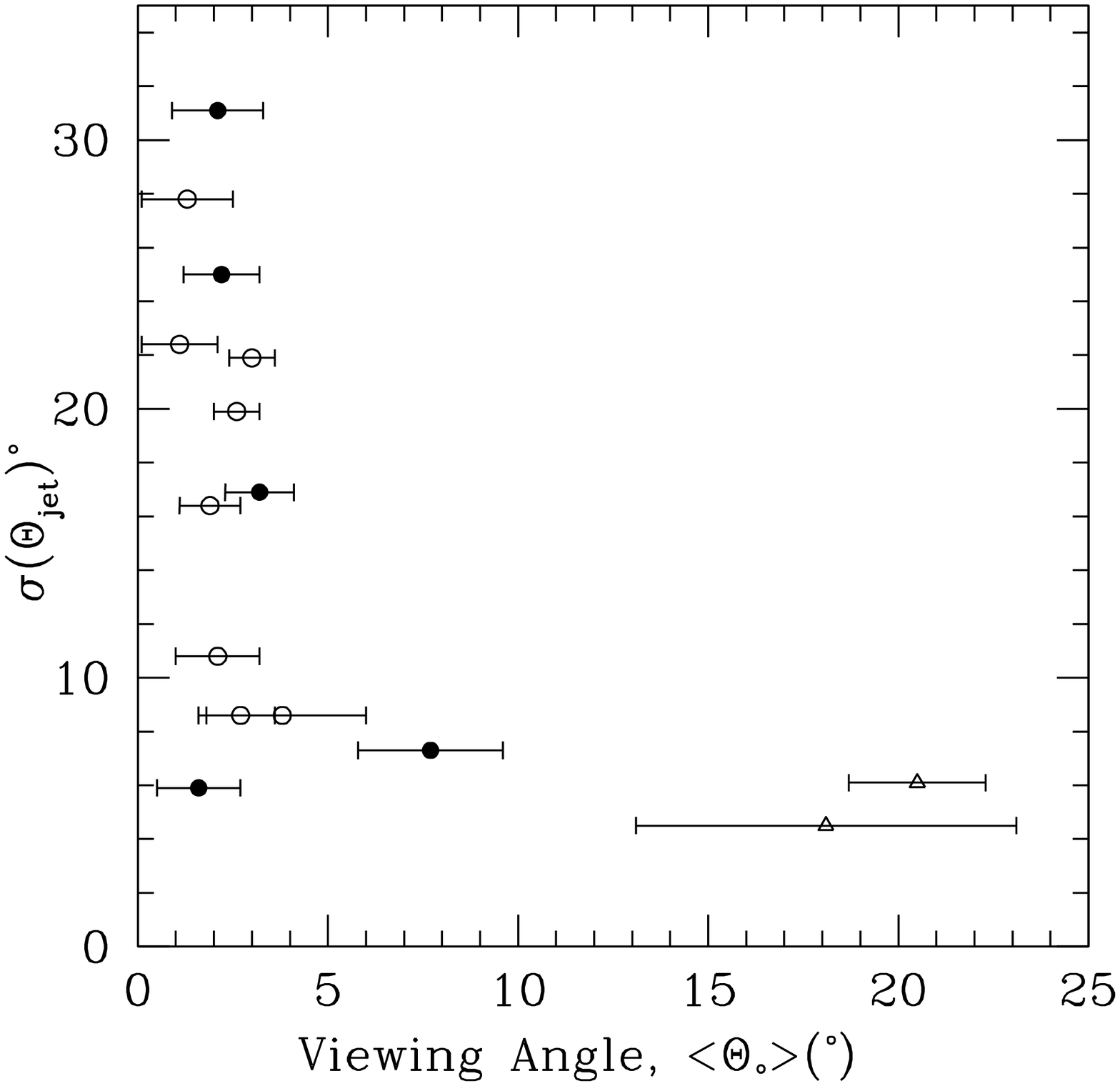}{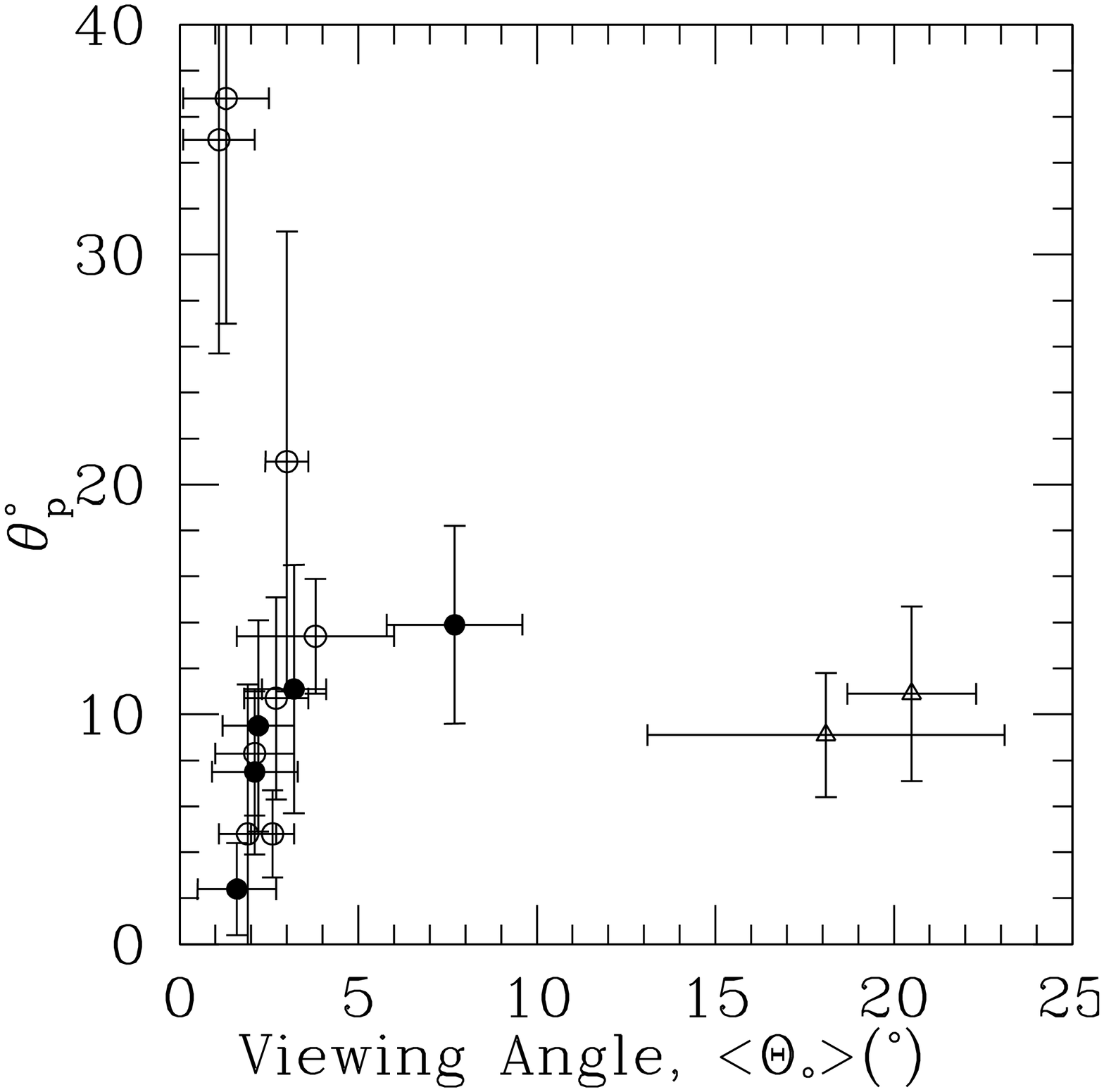}
\caption{{\it Left panel:} Scatter in the projected jet position angle vs.
the average viewing angle of jets. {\it Right panel:} Projected 
half opening angle vs. average viewing angle.  \label{SigmaT}}
\end{figure}

Monte-Carlo simulations for flux limited 
samples of radio sources predict a connection - although not direct correlation - 
between the Lorentz factor and viewing angle of the jets: the higher the Lorentz factor, 
the smaller is the viewing angle \citep{LM97}.
Our sample of the blazars and blazar-like sources shows a significant
correlation between 1/$\Gamma$ and $\Theta_\circ$ (coefficient of correlation 0.83, 
Fig. \ref{GT}); this supports the above prediction. Small intrinsic bends in jets 
oriented very close to the line of sight should be greatly amplified in 
projection on the sky \citep{R78}. 
The expected morphology has not been confirmed by VLBI surveys. For example, 
$\gamma$-ray blazars, which exhibit the highest 
apparent speeds \citep{J01,KL04}, do not possess more pronounced bends than 
sources not yet detected in $\gamma$-rays \citep{KL98,J01}. Estimates of the viewing angle 
obtained for our sample allow us to test the relation between morphology
and viewing angle of the jets more directly. We have 
determined the average projected position angle of the jet, $\Theta_{\rm jet}$, and 
its standard deviation, $\sigma(\Theta_{\rm jet})$, for each source (see Table 
\ref{Source_par}). In sources with jet direction close to 
the line of sight, a small intrinsic change of the component trajectory 
should result in a significant bend in the projected path, leading 
to a large scatter in $\Theta_{\rm jet}$ and yielding a high value of 
$\sigma(\Theta_{\rm jet})$. A similar test is to compare the projected opening angle 
with the viewing angle of the jets: jets viewed very close to the 
line of sight should have a wide projected opening angle on VLBI maps.
Figure \ref{SigmaT} presents 
both plots. The left panel shows the relationship between $\Theta_\circ$ and 
$\sigma(\Theta_{\rm jet})$. Although there is a general decrease 
in $\sigma(\Theta_{\rm jet})$ with increasing angle between the jet axis and line 
of sight (coefficient of correlation $-$0.58), the relationship is very weak
or absent for viewing angles $\lesssim 5^\circ$ (coefficient of correlation $-$0.18).
The right panel shows that there is no 
connection between the projected opening angle and viewing angle of the jet
(coefficient of correlation $-$0.16). This result
implies that amplification of the projected size by a small angle between the 
jet axis and line of sight is partly (or sometimes completely) canceled by
the proportionality between opening angle and $1/\Gamma$ (see below). Possible
reasons behind the lack of correlation between degree of apparent bending
and small viewing angle include: (1) an inverse relationship between bulk
Lorentz factor and degree of intrinsic bending, given that the momentum of
the jet is proportional to $\Gamma$, (2) broader opening angles in slower
jets, with filamentary structure appearing similar to bending of the axis, 
and (3) the resolution available transverse to the jet not being sufficient to
detect bending in many high-$\Gamma$ objects.   
 
\begin{figure}
\epsscale{1.0}
\plotone{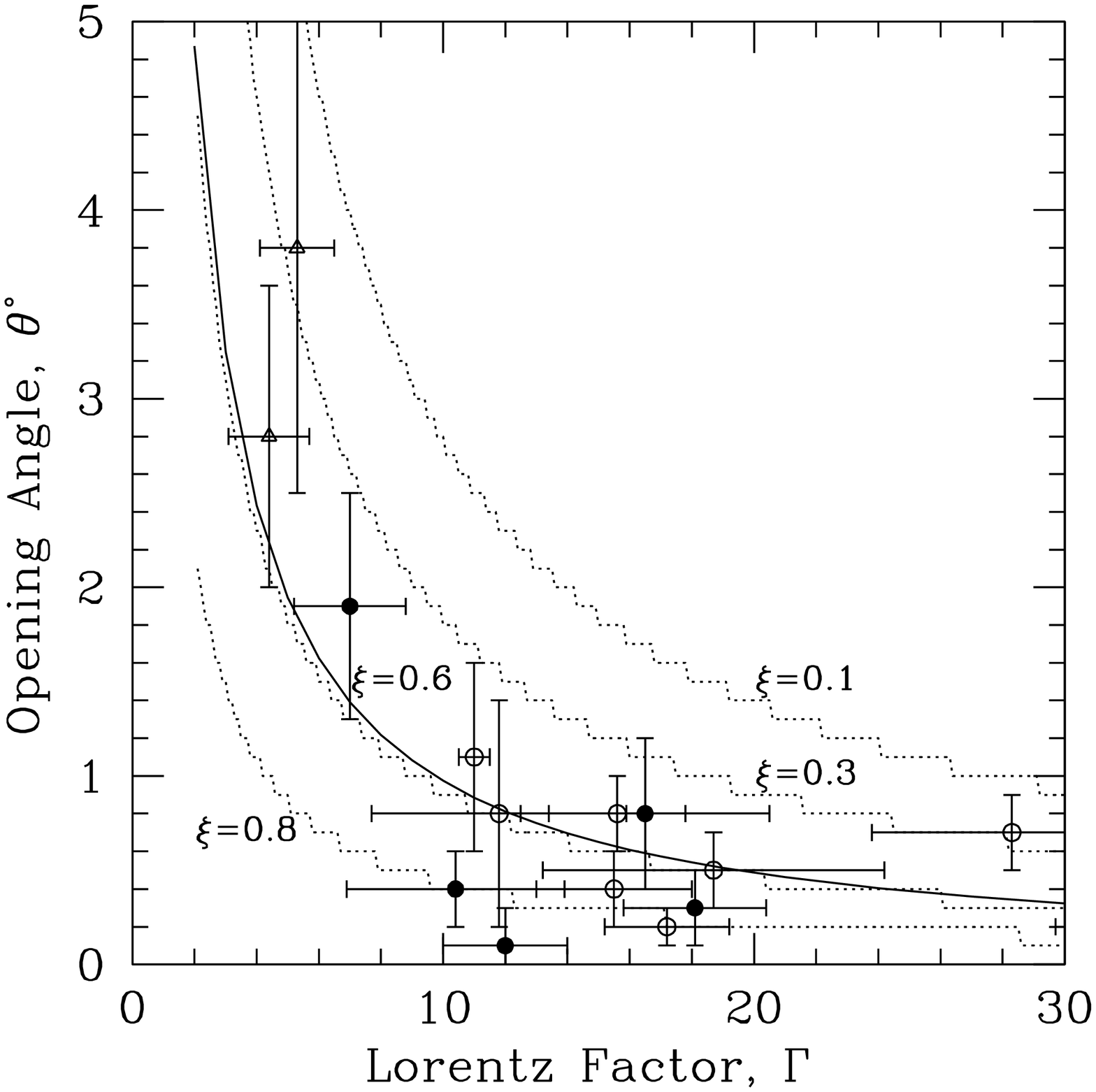}
\caption{Half opening angle vs. Lorentz factor of the jets. The solid curve
corresponds to the best approximation according to the $\chi^2$ test 
of the {\it observed} dependence within the assumed law $\theta=\rho/\Gamma$, where
$\rho$ is a constant. The dotted curves represent relationships between the half
opening angle and Lorentz factors expected in the gas-dynamic model
for different values of the parameter $\xi$.
\label{Open_G}}
\end{figure}

The Lorentz factor plays a significant role in determining the jet geometry. 
According to standard models of relativistic jets, the opening angle of the jet 
should be inversely proportional to the Lorentz factor \citep[e.g.,][]{BK79}. 
In Figure \ref{Open_G}, the estimated half opening angles are plotted versus 
the derived Lorentz factors. There is an obvious decrease of $\theta$ toward higher 
values of $\Gamma$. According to a $\chi^2$ test the {\it observed} dependence 
can be described by the relation $\theta\approx\rho/\Gamma$~rad, where 
$\rho=0.17\pm 0.08$ (solid line in Fig. \ref{Open_G}) at a 2.5\% level
of significance ($\chi^2$=6.875, f=2). The current 
models for the formation of relativistic jets that employ confinement of
a jet by magnetic forces \citep[e.g.][]{Meier00,VK04} do not yet specify
the opening angle of the jet as a function of the model parameters. 
Such an expression, however, has been derived based on
gas dynamics for a 
relativistic jet confined by pressure equilibrium with its surroundings 
\citep{DM88}. In this model the opening angle depends on the Lorentz 
factor of the flow and the ratio of the external pressure, $P_{ext}$,
to the initial pressure, $P_\circ$, of the plasma in the core region, 
$\xi=\sqrt{P_{ext}/P_\circ}$.
Using equations (18), (19a), and (19b) in \citet{DM88}, we calculate 
dependences of the opening angle on
Lorentz factor for different values of $\xi$ (dotted curves in Fig. \ref{Open_G}). 
The best fit of the {\it observed} dependence (5\% level of significance)
coincides with the model in which $\xi=0.6$ ($\chi^2$=4.586, f=2), corresponding 
to a pressure ratio in the core region $P_{ext}/P_\circ\approx 1/3$. 

\begin{figure}
\epsscale{1.0}
\plottwo{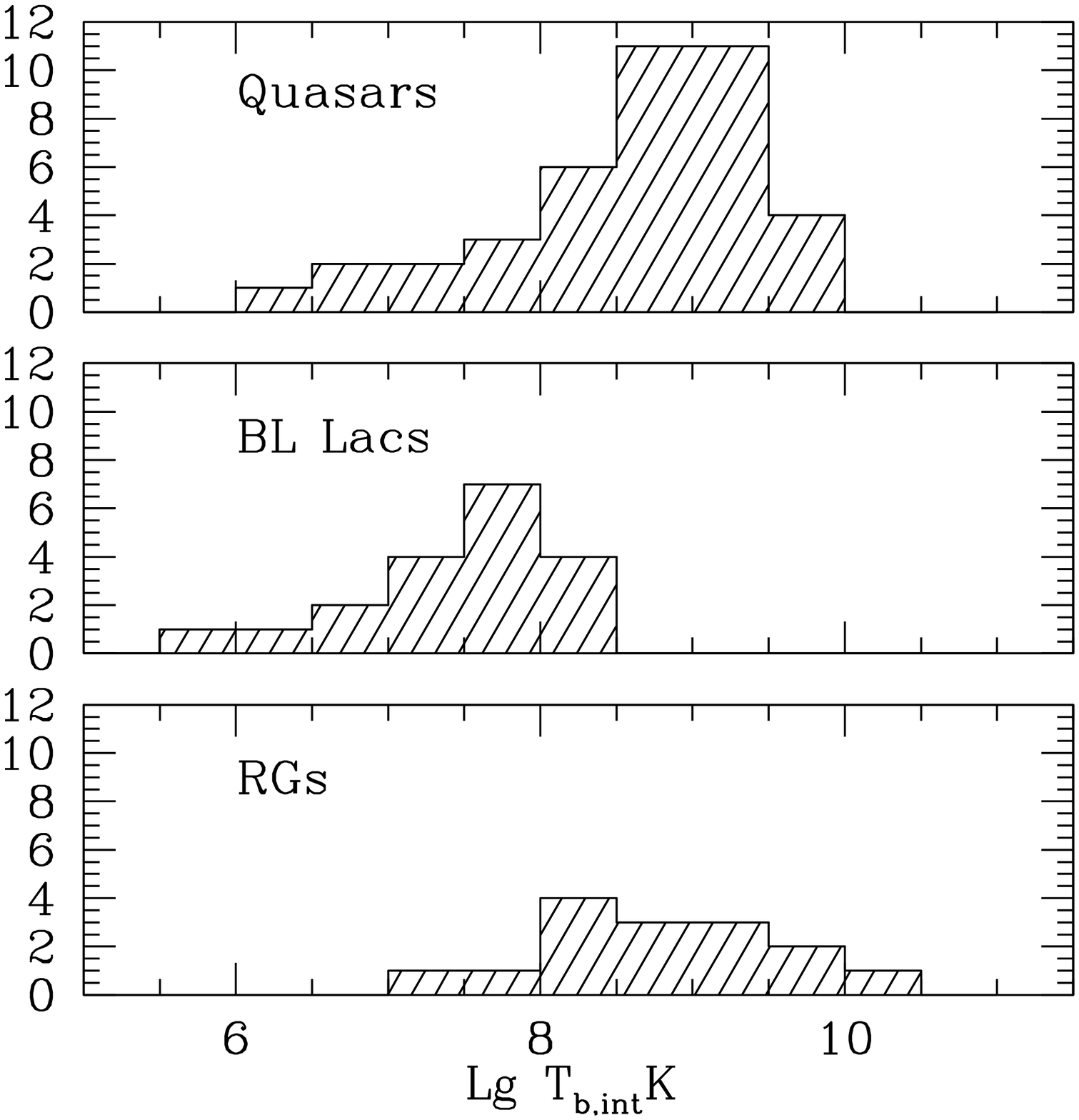}{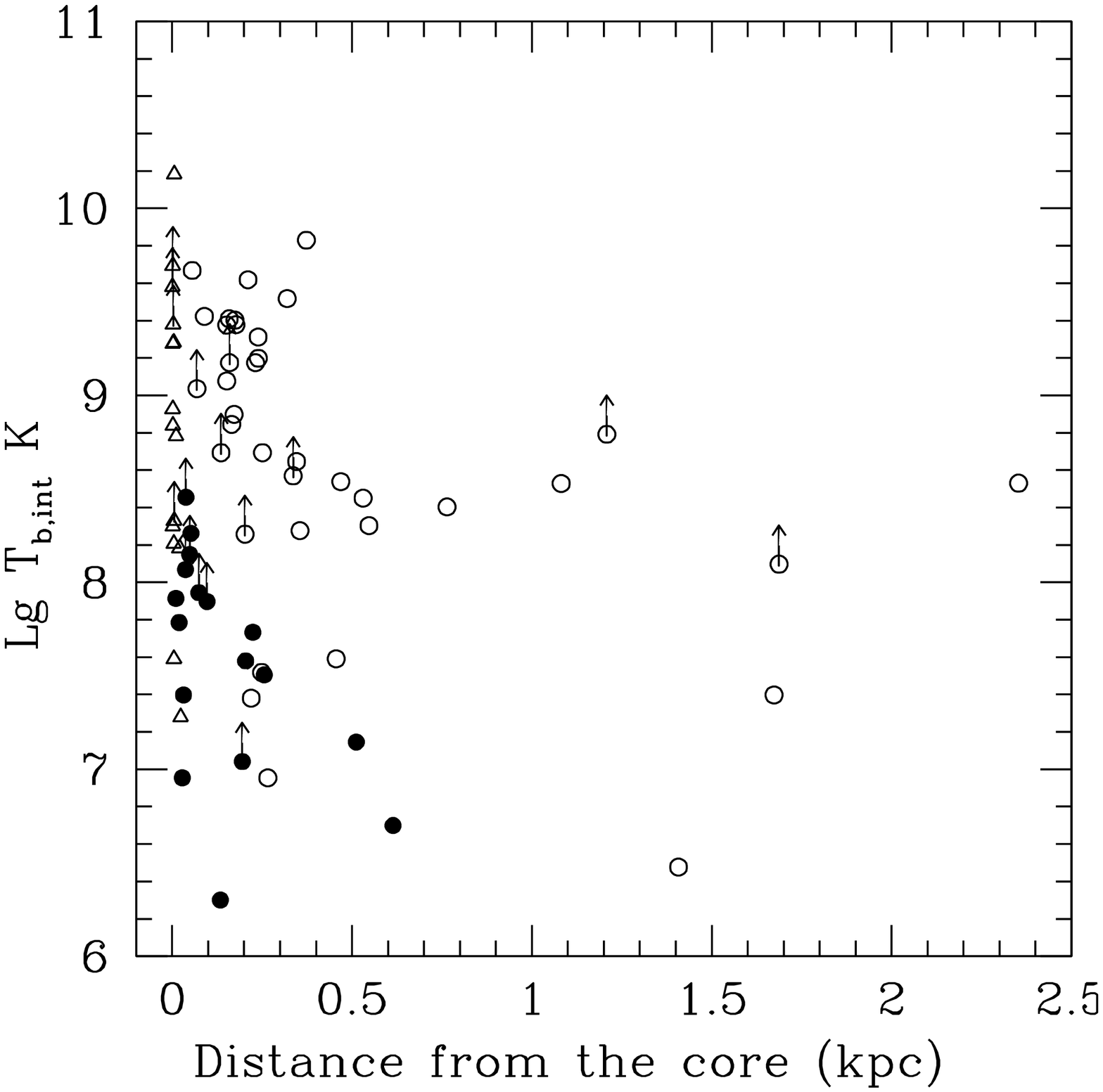}
\caption{{\it Left panel:} Distribution of intrinsic brightness temperature
of the superluminal jet components in 8 quasars, 5 BL~Lac objects,
and 2 radio galaxies. {\it Right panel:} Intrinsic brightness temperature
vs. deprojected distance from the core.  \label{TI}}
\end{figure}

Figure \ref{TI} presents the distribution of intrinsic brightness temperature of 
the jet components. Note that $T_{\rm b,int}$ is measured at distances ranging
from several parsecs to a few kiloparsecs from the core (see Fig. \ref{TI},
{\it right panel}), where the knots are generally optically thin. The distribution
peaks at $T_{\rm b,int}\sim 2\times 10^9$~K for the quasar knots and $T_{\rm b,int}
\sim 6\times 10^7$~K for the BL~Lac knots, while for the radio galaxies
the temperatures are evenly distributed between these values.
The maxima should be adjusted to slightly higher temperatures since for 30\%
of the quasar and BL~Lac components we have obtained only upper limits to the Doppler
factors. Comparison of these brightness temperatures with
the intrinsic equipartion brightness temperature of the optically thick part 
of the jet ($T_{\rm b,int}\sim 2-5\times 10^{10}$~K), thought to depend
weakly on source parameters \citep{Readhead,LV99}, implies a faster drop of the
intrinsic temperature from the compact core region to the more extended structure 
in the BL~Lac jets. One possible explanation for a lower intrinsic brightness
temperature is the presence of a stronger magnetic field \citep{Readhead}. This
would lead to more severe radiative losses and weaker jet components relative to 
the core in the BL~Lac objects, and perhaps a lower intrinsic brightness temperature 
in the cores as well.  

\begin{figure}
\epsscale{1.0}
\plotone{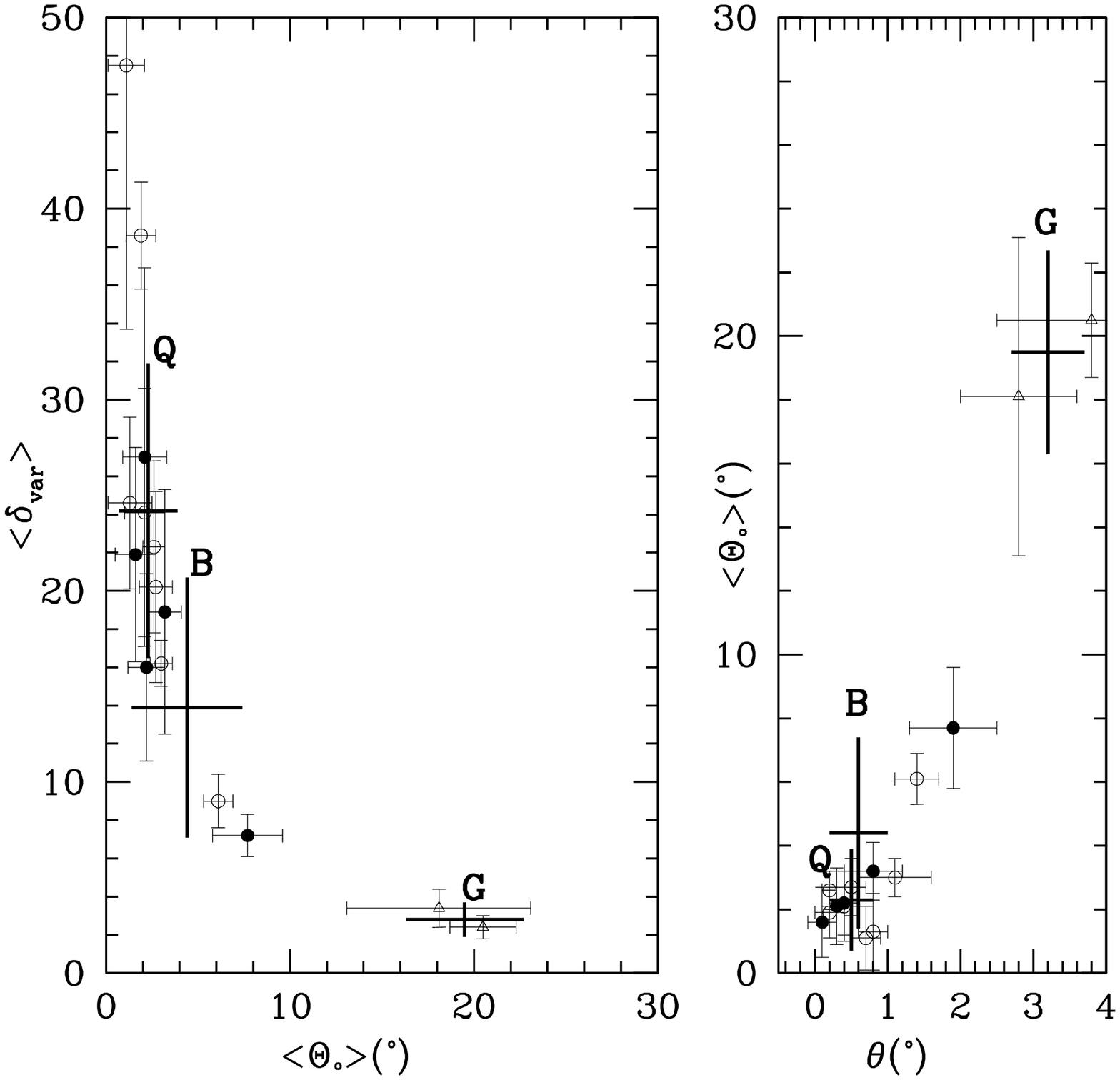}
\caption{{\it Left panel:} Average Doppler factor vs. average viewing angle of the jets.
{\it Right panel:} Average viewing angle vs.
intrinsic half opening angle of the jets. In both
panels the bold crosses indicate average parameters and their standard deviations for each
subclass.\label{DT}}
\end{figure}

In Figure \ref{DT} we plot the average Doppler factor vs. average viewing angle 
({\it left panel}), and the average viewing angle vs.
the intrinsic half opening angle of the jet for each source ({\it right panel}).
The bold crosses indicate the average position of the quasars, BL~Lac objects,
and radio galaxies, with the size of each cross equal to the 1$\sigma$ uncertainty
of the parameters. Although each subclass has a large scatter around the average 
values and statistically the difference in the parameters of the quasars and 
BL~Lac objects is negligible, the points form a continuous sequence on a 3-D plot of
$\delta$, $\Theta_\circ$, $\theta$ with:
quasars ($<\delta>$=23$\pm$11, $<\Theta_\circ>$=2.6$^\circ\pm$1.9$^\circ$, $\theta$=0.5$^\circ
\pm$0.3$^\circ$) 
$\longrightarrow$ BL~Lacs ($<\delta>$=13.5$\pm$6.7, $<\Theta_\circ>$=4.4$^\circ\pm$3.0$^\circ$,
$\theta$=0.6$^\circ\pm$0.4$^\circ$) $\longrightarrow$ radio galaxies
($<\delta>$=2.8$\pm$0.9, $<\Theta_\circ>$=19.5$^\circ\pm$3.2$^\circ$, $\theta$=3.2$^\circ
\pm$0.5$^\circ$).

\subsection{Accelerating/Decelerating Flow}
\begin{figure}
\epsscale{1.0}
\plotone{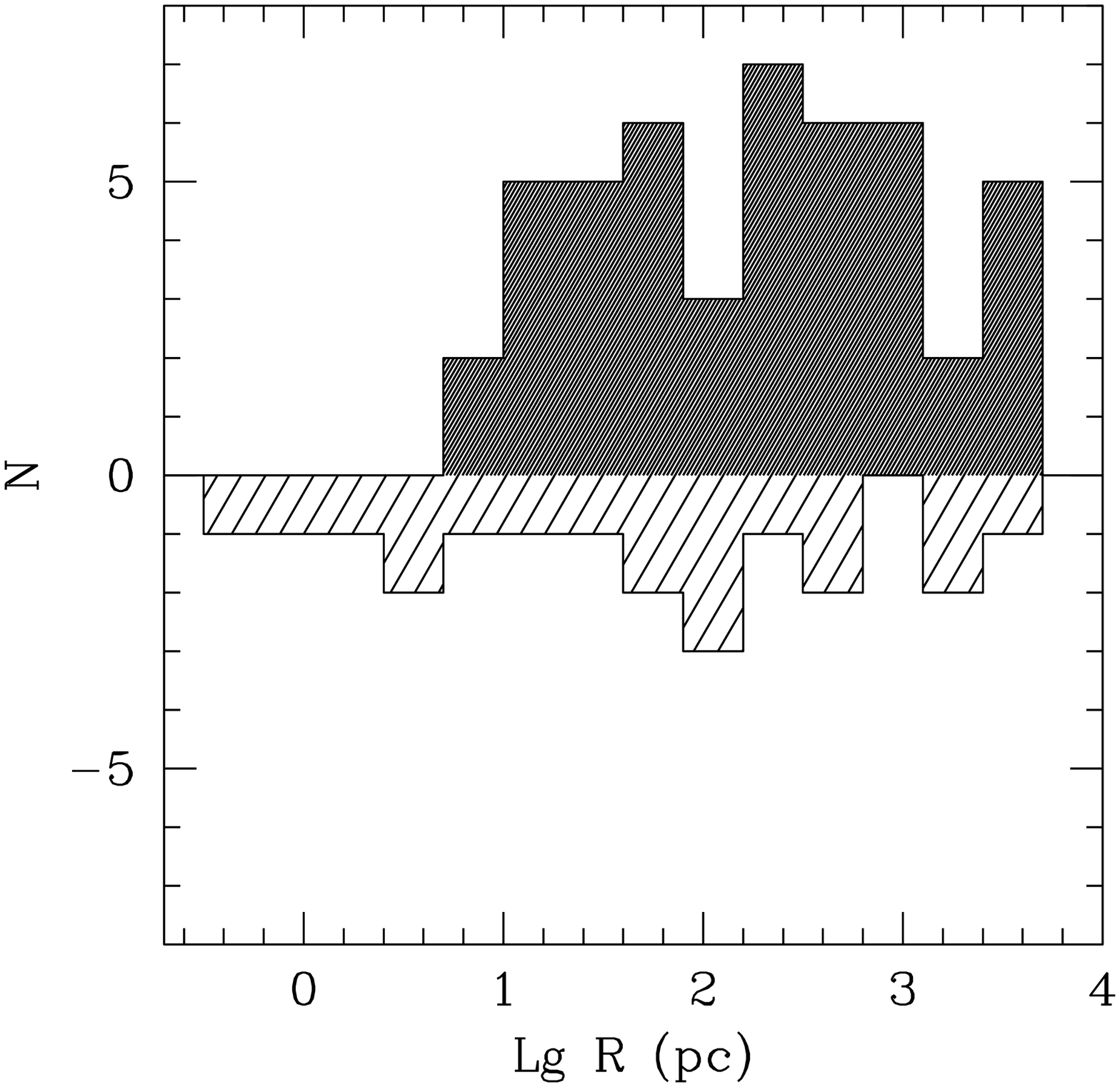}
\caption{Distribution of change of the instantaneous apparent speed relative to 
the average apparent speed with distance from the core (deprojected). A value of +1 is
assigned if the instantaneous apparent speed is higher than the average, and
$-1$ if it is lower (see text).\label{h_Vchange}}
\end{figure}

According to the $\chi^2$ test, one component in the radio galaxy 3C~120, 
five components in two BL Lac objects, and 13 components in six quasars  
exhibit a statistically significant change in the proper motion
with time corresponding to acceleration/deceleration with distance from the core. 
Therefore, in 9 out of 15 sources the apparent speed of individual
components varies. Moreover, although 86 knots are classified
as moving ballistically, $\sim 38\%$ have trajectories that we
are unable to fit well by a polynomial of any order, causing
us to suspect non-ballistic motion for these as well. Figure \ref{V_Change} 
shows the evolution of apparent velocity, including direction along the jet, 
derived from the best polynomial approximation of component positions across 
epochs for knots with a detected change in apparent speed.
We have analyzed the results to search for a general trend in the variation 
of apparent speed with distance from the core. For each component, we compare 
the instantaneous apparent speed at different distances from the core with the 
average (indicated in Table \ref{Speed}), and assign a value representing the 
apparent speed at each angular distance of $+1$ if the velocity is 
higher than the average, $-1$ if it is lower than the average, and $0$ if it 
equals the average. Then we construct the distribution of such changes along 
the jet using the derived viewing angle for each component 
(Tables \ref{Q_Param}-\ref{G_Param}), which allows deprojection of observed 
distances from the core.
The result is shown in Figure \ref{h_Vchange} where the distribution above the
$x$-axis represents accelerating components and the distribution under the $x$-axis
indicates decelerating components at corresponding deprojected distances
from the core. Figure \ref{h_Vchange} reveals that at distances larger than
$\sim$5~pc from the core an increase of apparent speed is more common.  
However, there are some 
jet components that undergo alternating periods of both acceleration and deceleration. 
The latter could be a signature of helical motion
\citep[e.g.,][]{D00,TK04} or pinch instabilities that cause the jet 
cross-section, pressure, and Lorentz factor to oscillate with distance
from the core \citep[e.g.,][]{GO97}.

The cause of the accelerations could be bending combined with selection of
objects with high Doppler factors such that the mean angle to the line of 
sight of the region near the core is less than optimal for
superluminal motion. Statistically, such jets are more likely to bend 
away from the line of sight, thus increasing their apparent speeds with
distance from the core. Alternatively, the acceleration could be physical, 
caused by considerably higher energy density in relativistic particles than 
in rest mass \citep{DM88} or magnetic acceleration \citep{VK04}.

\subsection{Forward and Reverse Shocks}
Our data reveal the coexistence of both very fast and much slower (but moving) knots 
in the jets of the majority of the sources in the sample (see Fig. \ref{Ident}).
Diversity in the apparent speeds of jet features might reflect intrinsic variations
in the pattern speed of disturbances in the jet flow. In the shock wave model for interpretation
of the radio light curves and features propagating down a relativistic jet,
a disturbance such as an increase in velocity or energy flux in a jet 
can create both forward and reverse shocks \citep[e.g,][]{HAA91}. Both move away from 
the central engine, 
but the reverse shock has a lower velocity than does the forward shock. 
In this context fast jet features can be associated with the 
forward shock and slow moving knots with the reverse shock. 
Analysis of the brightness of these features might reveal a prevalence
of different types of shock waves in the jets for different classes of AGNs. 

\begin{figure}
\epsscale{1.0}
\plottwo{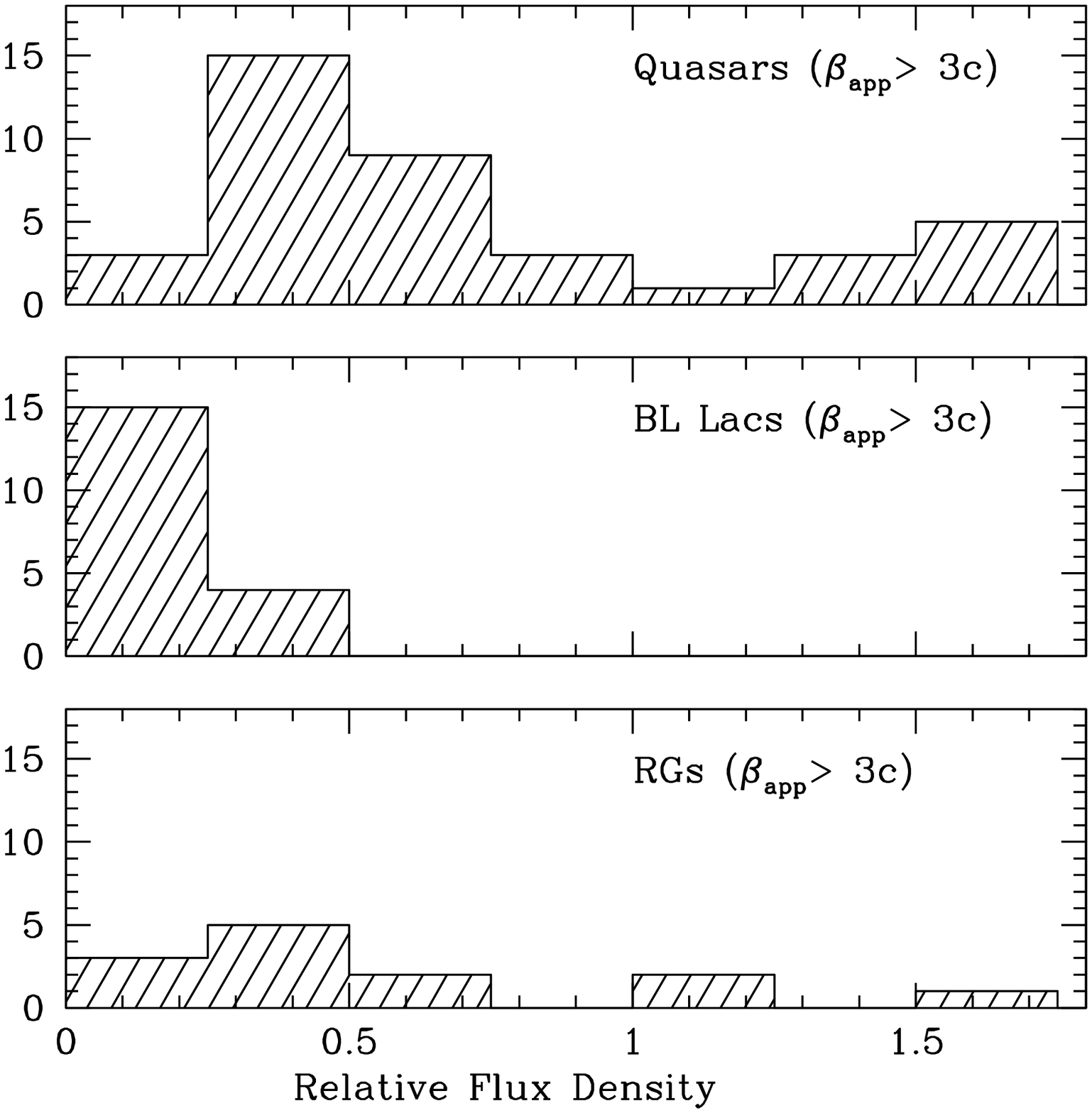}{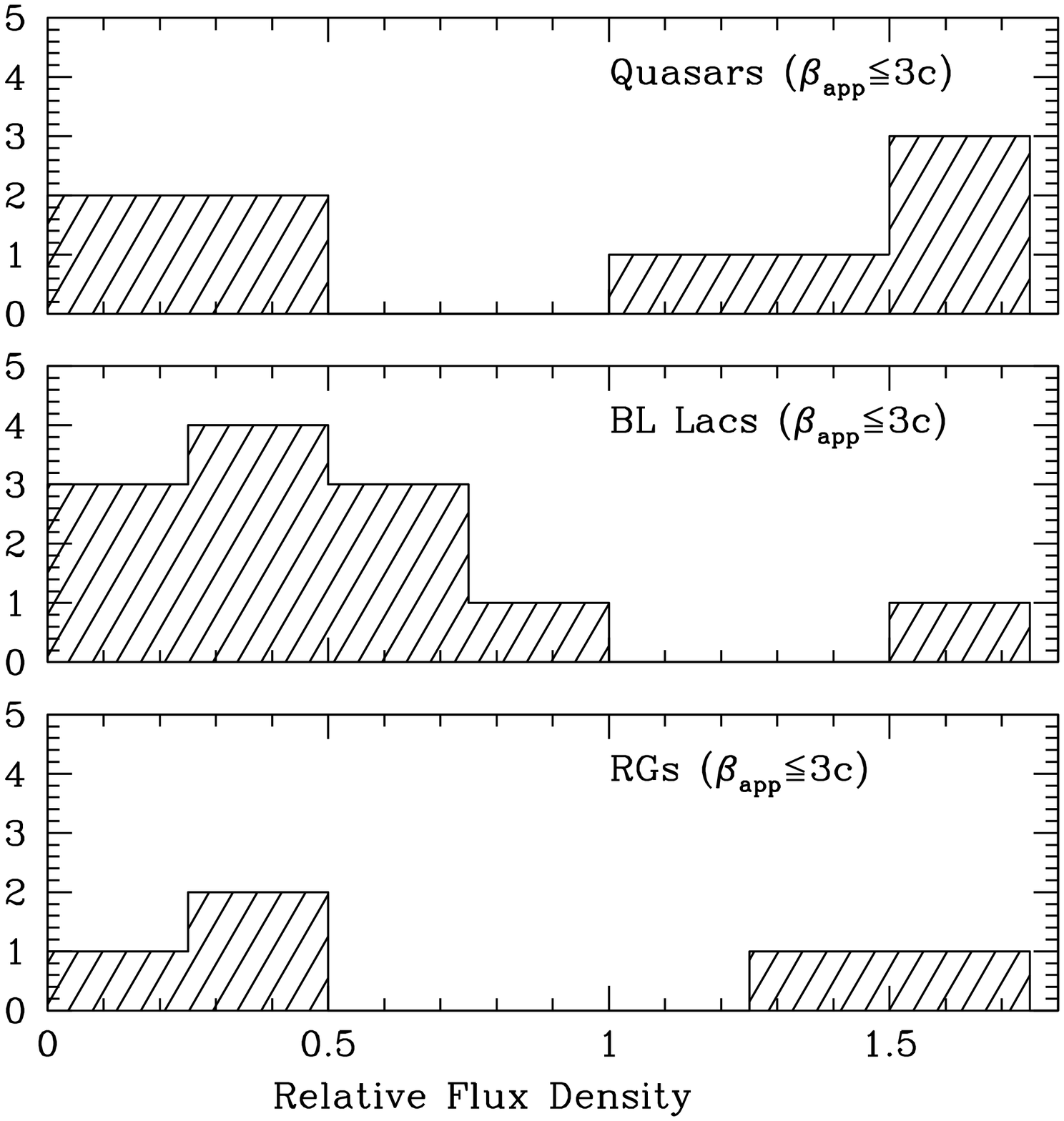}
\caption{Distributions of the flux density of the jet components relative to the
core - {\it left panel:}
for fast moving knots; {\it right panel:} for slow moving knots. \label{h_FR}}
\end{figure}

For each knot in Table \ref{Speed} (except knots classified as trailing features; see \S 6.6)
we have computed parameter $F_{\rm rel}=S_{\rm max}/<S_{\rm A0}>$,
which characterizes the flux of a knot relative to the core, where $S_{\rm max}$ is
the maximum observed flux density of the knot and $<S_{\rm A0}>$ is the average flux density 
of the core over epochs. Figure \ref{h_FR} shows the distributions of the derived
values of this parameter for fast ($\beta_{\rm app}> 3c$) and slow 
($\beta_{\rm app}\leq 3c$) moving knots in the quasars, 
BL~Lac objects, and radio galaxies. The separation into fast and slow features
is not strictly defined, and the distributions do not change significantly if the dividing
velocity is in the range from 2$c$ to $4$c. The distributions of $F_{\rm rel}$
of fast knots in the quasars
and BL~Lacs are different at 99.5\% level of confidence ($f$=5) according to 
the $\chi^2$ test. The brightness of $\sim$50\% of the fast knots in the quasars is 
comparable to the brightness of the core ($F_{\rm rel}\gtrsim$0.5), while 
the brightness of fast knots in the BL~Lac
objects never exceeds half the brightness of the core, and the distribution peaks at 
$F_{\rm rel}\leq$1/4. The distributions of $F_{\rm rel}$ 
in the radio galaxies cannot be classified due to the small number of objects. 

The distributions for slow moving knots indicate the presence of two populations:
bright ($F_{\rm rel}\geq$1) and faint ($F_{\rm rel}\leq$0.5 for the quasars  
and radio galaxies and $F_{\rm rel}<$1 for the BL~Lacs).
The population of bright slow knots consists of jet features with
subluminal apparent speeds that most likely represent stationary shocks in the jets.
This population is most prominent in the quasars. The population of faint slow
knots is prominent in the BL~Lac objects, including 39\% of all identified jet
features, while in the quasars such features comprise only 7\% of detected knots. 
About 70\% of faint slow knots in the BL~Lac objects
have flux within 25\% to 100\% of the flux of the core, significantly brighter 
than fast knots in these objects. 

If fast knots represent forward shocks and slow knots correspond to reverse
shocks, then the forward shock is stronger in the quasars and, perhaps, 
the radio galaxies, while the reverse shock dominates in the jets of BL~Lac objects.
A reverse shock will be strong relative to its corresponding forward shock
when the disturbance is prolonged such that the faster flow enters the rear of the shock 
structure over an extended period of time. Otherwise, only the forward shock
will be prominent and the knot will have a Lorentz factor only slightly ($\lesssim$6\%)
less than that of the forward shock front \citep[e.g., ][]{SOK04}. BL~Lac objects
might therefore have more prolonged disturbances of lower amplitude than those 
in quasars. If jets possess spine-sheath structure \citep{L99}, then an
alternative explanation for the above effect could be that the power
of high-$\Gamma$ spines in BL~Lacs is lower than in quasars.  
The proposal that the
nature of shock waves in quasars is different from that of BL~Lac objects has been 
suggested by \citet{W94}, who interpreted the evolution of the flux and
polarization of the quasar 3C~345 as forward shocks that have greater speed
than the underlying jet. This is in contrast to the BL~Lac object OJ~287, studied by \citet{CW88},
who modeled the polarization, kinematics, and X-ray variability as reverse shocks,
which are slower than the underlying jet. Our data support the idea
that this could be a key difference between the two classes of blazars.

\subsection{Trailing Shocks} 
\begin{figure}
\epsscale{1.0}
\plotone{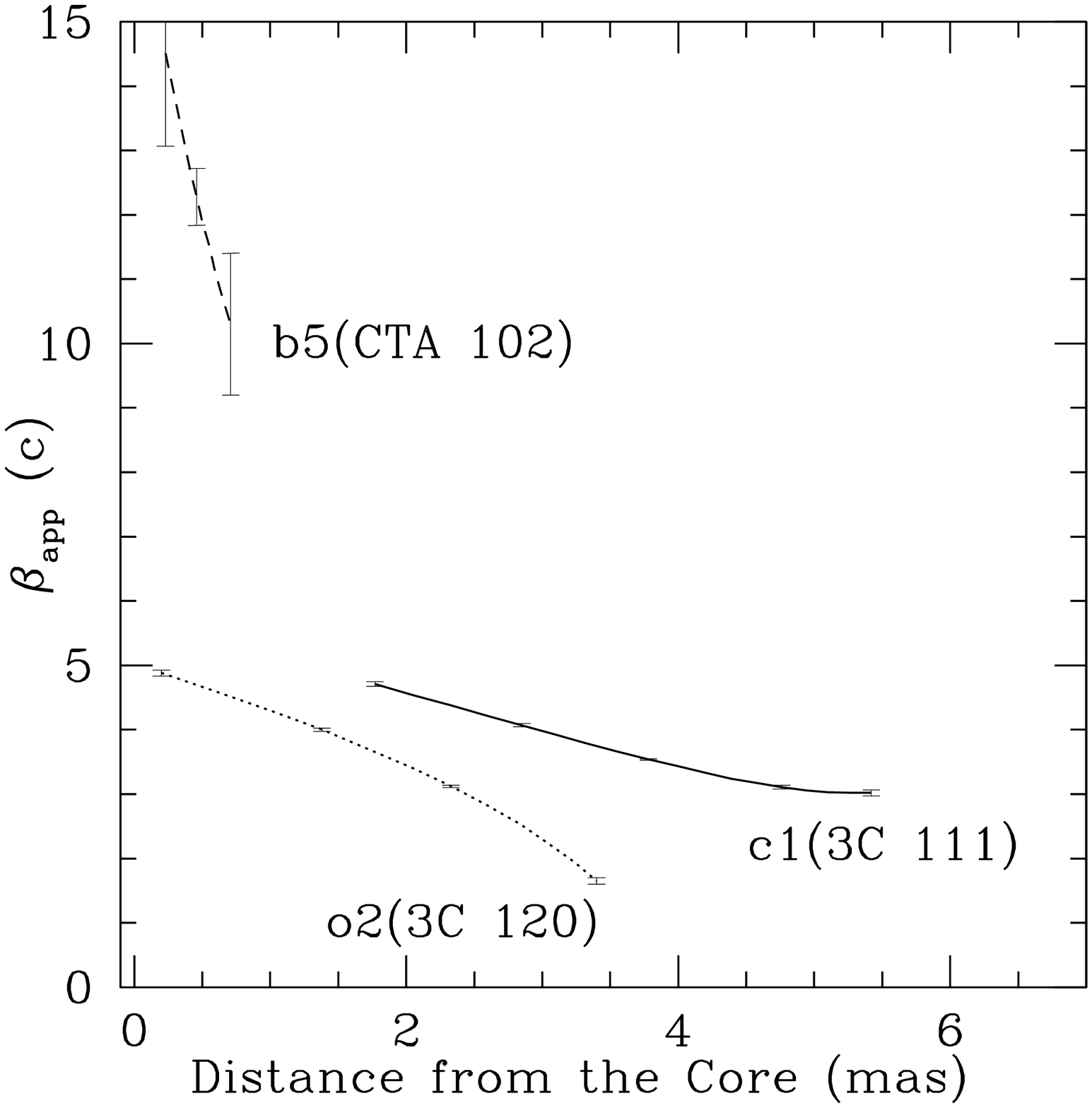}
\caption{Apparent speed vs. distance from the core
as defined by a polynomial approximation, for trailing components 
observed at $>$10 epochs; error bars for the apparent speed 
are computed using uncertainies of polynomial coefficients.
\label{V_trail}}
\end{figure}

Numerical hydrodynamical simulations \citep{A01,A03} show that the interaction
of disturbances in the jet flow with the underlying jet and/or the external medium 
can play a significant role in the variability of the jet emission. In particular,
multiple conical shocks can form behind a strong shock wave propagating down the jet. 
These {\it trailing} shocks appear
to be released in the wake of the primary superluminal component rather than ejected
from the core. The simulations show that the ratio between the apparent velocity of the main
and trailing components is a function of distance along the jet: the closer to the core
a trailing component first appears the slower it moves, although for a given trailing 
component the simulations predict deceleration with distance from the core. 
Trailing components are oblique shocks in three-dimensional models and
should possess different polarization properties than the leading shock. 
\citet{G43} identified several knots in the jet of the radio galaxy
3C~120 as having the characteristics of trailing shocks. Our sequences of
high resolution VLBA images reveal a number of jet features in different sources 
that have properties matching those expected for this phenomenon.
These are (see Fig. \ref{Ident}): $c1,c2$ behind knot $C1$ in the 
radio galaxy 3C~111; $o2$ behind $o1$ found previously by \citet{G43} in 3C~120; 
$b1, b2$ following $B1, B2$, respectively, in the quasar
3C~273; $c9$ behind $C9$ in 3C~345; $b5,b6$ following $B5,
B6$, respectively, in CTA~102; and $b3$ behind $B3$ in 3C~454.3.
Components $c2$ (3C~111), $b2$ (3C~273), and $b5$ and $b6$ (CTA~102) 
emerge from bright knots at some distance from the core 
(see Fig. \ref{Ident}), while the remainder are already trailing bright features 
from the first epoch of our
observations. All of them have apparent speeds less than the corresponding
main component and exhibit different EVPAs than the fast feature (see Fig. \ref{Map}).
Although our observations do not allow us to check every trailing component
for deceleration, a decrease in apparent speed is 
pronounced for $c1$ and $o2$ in 3C~111 and 3C~120, respectively,
and can be inferred for $b5$ in CTA~102 (see Fig. \ref{V_trail}).
For $b5$ in CTA~102 a second-order polynomial does not satisfy the $\chi^2$ 
test but improves goodness-of-fit to the data significantly.
   
\begin{figure}
\epsscale{1.0}
\plottwo{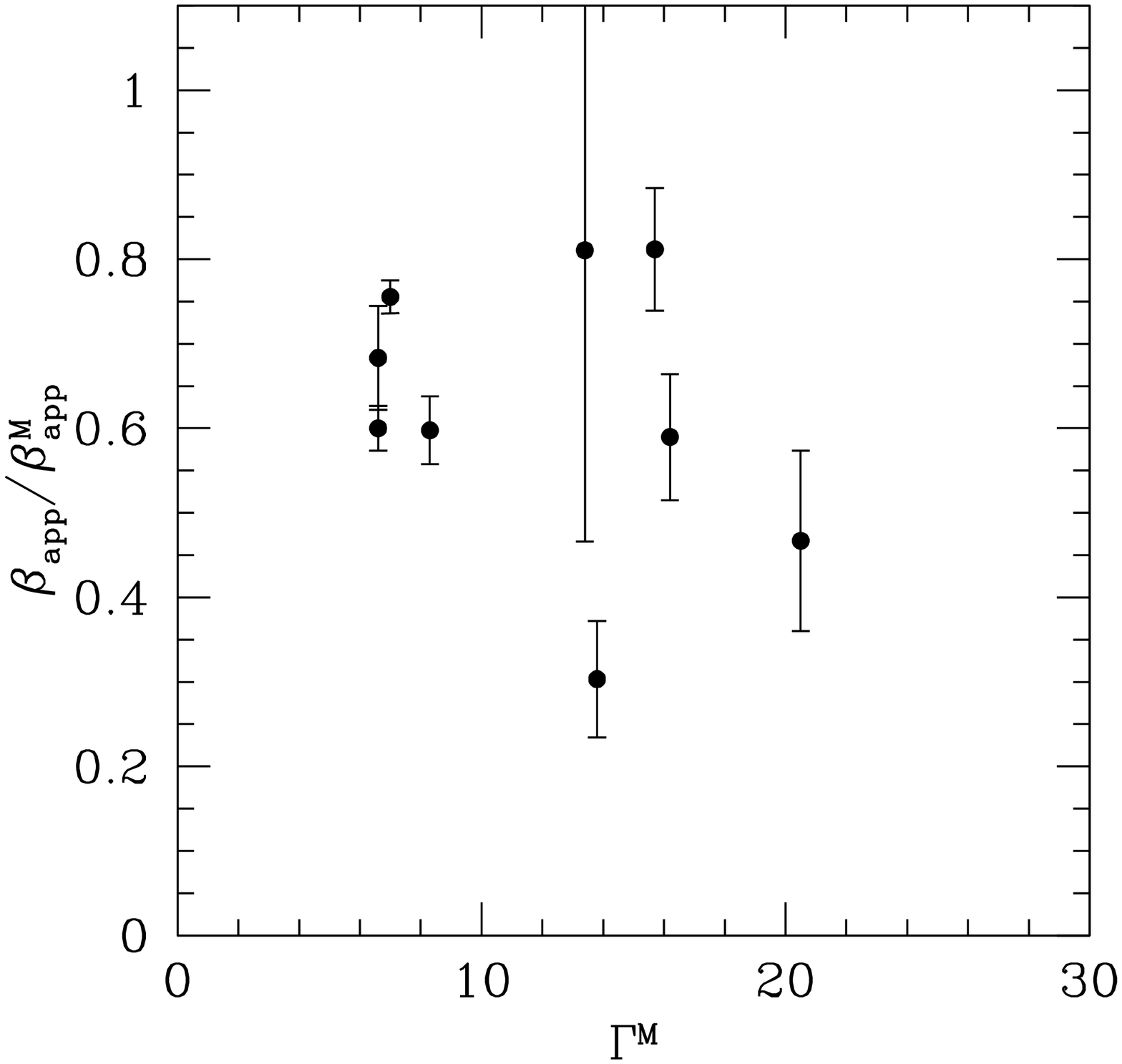}{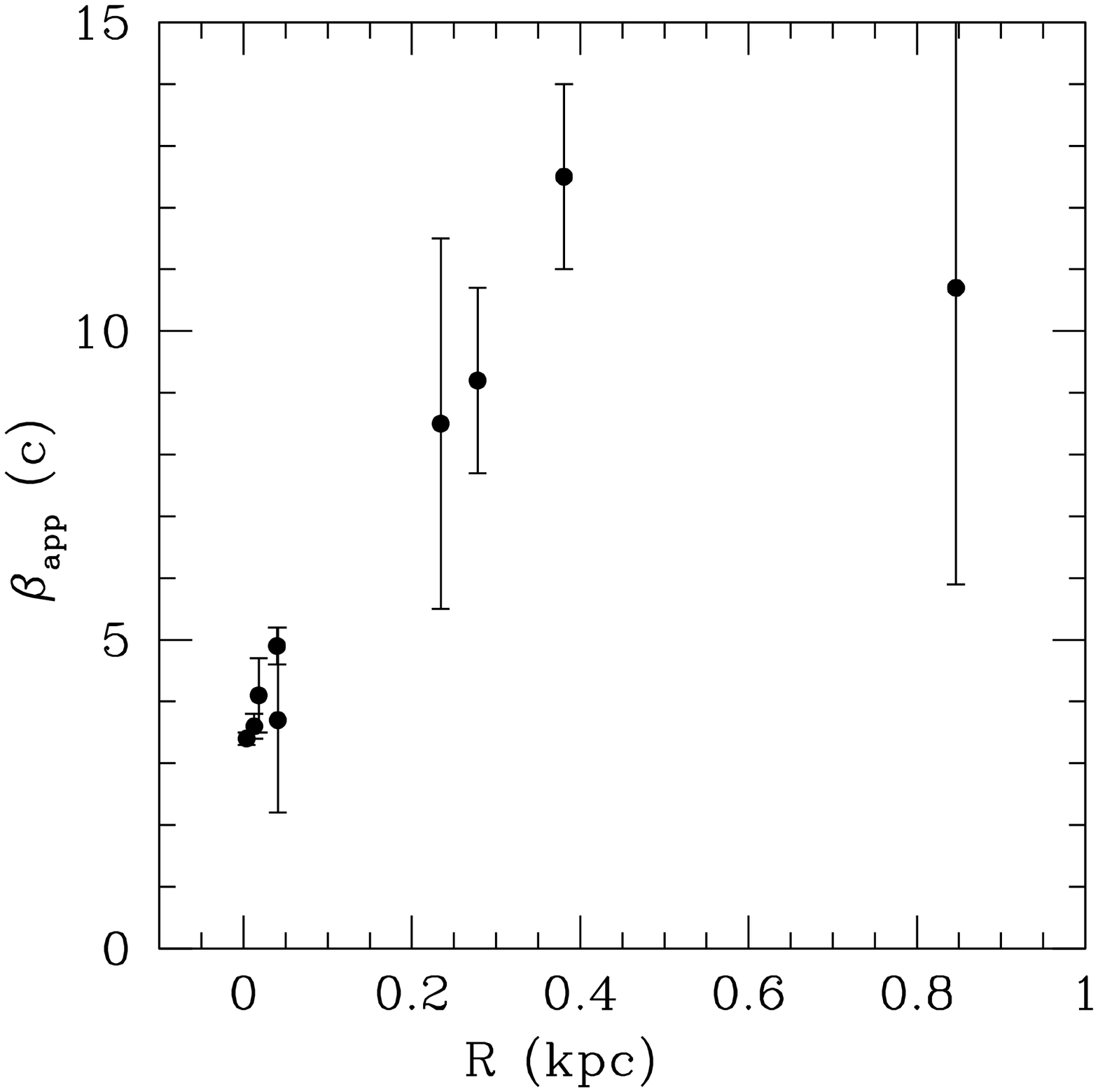}
\caption{{\it Left panel:} Relative apparent
speed of trailing components vs. Lorentz factor of the main disturbance. 
{\it Right panel:} Apparent speed of trailing components vs.
deprojected distance from the core. 
\label{Trail}}
\end{figure}

We characterize each trailing component by three parameters: 
$\beta_{\rm app}$ is the average apparent speed listed in Table \ref{Speed},
$\beta_{\rm app}/\beta_{\rm app}^M$ is the ratio of apparent velocities of trailing component
to corresponding main component, and $R$ is the average deprojected distance from the core 
at which a trailing component is detected in our data ($R$ is 
calculated using $<R>$ indicated in Table \ref{Speed} and jet parameters 
given in Table \ref{Source_par}).
Figure \ref{Trail} ({\it left panel}) demonstrates that trailing components form in the wake 
of bright knots having values of $\Gamma$ extending from 5 to $>$20 and 
different values of $\beta_{\rm app}/\beta_{\rm app}^M$ (all of them $<$1)
are observed for similar Lorentz factors. Figure \ref{Trail} ({\it right panel}) 
shows that there is an increase 
of the apparent speed of trailing components with
distance from the core (coefficient of correlation 0.71). 
According to \citet{A01}, trailing components represent 
pinch waves excited by the main disturbance, and an increase of their speed at larger distance
reflects acceleration of the expanding jet. Our data shown in Figure \ref{Trail}
imply such an acceleration of the underlying jet that
appears not to depend strongly on the Lorentz factor of the main disturbance,
consistent with the conversion of internal energy into bulk kinetic energy that
accompanies expansion.
 
Figure \ref{Trail} ({\it left panel}) shows a possible inverse correlation between
$\beta_{\rm app}/\beta_{\rm app}^M$ and Lorentz factor of the main disturbance
(coefficient of correlation $-$0.40). The correlation is most likely
the result of selection effects. Detection of a trailing component close to 
a leading shock with a high Lorentz factor is complicated by stretching of the 
longitudual size of the main component in the observer's frame by a factor 
$\propto\Gamma$ owing to light-travel time delays. Detection of trailing components
that lag greatly behind a main disturbance with a low Lorentz factor is hampered 
by difficulties in associating with confidence trailing components to the leading knot.  
For this reason, many of the subluminal or quasi-stationary features detected 
near the core ($A$-components) in the majority of sources in the sample
could represent a superposition of many trailing components formed behind a 
number of superluminal features. This possibility is valid only if the direction
of the jet ``nozzle'' changes between ejections, since a second major
disturbance passing through a trailing shock would destroy it.

\citet{KL04} have found that there is a systematic decrease in $\beta_{\rm app}$ 
with increasing
wavelength, which they suggest results from sampling different parts
of the jet structure at different frequencies. Although this might be the case
for average apparent speeds derived from surveys at different wavelengths,
for individual sources with superluminal components detected at
similar distances at different frequencies, the effect
can be caused by the structure containing the leading perturbation plus slower trailing shocks
being unresolved at the longer wavelengths. Perhaps the difference in 
the proper motions of the polarized and total intensity components $B8p$ and $B8$
noted in the quasar 0528+134 (see Table \ref{Speed}) is an example
of this effect, with the polarized intensity image (which reveals the leading
compact component) playing the role of the finer resolution of 
shorter wavelength observations.  
    
\subsection{Frequency of Superluminal Ejections}
\begin{figure}
\epsscale{1.0}
\plotone{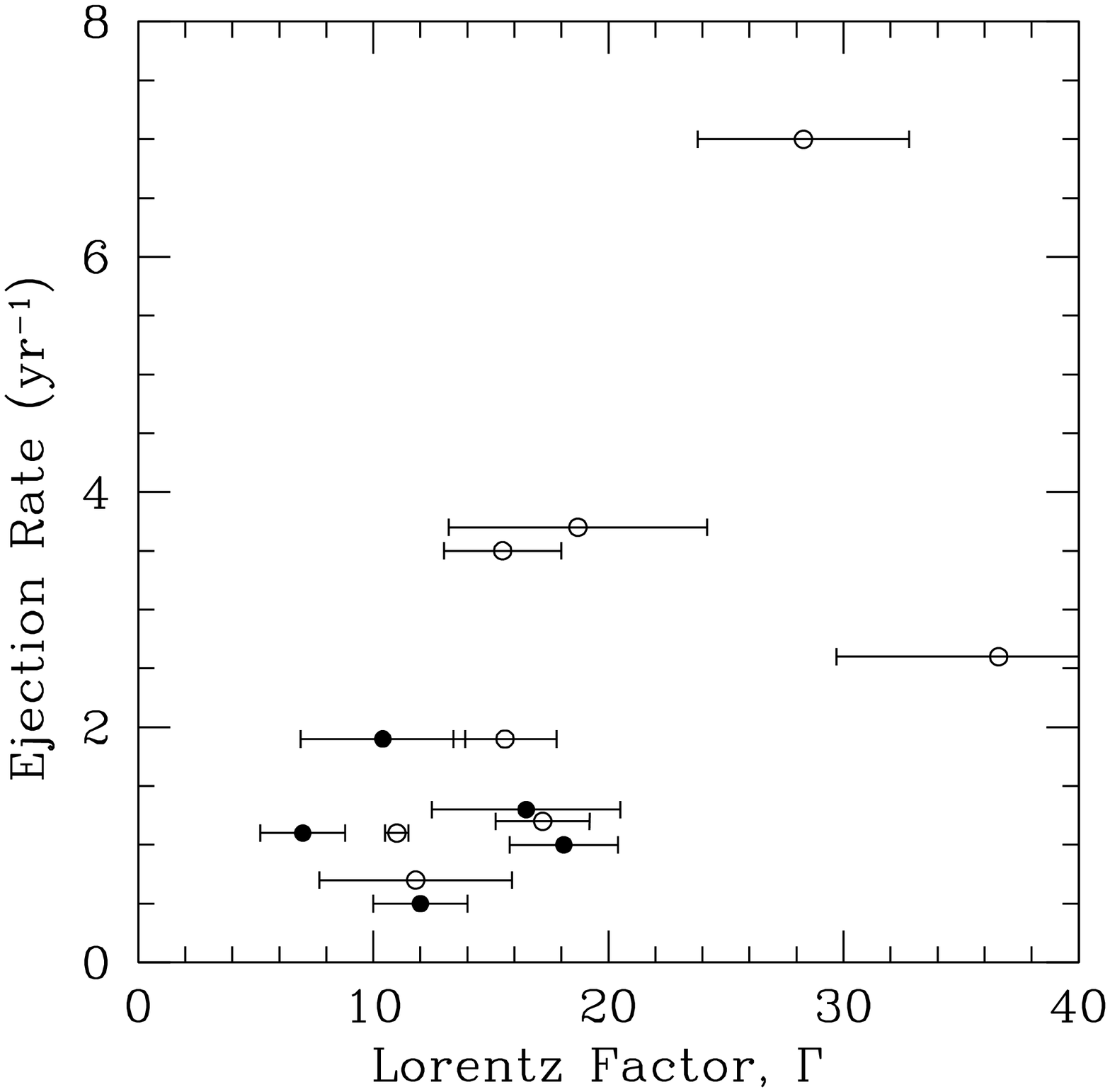}
\caption{Rate of superluminal ejections in the rest frame of the AGN
vs. average jet Lorentz factor. \label{Eject}}
\end{figure}

Our monitoring is unique in terms of the number of blazars 
observed in a regular manner at high angular resolution over 3 years. 
The excellent time coverage allows us to determine the rate of superluminal ejections,
thought to be controlled by activity in the cental engine.
We list in Table \ref{Source_par} the rate, $f_{\rm ej}$ 
(multiplied by the time dilation factor 1+z), of superluminal ejections 
over three successive years based on the results given in Table \ref{Speed}.
In Figure \ref{Eject} we plot the ejection rate versus the average jet 
Lorentz factor for blazars. There is a trend suggesting a positive correlation
between $f_{\rm ej}$ and $\Gamma$ (coefficient of correlation 0.5)
The formal $t_{\zeta,\nu}$ test rejects the hypothesis that there is 
no relation between $f_{\rm ej}$ and $\Gamma$ at a confidence level $\zeta$=0.05. 
However, the trend could be an artifact of the bias toward highly Doppler 
boosted objects in our sample. Nevertheless,
blazars form the largest class of identified $\gamma$-ray sources \citep{H99},
for which a connection between $\gamma$-ray events and radio jet activity
has been found \citep{J01g}. If nonthermal flares are associated with ejections
of new superluminal knots, as seems to be the case at mm wavelengths \citep{SAV02},
then we would expect $\gamma$-ray light curves to reflect the rate of ejections.
In this case, the possible correlation between
the Lorentz factor and ejection rate would imply that
the most variable $\gamma$-ray sources (in terms of major flares per year)
should possess the highest Lorentz
factors. This can be tested by $\gamma$-ray light curves 
of blazars obtained by the Gamma Ray Large Area Space Telescope (GLAST,
expected to begin operation in 2007), along with VLBI monitoring of the
radio jets.  

\section{Summary}  
In this paper we have presented the entire data set collected 
during a three-year program of monitoring AGNs using the VLBA,
and have discussed the main results 
relating to the kinematics of jets. The sequences of images reveal
short timescales (for some sources shorter than 2 months) of 
variability of the jet structure and even more rapid variability in 
polarization. This program illustrates the importance of 
intensive monitoring for understanding jet physics in 
superluminal radio sources. 

We have measured the apparent speed of 106 features in the inner jets 
(within 4~mas of the core) of two radio galaxies, five BL Lac objects, 
and eight quasars from 1998 March to 2001 April. Superluminal apparent speeds 
occur in 80\% of the knots, 26\% of which show statistically significant
deviations from ballistic motion. The majority of non-ballistic
components undergo an increase of apparent speed with distance from
the core, although local decelerations are observed in some cases 
alongside the general acceleration. This could be the result of physical
accelerations or from selection of sources whose angles to the line of
sight $< {\rm sin}^{-1}(1/\Gamma)$ near the core and closer to this value 
farther out. Many of the jets contain both very fast and much slower moving features,
which might be explained as forward and reverse shocks, respectively.
It appears that fast features are pronounced in the jets of the
quasars, while slower moving knots dominate in the BL~Lac jets. 
This suggests a different nature of the main disturbances seen in the jets 
of the two classes of blazars.

The properties of 11\% of the 
superluminal components are consistent with the characteristics of trailing 
shocks expected to form in the wake of strong disturbances in the flow.
Trailing components emerging from bright knots farther down the jet
have a faster apparent speed than those generated near the core while a given 
trailing feature most likely decelerates along the jet.
According to the numerical simulations by \citet{A01}, the trailing
components are caused by the triggering of pinch modes by the main disturbance.
The jet accelerates close to the core as internal energy is converted into 
bulk kinetic energy, leading to trailing shocks being faster farther downstream. 

In four sources (3C~120, OJ~287, 3C~345, and CTA~102)
the behavior of jet components appears to be affected by the 
interaction with the external medium (see  \S A). For each source a number 
of components decelerate, change trajectory, brighten
in total and polarized flux, and undergo a rotation of the EVPA at the
same distance from the core along an edge of the jet. 
This implies the existence of gas clouds 
a few parsecs to $>$1~kpc (deprojected) from the central engine,
intermediate between the locations of the very dense clouds of the broad line 
regions and more rarefied clouds of the narrow line regions. 

Using measurements from the VLBI images of parameters such as flux density, 
apparent speed, and size of components, we have estimated Lorentz and Doppler
factors as well as viewing angles for superluminal knots and the opening angle
of each jet. This is a new method to define jet parameters, based on
the assumption that the decay in flux of the superluminal components is caused 
by radiative losses rather than by cooling from expansion, and is subject
to light-travel delays.
We demonstrate that at high radio frequencies these assumptions are
most likely correct. The derived parameters of the jets indicate
that in our sample the quasars have the highest Doppler factors ($\delta$) and 
smallest viewing ($\Theta_\circ$) and opening ($\theta$) angles, while the 
two radio galaxies possess significantly lower Doppler factors, larger viewing 
angles, and wider opening angles despite their ``blazar-like'' radio properties. 
This implies that in the 3-dimensional parameter space 
($\delta$,$\Theta_\circ$,$\theta$) the radio galaxies, BL~Lacs, and quasars in 
our sample occupy different regions. The regions of  
the quasars and BL~Lac objects partly overlap, while the radio galaxies 
are significantly distinguished from the blazars in all three parameters.
Since our sample is not a complete 
one, the major differences suggested by these results need to
be investigated further with larger, complete samples.
The inferred relationship between the half opening angle and the Lorentz factor agrees 
with the expectation of gas-dynamic models that predict smaller values of $\theta$
for higher Lorentz factors and a dependence of $\theta$ on the ratio
of the external and internal jet pressure. The best 
approximation to the relation is very close to $P_{\rm ext}/P_\circ\sim 1/3$.
This does not, however, exclude collimation by magnetic pinching, 
which might produce a similar $\theta-\Gamma$ relation.
 
We have estimated the intrinsic brightness temperatures of jet components in 
the quasars, BL~Lacs, and radio galaxies on parsec scales, obtaining averages of 
$1.1\times 10^9$~K, $5.5\times 10^7$~K, and $3.5\times 10^9$~K, respectively. 
Comparison of these values with the 
equipartition brightness temperature of the optically thick part of the jets,
$T_{\rm b,int}=2-5\times 10^{10}$~K, suggests a stronger magnetic field in the
BL~Lac objects.   

There is a possible positive correlation between the Lorentz factor of the jet 
and ejection rate of superluminal components for the blazars. This can be
tested with the $\gamma$-ray light curves that will be measured by the
GLAST mission.

\appendix
\section {Notes on Individual Sources}

\begin{figure}
\epsscale{1.0}
\plotone{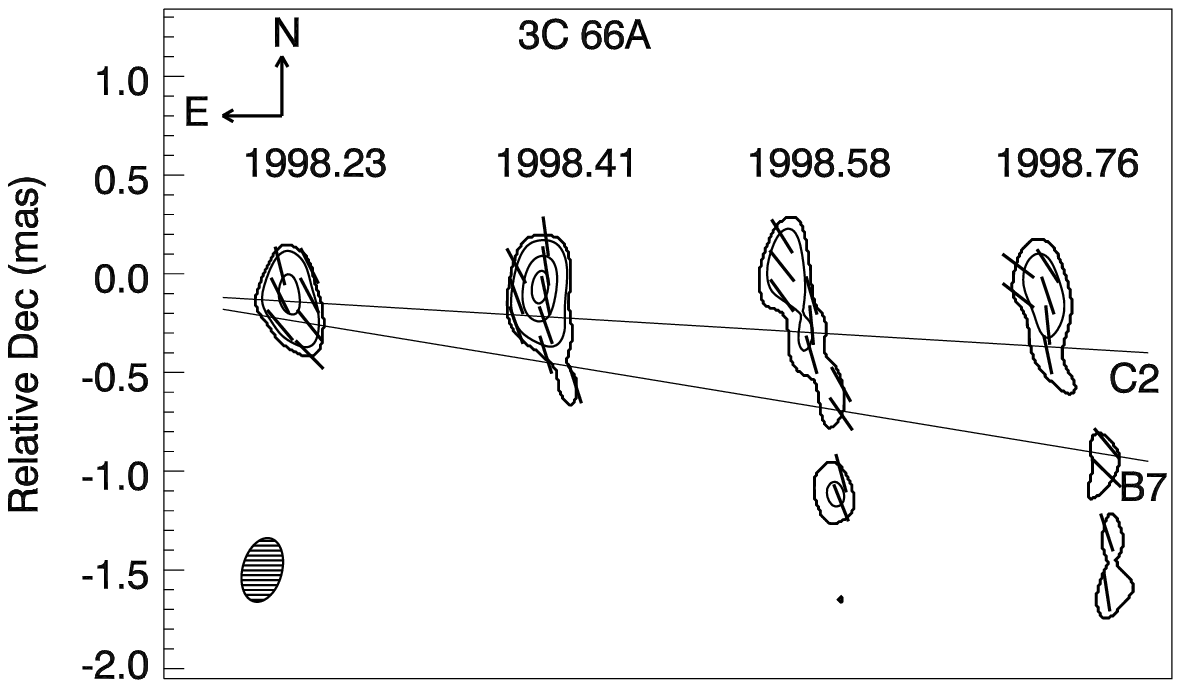}
\caption{Polarized intensity images of 3C~66A at four successive epochs when
highly superluminal component $B7$ is detected. The polarized intensity peak is 
19.3~mJy~beam$^{-1}$. The size of the beam is indicated in Table \ref{Sample}.
Contour levels increase by a factor of 2, with the lowest contour corresponding
to 8\% of the peak polarized intensity. The line segments show the direction 
of the electric vectors. The solid lines indicate motion
of component $B7$ and $C2$ on the total intensity images. \label{3c66a_pol}}
\end{figure}

{\bf 3C~66A:} The radio jet consists of a bright, polarized core with weak,
diffuse extended structure (Fig. \ref{Alla}). The set of images allows us to trace
components $C1$, $C2$, $C3$, and $C4$ at many epochs (Fig. \ref{Ident}). They 
are characterized by moderate velocity, 2-5c, typical of those measured in BL Lac
objects at lower frequencies \citep{GD94}. 
Component $C1$ is probably feature $C$ from \citet{J01}, where it was thought to be 
stationary. The centroids of $A1$ and $A2$ appear to move toward the
core at mildly superluminal speeds; however, these are extended features, hence the
shifts in position could be due to changes in their brightness distributions. This is
supported in the case of $A2$, whose proper motion makes an angle of
$45^\circ$ to the line between the core and the feature's centroid. Figure \ref{Ident}
shows possible identifications of very weak components $B4, B6$, and $B7$. The designation 
of knots follows that of \cite{J01}; Figure \ref{Ident} includes their data, marked
by triangles. $B$ knots move at a high apparent speed, $> 20c$. 
The reality of this faster motion is supported by a region
of enhanced polarized intensity that progresses rapidly
downstream at the same epochs as when the high apparent speed is detected in the total
intensity images (see Fig. \ref{3c66a_pol}). It appears that the more slowly moving 
$C$ components and highly polarized, fast $B$ components coexist in the jet.
One of possible physical explanations for this dichotomy is a double shock structure, 
with the faster forward shock being weaker than slower reverse shock (see \S 6.5).

{\bf 3C~111:} The innermost jet of this galaxy exhibits strong activity. 
At least 7 moving components ($B1-B7$, see Fig. \ref{Ident}) can be identified inside
1~mas of the core. These knots move ballistically at a typical apparent 
speed $\sim 3$c. There are two bright features near the core, $A1$ and $A2$, that
are stationary according to the $\chi^2$ test. The most prominent feature of the jet is an
extended, polarized blob northeast of the core. The feature appears to be
associated with a major outburst observed at 90~GHz in 1996 January \citep{APKG98}, 
at 37~GHz in 1996 March, and at 22~GHz in 1996 August \citep{T04}. 
The first ejection around this time appears to have occurred
earlier than the derived epoch of ejection $T_0$ 
(see Table \ref{Speed}), since complex structure of the core region was detected already
in a 43~GHz VLBI map obtained in 1996 July \citep{APKG98}. This implies slower motion
of the disturbance near features $A1$ and $A2$, which agrees with the
description of the structure by \citet{APKG98} as ``a mix of dimming and flaring
of stationary and moving emission components.'' In our observations we identify the
brightest part of the outburst remnant as  $C1$, which moves ballistically at a higher 
apparent speed than that of the
innermost components. Knots $c1$ and $c2$ detected behind
$C1$ seem to belong to the same disturbance. At the last 3 epochs the parameters
of $C1$, $c1$, and $c2$ are obtained after tapering the images by reducing the weighting
of the data from the longest baselines, and are not used to compute the jet velocities.

\begin{figure}
\epsscale{1.0}
\vspace{-1cm}
\plotone{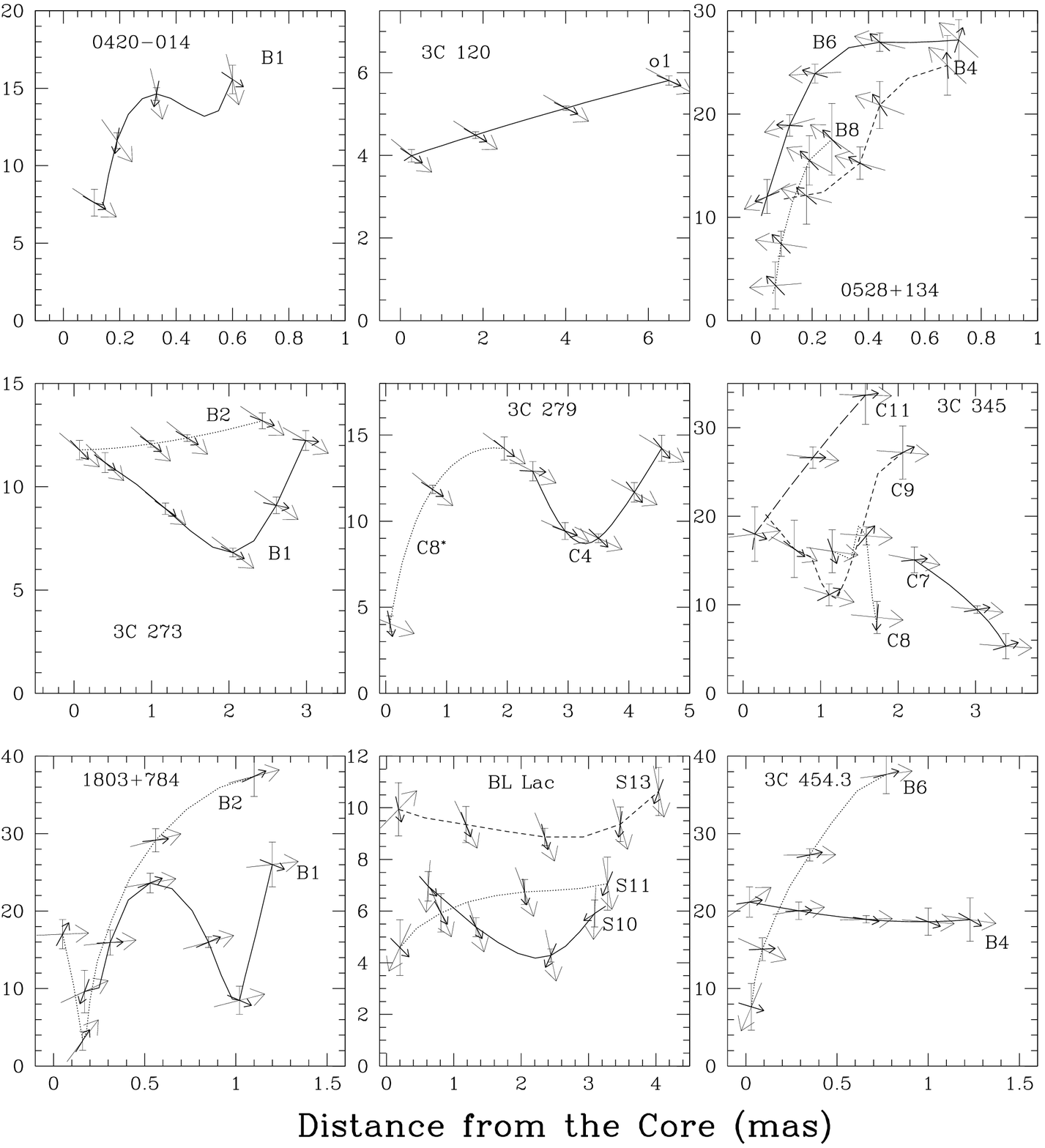}
\vspace{-3cm}
\caption{Apparent speed vs. distance from the core, 
as defined by a polynomial approximation, for accelerating/decelerating knots 
(Table \ref{Accel}). Short arrows represent the direction of the apparent velocity,
long arrows indicate the local jet direction; both are derived from the best-fit 
polynomial. Error bars for the apparent speed 
are computed using uncertainies of polynomial coefficients.  \label{V_Change}}
\end{figure}

{\bf 0420$-$014:} The parsec-scale jet of this quasar is strongly dominated
by the VLBI core, which has a complex polarization pattern
(see Figs. \ref{Allc} and \ref{Map}), implying multi-component substructure
in the core region.
We detect two moving components in the innermost part of the jet.  
Knot $B1$ has slower motion near the core, $\sim$7$c$ (Fig. \ref{V_Change}),
and accelerates significantly beyond 0.2~mas as its trajectory 
turns from $-100^\circ$ to $-175^\circ$ (Fig. \ref{traj}). 
The bright component $B2$ traced inside 0.15~mas from the core
moves ballistically along $\Theta\sim -72^\circ$ at a similar apparent speed 
as that derived for $B1$ near the core. The behavior of the knots is consistent with 
the common curved trajectory for jet components found by \citet{B00}.

\begin{figure}
\epsscale{1.0}
\plotone{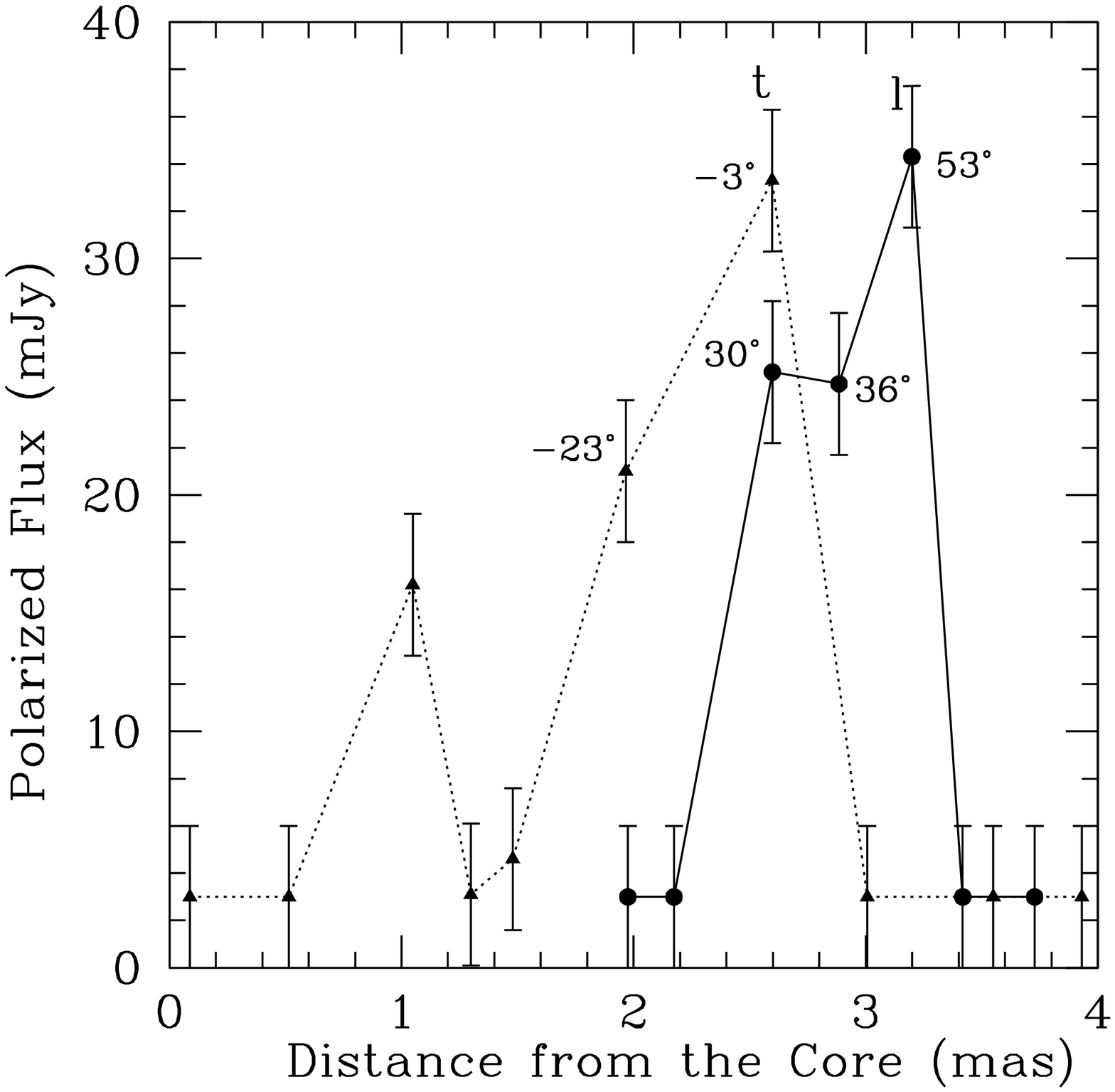}
\caption{Polarized flux density of the knots $l$ (solid line) and $t$ (dashed line) 
in the radio galaxy 3C~120 at different distances from the core. The numbers
indicate values of the EVPA. \label{3c120_cloud}}
\end{figure}

{\bf 3C~120:} Our observations continue to trace components $d$, $h$, $l$, $o1$, and $o$
detected in \citet{G43}, and we identify new components $t$, $u$, $u1$, and $v$
(Fig. \ref{Ident}). Components $d$ and $h$ move ballistically with the same 
apparent speed found previously, while $l$ has decelerated since
earlier epochs. Components $o1$ and $o2$ represent the front section of the major 
disturbance designated in \citet{G43} as {\it complex o}. Component $o1$ 
brightens in total and polarized flux at $\sim$3.4~mas, similar to the behavior
seen for $h$ and $l$ \citep{G43}. After flaring,
$o1$ and $o2$ diverge in their behavior: $o1$ accelerates and $o2$ decelerates
(Fig. \ref{Ident}). Such a development is expected if $o2$ is a trailing component
that forms in the wake of the accelerating knot $o1$, which represents
the major disturbance in the flow \citep{G43}.
The most prominent feature at later epochs is $t$, which 
evolves in the same manner as $l$. The components 
have nearly equal proper motions, 2.17$\pm$0.06 ($l$, \citealt{G43}) and 
2.04$\pm$0.08 mas~yr$^{-1}$ ($t$), and similar trajectories (Fig. \ref{traj}). 
They both undergo a flare in
total and polarized flux at a location between 2.5 and 3.5~mas 
from the core and the EVPAs of both rotate during the flare
(Fig. \ref{3c120_cloud}).  \citet{G00} explained these properties
as the results of interaction between the jet and an interstellar cloud. 
The behavior of component $t$ is consistent with this, which suggests that  
it interacts with the same cloud. However, there is a significant shift 
($\sim 0.6$~mas) between the projected positions of the peaks of the flares 
(see Fig. \ref{3c120_cloud}). Most likely, the shift is a consequence
of the slightly more southern trajectory of $t$ (Fig. \ref{traj}). 
For a viewing angle $\sim 21^\circ$ and opening angle $\sim 4^\circ$ 
(see Table \ref{Source_par}), the shift by 0.6~mas gives a lower limit to
the size of the cloud of 1~pc at a deprojected distance from the VLBI core 
of $\sim$7-8~pc. 
   
{\bf 0528+134:} Components $B3$ and $B4$ detected 
in \citet{J01} are seen during the early epochs of our monitoring and four new
components are ejected later (Fig. \ref{Ident}). All components have similar trajectories, 
with
a sharp bend to the north beyond 0.5~mas from the core, as previously noted by other
authors \citep[e.g.,][]{B99}. The projected trajectories of the knots
are shifted relative to each other (see Fig. \ref{traj}), which suggests  
different angles of ejection. This is supported by a difference in the apparent speed 
near the core, from $>10c$ for the fastest and most eastern components to $\sim 3c$ 
for the most northern component. We find progressive acceleration in at least 
three components (Fig. \ref{V_Change}). The most northern knot, $B8$,
which is strongly polarized and the brightest feature in the jet (Fig. \ref{Map}), shows 
significantly different motion of the total and polarized intensity peaks.
The polarized region, $B8p$, moves ballistically ahead of the main concentration of
total intensity, $B8$. Table \ref{Speed} lists the parameters of $B8$ and $B8p$ separately.
We suggest that $B8p$ represents a leading shock of a strong disturbance with
complex structure (see \S 6.6).   

\begin{figure}
\epsscale{1.0}
\plotone{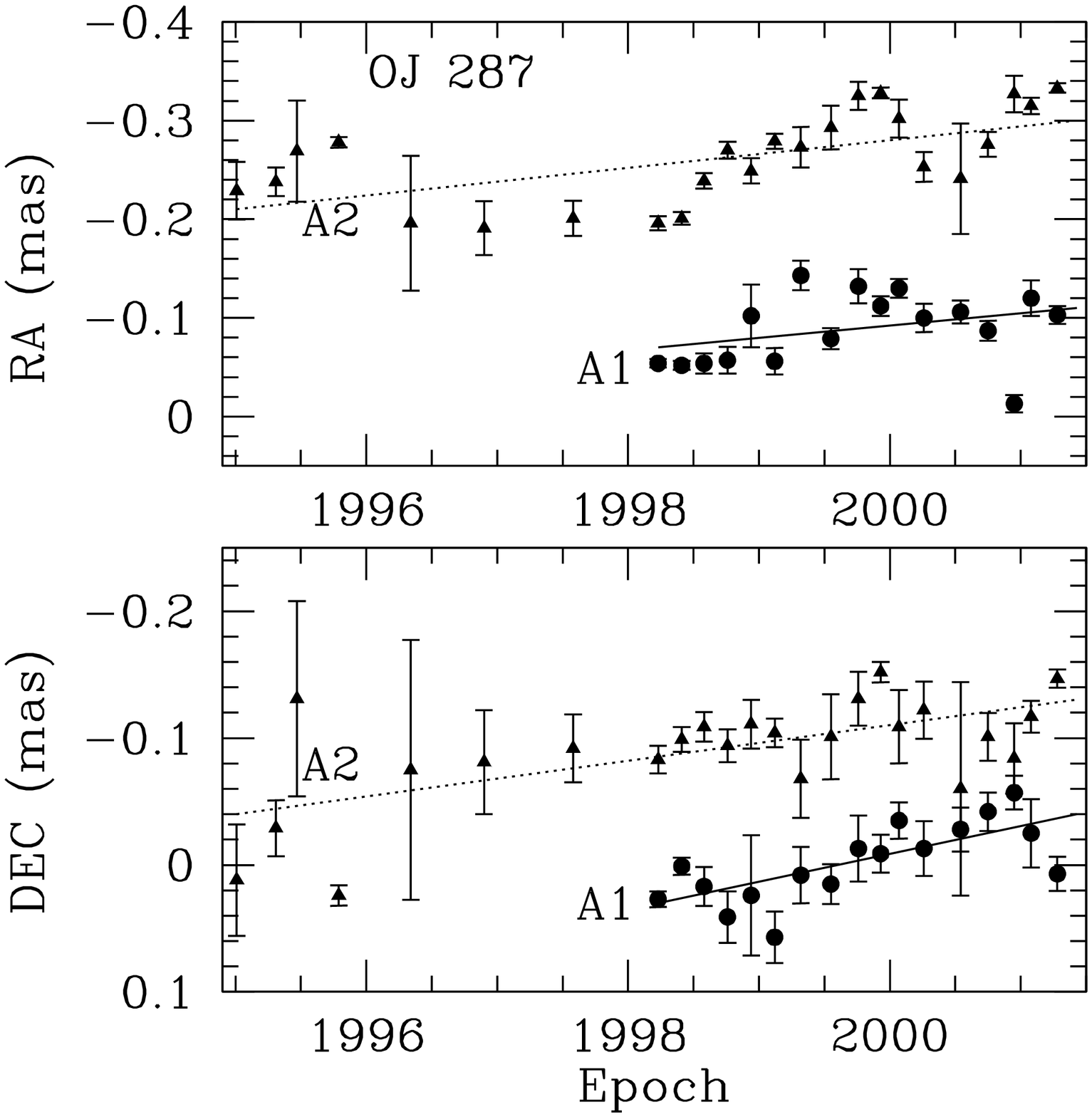}
\caption{Relative right ascension and declination vs. epoch for components $A1$
(circles) and  $A2$ (triangles) in OJ 287. The lines show a straight-line  
approximation to the temporal behavior. \label{OJ 287_A}}
\end{figure}

{\bf OJ 287:} The jet appearance changes considerably during the observations 
(Fig. \ref{Allf}), largely due to brightness variability of three 
prominent jet features, $A1, A2$, and $A3$ (Fig. \ref{Map}). On average, 
$A1$ and $A2$ separate from the core at subluminal speeds, 
although significant fluctuations relative to linear motion are observed 
(see Fig. \ref{OJ 287_A}). Component $A2$ might be associated with  
$C1$, classified by \citet{J01} as a stationary feature. 
Table \ref{Speed} gives the parameters of $A2$ based on our data alone,
and also based on these data combined with those of \citet{J01} under the assumption 
that $A2$ and $C1$ are  the same feature. The latter is listed as $A2*$ in 
Table \ref{Speed}. The addition of the previous data decreases the proper motion by a 
factor of $\sim$2 and in both cases the proper motion does not exceed zero at the 2$\sigma$ 
level. This suggests that the feature may fluctuate about a stationary position 
rather than translate downstream. 
We identify several fast moving components in the jet. Knots $B4-B7$ \citep[the
designation continues the scheme of][]{J01} have very short lifetimes of $\sim$1~yr
at 43~GHz, and can be resolved only after they have passed $A2$ 
(see Fig. \ref{Map}). However, a notable brightening of $A2$ occurs at the epochs 
when a $B$ component likely reached $A2$ (see Fig. \ref{Ident}). The $B$ components show high 
superluminal speeds that are slightly slower than those of knot $K3/U3$ in \citet{HO01}. 
Knot $K3/U3$ is observed to accelerate and its motion at 22/15~GHz is measured
$\sim$2 times farther from the core than the locations of the $B$ components in
our images. 
 
A distinguishing feature of the jet is component $A3$, the formal solution
for the motion of which is upstream, toward the core. This motion is especially pronounced in the
likely event that $A3$ is associated with $C2$ identified by \citet{J01}, in which case $A3$ 
also undergoes a significant change in position angle relative to the core (Fig. \ref{traj}).
The latter can be explained by swinging/precession of the jet nozzle or by a translation 
of the core, either of which agrees with changes in trajectory of components near the core 
from $\Theta\sim -90^\circ$ for $B1$ \citep{J01} to $\Theta\sim -115^\circ$
for $B6$. The alignment between the innermost jet direction and the position 
of component $A3$ suggests that the feature is the result of interaction between the 
jet flow and the external medium. The upstream apparent motion of $A3$
might then trace the location of an interstellar cloud (projected on the sky) 
when the intrinsic position angle of the jet changes. In this case, the dramatic fading 
of the feature toward the end of our observations (see Fig. \ref{Allf}) could be 
connected with an intrinsic change of the inner jet direction (component $B6$ has 
the most southern projected position angle during 1995-2000) so that the jet flow 
misses the cloud. This interpretation agrees with the findings of \citet{TK04}.
These authors used the Radio Reference Frame Image Database data 
and determined that the projected position angle of the inner jet of OJ~287
rotated clockwise by $\sim 30^\circ$ in 8~yr, in agreement with
a ballistic precession model if the velocities of the components decrease
beyond $\sim$1~mas from the core. We suggest that the interaction with an extragalactic
cloud might cause the deceleration of components. From the derived parameters of the jet 
(see Table \ref{Source_par}), we place the cloud at a deprojected distance of $\sim 90$~pc from 
the core. Acceleration of $K3/U3$ detected by \citet{HO01} takes place between 0.2 
and 1~mas from the core and reflects an increase in the apparent speed beyond  
$A1$ and $A2$ and before $A3$ located at $\sim$1~mas.
   
\begin{figure}
\epsscale{1.0}
\plotone{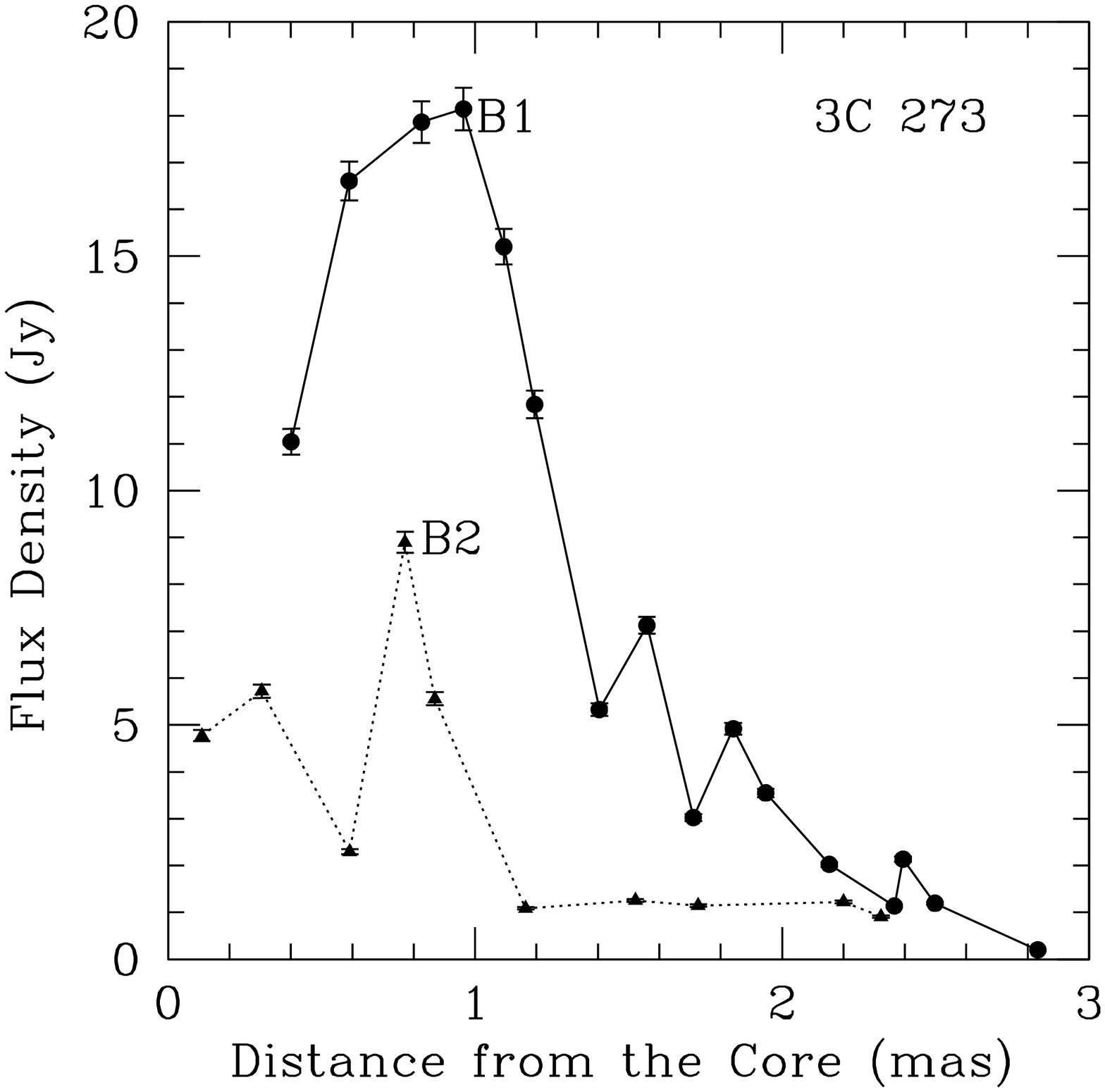}
\caption{Total flux density of knots $B1$ (solid lines) and $B2$ (dashed lines) in 
the quasar 3C~273 at different distances from the core.\label{3c273_fluxB}}
\end{figure}

\begin{figure}
\epsscale{1.0}
\plotone{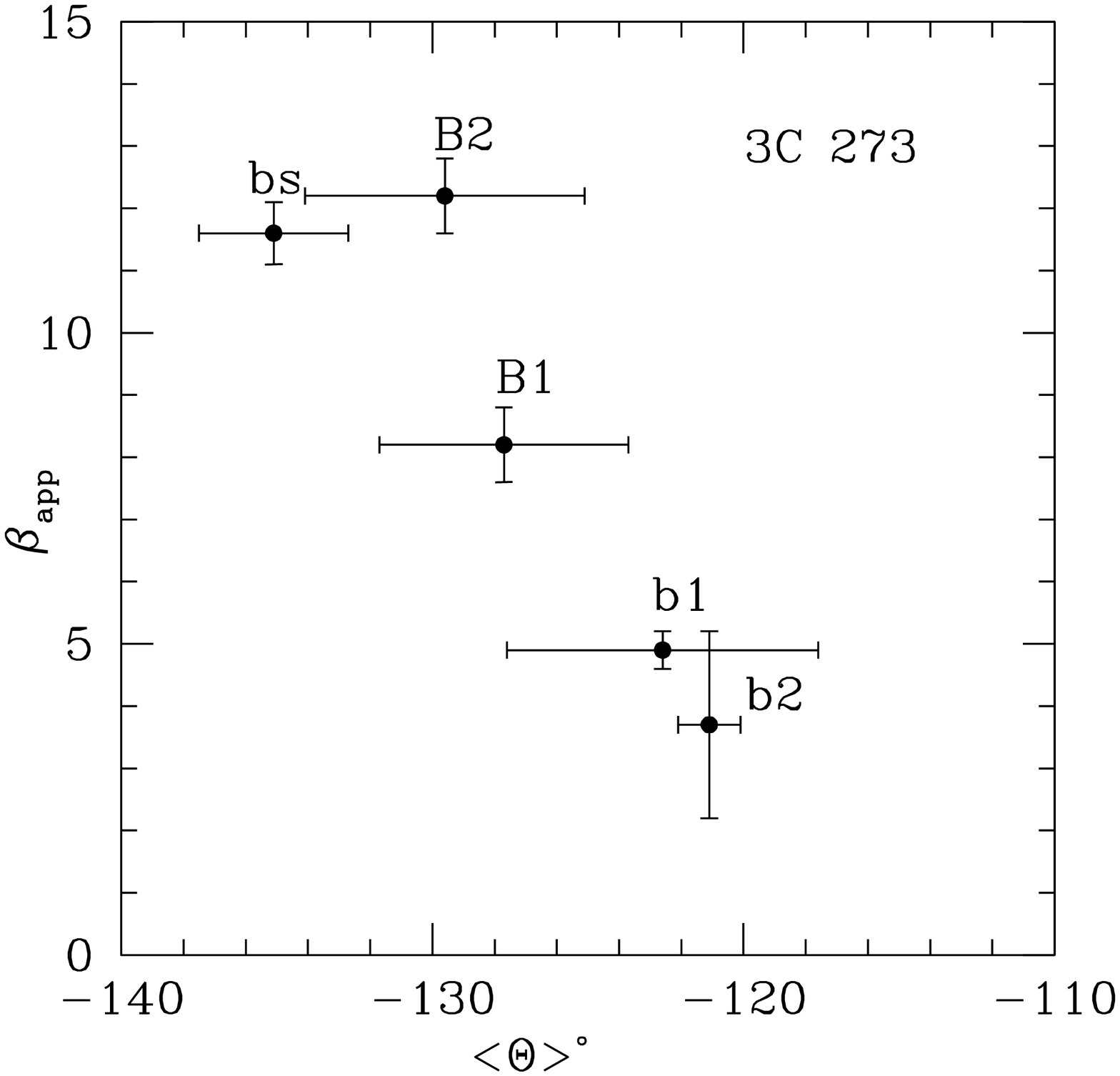}
\caption{Apparent speed vs projected position angle of superluminal  
components in the quasar 3C~273. \label{3c273_cross}}
\end{figure}

{\bf 3C 273:} The most prominent features in the images (see Fig. \ref{Allg}) 
are an unpolarized core and strongly polarized moving
knots $B1$ and $B2$ that both brighten at $\sim$0.8~mas (Fig. \ref{3c273_fluxB}). 
Each component becomes 
strongly polarized just as it leaves the core region
($\sim$0.3~mas from the core, Fig. \ref{Map}) while the core has a low
polarization (comparable with the noise level). 
The similarity of the EVPA and fractional polarization
at 7 (brighest polarized feature), 3, and 1.3/0.85~mm suggests a low fractional
polarization in the core at the high frequencies, where Faraday effects 
are not important. This implies that the core has intrinsically low polarization,
which might be caused by a strongly turbulent magnetic field on scales smaller 
than the 7~mm synthesized beam. 
After they flare, $B1$ and $B2$ fade dramatically and 
expand both along and transverse to the jet direction. This results in
the formation of new features $b1$ and $b_s$ related to $B1$,
and $b2$ related to $B2$ (Fig. \ref{Ident}). The $b$ components
might form in the interaction of relativistic 
shocks connected with features $B1$ and $B2$ and the underlying
jet flow. These components should have lower apparent speed than 
the main disturbances \citep{A01}, but this is not the case for $b_s$. 
Alternatively, the $B$ components are fairly 
extended features in a broad jet and the $b$ knots might represent sections of the jet
where the velocity vectors lie at different angles to the line of sight. The latter 
interpretation is supported by a difference in the projected 
trajectories of the $B$ and $b$ components (Fig. \ref{traj}) and the strong
dependence between the projected position angle and apparent speed of components, 
with the more northern knots being slower (Fig. \ref{3c273_cross}). 
A substantial velocity gradient across the jet is also expected according to
a model where the underlying jet flow has a double-helix structure
and a slower Lorentz factor than that of disturbances \citep{LZ01}. 
The brightening of $B1$ and $B2$ at $\sim$0.8~mas from the
core might be connected with either (1) a change of the jet direction (a slight
curvature of the projected trajectories of  both components is
seen in Fig. \ref{traj} at this distance), (2) an intensity peak of a
threadlike pattern of the underlying flow where the local velocity
vector bents toward the line of sight \citep{LZ01},
or (3) interaction with the external medium. However, a stable
polarization direction in $B2$ before, during, and after
the flare (see epochs 1999.76, 1999.93, and 2000.07 in Fig. \ref{Allg})
does not support (3).

At many epochs the images contain
a diffuse, highly polarized (up to 50\%) feature, $C1$, farther down the jet 
(see Fig. \ref{3c273_C1}), moving at an apparent speed of $\sim 7c$. According 
to the time of ejection, $C1$ corresponds to knot $G1$ found by \citet{J01}. 
If $C1$ and $G1$ are the same feature, it has maintained a nearly constant proper 
motion over $\sim$5~yr.

{\bf 3C~279:} Our data reveal motion of at least 8 components, $C9-C16$ (our designations
continue those of \citealt{W01}), with apparent speeds up to 17$c$, three times  
higher than previously reported velocities measured since 1991 \citep{W01,HO01,J01}. 
The identification of components in Figure \ref{Ident} includes
results at 43~GHz from \citet{W01} (triangles) and \citet{J01} (squares). Components 
$C9-C16$ are shown in Figure \ref{Map} at the last epoch of monitoring. \citet{J04}
suggest that the remarkable change in apparent speed is caused by a shift in
the direction of the jet ``nozzle'' while the Lorentz factor remains constant. 
This conclusion is supported by a gradual decrease of the projected
position angle $\Theta$ of each successively ejected component (starting with $C9$).
In this interpretation, the jet flow is characterized by $\Gamma\geq 20$  while 
$\Theta_\circ$ changes from 0.5$^\circ$ for components with the lowest speed 
(5.5c for $C8$ found by \citealt{W01}) to 1.6$^\circ$ for components 
with the highest speed. It is also possible that component $C8$ accelerates
(see Fig. \ref{Ident}) due to an increase in the Lorentz factor of the flow. We designate
this possible component as $C8*$; it corresponds to $C8$ before the acceleration and 
to $C9$ after the acceleration. As indicated in Table \ref{Accel}, the change
in the velocity component parallel to jet, $\dot{\mu}_\parallel$, is significant. 
Another method, which we apply in this paper to estimate Doppler and Lorentz 
factors and viewing angle for each superluminal knot (see \S 6), suggests that 
both the Lorentz factor and the angle between the jet axis and line of sight 
for $C9-C16$ are different from those of $C8$. 
In any event, these results are generally consistent with viewing angle playing 
the major role in the change of apparent speed. 

Contemporaneous to these changes in the inner jet, an increase in the apparent speed 
is observed for $C4$ at $\sim$3~mas from the core at epoch $\sim$1998.2 
\citep{H03}. Our data, combined with previously published results at 43~GHz 
\citep{W01,J01}, show that variations in the trajectory and apparent speed of $C4$
occurred at least twice between 1995 and 2001. This identifies three segments ($a$, $b$, and $c$)
of the trajectory (denoted by dotted lines in Fig. \ref{traj}) where the motion is ballistic
according to a $\chi^2$ test. The ballistic sections
of the trajectory define two epochs, 1996.1$\pm$0.1 and 1998.07$\pm$0.03, when 
the apparent speed changes. The values of $\beta_{\rm app}$ corresponding to these paths 
are given in Table \ref{Speed} using designations $C4a, C4b$, and $C4c$, respectively.
Table \ref{Speed} shows that $C4$ had a high apparent speed ($\beta_{app}\sim 13c$) in
1995 and decelerated dramatically in 1996, then accelerated again from 1998 until 
at least 2002 according to \citet{H03}. Both periods 
of acceleration correspond to fading of the component, while during times of lower
apparent speed $C4$ brightened. Such behavior may be the result of 
helical motion, with the radius of the helix increasing outward. 
  
{\bf 1510$-$089:} The radio emission of this quasar is strongly core dominated
with usually weak extended jet structure. Our data indicate
that the quasar possesses an ultra-relativistic jet with apparent speed
up to $\sim$46$c$. However, even bimonthly observations are not sufficient to state 
that the identification of components shown in Figures \ref{Alli}, \ref{Ident}, and
\ref{1510_bw4} is unique. On the other hand, this identification is supported by 
the total and polarized intensity images along with the total intensity
light curves at 37~GHz \citep{T04} and 14.5~GHz (UMRAO data base).
We detect three moving components, $B1$ (Fig. \ref{Alli}), $B2$, and $B3$ (Fig. \ref{1510_bw4}).
In all three cases we observe a similar polarization behavior in the core region
when the component is emerging from the core - the core region is strongly polarized
with EVPA along the jet direction (see Fig. \ref{Alli}).
In the cases of components $B2$ and $B3$ polarized features with similar EVPA 
are subsequently observed propogating down the jet (Fig. \ref{1510_bw4}). 
For all three components the time of the ejection
falls on the rising branch of an outburst in the total intensity light curves
(an especially prominent flare is connected with the ejection of component $B2$). 
The modeling of the jet structure shows two features, $A1$ and $A2$, that
fluctuate near positions 0.14 and 0.54~mas, respectively (Fig. \ref{Ident}).
Component $A1$ is usually canot be resolved from a moving component
at epochs when the latter is close to $A1$, while $A2$ shifts
downstream when a moving component is approching and returns back
when the moving knot has passed by (see Fig. \ref{Ident}). Such a behavior
of jet features formed by stationary shocks is seen in numerical 
simulations of jets with variable input flow velocities \citep{A03}.
At many epochs, a feature, $S$, with flux
significantly above the noise is seen southeast of the core 
(Fig. \ref{Map}), close to the position angle of the arcsecond scale jet 
\citep{H1510}. On the other hand, this component is located at a similar distance 
from the core as $A1$ but in the opposite direction. This suggests
possible artificial generation of the knot owing to modeling of an actually slightly
extended core using a point-like source. However, there is no correlation between 
the positions and fluxes of $S$ and $A1$. The position angle of $S$ varies within
the range 71$^\circ$ to 168$^\circ$ and the flux varies from 0.03~Jy to 0.5~Jy,
while the values of $\Theta$ and the flux of $A1$ are fairly stable (parameters 
of $A1$ and $A2$ in Table \ref{Speed} are computed for
epochs when the features are not confused with moving knots). Differences in the
properties of $S$ and $A1$ suggest that these features along with $A0$
constitute a complex core region inside the 43~GHz synthesized beam. 

\begin{figure}
\epsscale{1.0}
\plotone{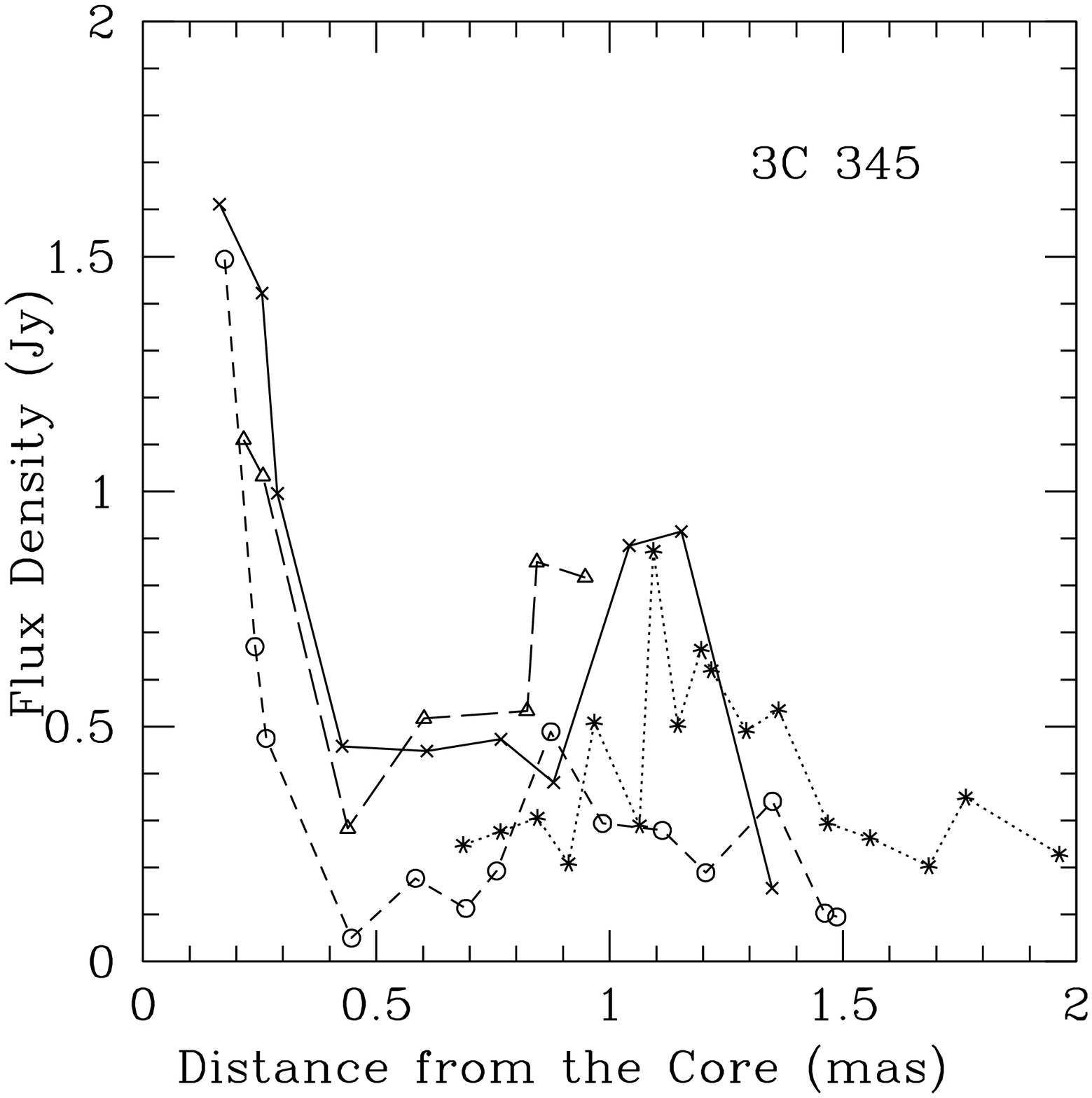}
\caption{Total flux density of knots $C9$ (asterisks), $C10$ (circles), $C11$ (crosses), 
and $C12$ (triangles) in the quasar 3C~345 at different distances from the core.
\label{3c345_Cflux}}
\end{figure}

{\bf 3C 345:} At many epochs the core of this quasar is not the brightest feature 
of the jet (Fig. \ref{Allj}). The core is usually unpolarized at 43~GHz 
(Fig. \ref{Map}), although the fractional polarization of the component nearest to 
the core, $A1$, can be as high as 20\%. The latter fact suggests the low
polarization of the core is intrinsic rather than caused by Faraday effects.
We detect seven superluminal 
knots with average apparent speeds in the range 9-24$c$ (Fig. \ref{Ident},
where the component designations follow \citealt{RZL00}), and at least four of 
these exhibit acceleration or deceleration (Figs. \ref{traj} and \ref{V_Change}). 
For knots $C10-C13$ statistically significant acceleration 
is found only for $C11$. Similar acceleration was reported 
for components $C4$ and $C7$ by \citet{BCUP83} and \citet{RZL00}, 
although beyond 2~mas from the core $C7$ decelerates as indicated by both 
$\dot{\mu}_\parallel$ and interpolation of its position relative to 1996.81 
as observed by \citet{RZL00} (asterisks in Fig. \ref{Ident}). 
It appears that $C8$ repeats the stages of $C7$, 
although it experiences an earlier and greater deceleration, possibly signifying a 
more complex trajectory (see Fig. \ref{traj}). The change in apparent speed and 
the twisted trajectory are also pronounced for $C9$. Although the 
historically extrapolated position of $C9$ (using the average apparent speed of $15.6c$) 
agrees well with the measured position at epoch 1996.81 \citep{RZL00}, it  
decelerates significantly from $\sim 20c$ near the core to $\sim 10c$ at 
1~mas, and then accelerates beyond 1.5~mas. The deceleration is
accompanied by brightening of the knot and rotation of the EVPA 
by $\sim 50^\circ$, making it almost perpendicular to the jet direction 
(see epochs 1998.76 and 1998.94 in Fig. \ref{Allj}). 
In addition $C9$ has the most southern position angle in the jet 
during the deceleration period ($\Theta\sim -116^\circ$), similar to that observed 
for $C8$ by \citet[][$\Theta\sim -122^\circ$]{RZL00}.
\citet{W94} interpreted a change they observed in apparent speed in terms of a flow with
constant Lorentz factor along a bent trajectory. 
Component $C9$ expands as it propagates down the jet and leaves behind 
a complex structure that consists of knot $c9$ moving at apparent 
speed $\sim 9c$ (Fig. \ref{Ident}) plus two extended features, $c9n$ and $c9s$, 
that form later at the northern and southern edges 
of the jet, respectively (Fig. \ref{Map}).  
The majority of moving features have a trajectory with zig-zags in the north-south 
direction (although the net motion is to the west), an increase in flux, and rotation 
of the EVPA at a distance $\sim$0.8-1~mas from the core 
(Fig. \ref{3c345_Cflux}). These variations could be the result of interaction 
with the external medium, as suggested by \citet{RZL00}. We estimate
the deprojected distance of this possible interaction as $\sim$300~pc. 

{\bf 1803+784:} The curved jet of this BL Lac object has been interpreted
as the result of helical structure \citep{BWK01,TKKL02}. 
Our high resolution 43~GHz images show at least two
superluminal components near the core, $B1$ and $B2$. These follow a similar 
curved path  (Fig. \ref{traj}) and have average apparent speeds of
$\sim 16c$, with acceleration and deceleration along the trajectory (Fig. \ref{V_Change}). 
A passage from deceleration to acceleration and vice versa is detected three times
for $B1$, and locations of these transitions along the jet coincide
with bends in the trajectory (compare Figs. \ref{traj} and \ref{V_Change}). 
This behavior supports the hypothesis of helical structure of the jet. 
\citet{GC03} classify the jet feature 
$\sim$1.5~mas from the core as a stationary knot, $S$. In our data this feature 
appears to be associated with $A2$, and has an upper limit to its apparent 
speed of $\sim 1.5c$ (Fig. \ref{Ident}). No motion is detected for bright knot 
$A1$ near the core. \citet{TKKL02} found that a superluminal component 
at 8.5~GHz only becomes visible at a distance $\geq$1~mas from the core.  
Moving component $C1$ is 
detected beyond $A2$ and has an apparent speed of $\sim 10c$, similar to that found by 
\citet{BK95}. The fractional polarization in the jet is high. For example, $A2$ is 
polarized up to 30\% with the EVPA aligned with the jet direction.
In contrast, the EVPA in the core region is essentiallyradial (see Fig. \ref{Map}),
supporting the inference of a toroidal magnetic field geometry \citet{GD00}, but only if 
the viewing angle is almost $0^\circ$ in the immediate vicinity of the core. It is
interesting to note that the viewing angle of the jet derived in \S 6 is 
indeed quite small.

{\bf 1823+568:} The 43~GHz images of this BL Lac object contain bright stable
features $A1, A2$, and $A3$ within 0.5~mas of the core (Fig. \ref{Ident}). 
All of these components are strongly polarized, especially $A3$, where
the fractional polarization reaches 25\% (Fig. \ref{Map}).  
Four components, $A2, A3, C2,$ and $C1$, detected in our images can be
identified with components $K4, K3, K2$, and $K1$, respectively, from \citet{GC00}.
The locations of  $A2$ and $A3$ are slightly farther from
the core than $K4$ and $K3$ (epoch 1995.58). This can be attributed to a shift 
in the core position at 43~GHz relative to 22~GHz due to opacity effects, in  
which case the positions of these features would be stable
over a period of 5 years. Despite excellent time coverage (Fig. \ref{Alll}), 
we do not detect motion 
inside 1~mas of the core in our observations. This has 
several possible explanations:  1) ejections of superluminal components are rare  
in this object, with mean time between ejections $\geq$ 20 years in the observer's frame 
(see Table \ref{Speed}); 2) moving components are much fainter than stationary 
features; or 3) the innermost part of the jet is very close to the line of sight. 
However, we do detect two moving components, $C1$ and $C2$, beyond 1~mas from 
the core where the jet bends slightly to the west (see Fig. \ref{Map}).  
Knot $C1$ has a low apparent speed of 3.8c that appears to be essentially constant, 
since the linear extrapolation of its motion back to epoch 1995.58 agrees 
well (within 0.03~mas) with the position of knot $K1$
in the image of \citet{GC00}. For the fastest moving feature in the jet, 
$C2$ ($\beta_{app}\sim 8c$), extrapolation from our sequence produces a
position of the component at epoch 1995.58 that is farther 
from the core by $\sim$0.1~mas than was observed for $K2$. This implies a 
slower apparent speed farther from the core. Motion of $C2$ is not detected
in our images after it passes $C1$ (Fig. \ref{Ident}).

{\bf BL Lac:}  Our data for BL~Lac have been analyzed in \citet{AS03}. At least
three new components are ejected during the monitoring program (Fig. \ref{Ident}).
These, together with previously ejected knots observed by \citet{D00}, form 
a sequence of superluminal features that appear to emerge from the core 
ballistically along different position angles, indicating a change in 
direction of the jet ``nozzle.'' The trajectories of the knots exhibit 
curvature that is especially pronounced for components $S10, S11$, and $S13$ 
(Fig. \ref{traj}). Despite different average apparent speeds, the components 
accelerate in a similar manner suggestive of a small change 
in the Lorentz factor and a substantial change in direction 
(see Table \ref{Accel} and Fig. \ref{V_Change}). 
The core region can be represented by two point-like features, 
$C1$ and $C2$ \citep[][and Fig. \ref{Map}]{AS03}. In the inner jet  
two bright stationary knots, $A1$ and $A2$, are 
observed at all epochs (see Figs. \ref{Ident} and \ref{Map}).
  
{\bf CTA 102:} The apparent velocity of jet components in this quasar is fairly high,
$\beta_{app}\sim 18c$, in agreement with previous findings \citep{J01,R03}.
We label newly detected moving components as $B5$ and $B6$ (Fig. \ref{Ident}) 
following the scheme of \citet{J01} and rename $J$ in \citet{R03} as 
$B4$ based on its estimated ejection time. Components $B5$ and $B6$ have ballistic 
motion. However, at $\sim$ 0.2~mas from the core both knots split, forming 
more slowly moving components $b5$ and $b6$ behind the main features. 
At some epochs (e.g., Fig. \ref{Map}) $B5$ is 
significantly brighter than the core both in total and polarized intensity. 
There are two features, $A1$ (near the core) and $C$ ($\sim$2~mas from the core), 
that do not move according to a $\chi^2$ test. Component $C$ was detected by 
\citet{J01} and classified as probably stationary, although \citet{R03} associated 
the feature with different moving components. As noted by \citet{J01}, $C$ is 
strongly variable in brightness and located where the jet bends sharply to the 
south. 
In 1998-1999, $C$ was bright, very diffuse, and showed a double
structure that includes knot $c$ closer to the core (see Fig. \ref{Map}). 
We detect no motion of $c$ (Fig. \ref{Ident}). Both components fade 
dramatically by the end of the monitoring (Fig. \ref{Alln}) despite the approach of
the growing inner part of the jet toward $C$. This behavior appears
similar to that observed for component $A3$ in OJ~287 (see above), where
we associate the fading of $A3$ with a change in the jet direction.
Figure \ref{traj} shows the trajectories of $B5$ and $B6$, as well as 
$B1$ and $B3$ from \citet{J01}, plus the position of $C$ during both 
sets of observations. The location of $C$ fluctuates over the range 
1-1.5~mas from the core in right ascension and within 1.4-1.6~mas (to the south) 
in declination. The innermost jet shows a significant ($\sim 20^\circ$)
difference in projected position angle of $B5$ and $B6$ relative 
to $B1-B3$, suggestive of wobbling of the jet ``nozzle'' and/or change of
the VLBI core position. These properties imply interaction between the moving 
flow and external medium at a deprojected distance $\sim$1.4~kpc from 
the core that results in the formation of a diffuse feature.

{\bf 3C 454.3:} This quasar has very complex structure at 43~GHz. 
Figure \ref{Ident} plots only bright components (flux density $>$0.1~Jy). One of the most
interesting features of the jet is highly polarized component $C$. At all epochs 
(see Figs. \ref{Allo} and \ref{Map}) the EVPA of the brightness peak of $C$ 
aligns with the jet axis. 
No change in separation of the component from the core has been seen during either
previous observations \citep{J01} or our current sequence of images. However,
combining the data from the two studies reveals a drift of $C$ 
to the southwest (see Fig. \ref{traj}) at an apparent speed of $\sim 1.6c$ between 
1995 and 2001 (the component defined in this way is referred to as $C*$ 
in Table \ref{Speed}), while the size of 
the knot remains fairly stable at $\sim 0.25\pm 0.10$~mas. The absence of proper 
motion within each data set suggests that the observed difference in position of 
the feature ($\sim$0.2~mas in declination and $\sim$0.1~mas in right ascension) 
could be caused by a displacement to the northeast of the VLBI core between 1997.6 
and 1998.2.  
Beyond $C$, trajectories of moving components separate into two groups: 
northwestern ($B3,B6$) and southwestern ($B4$), with a difference between the
projected trajectories that can reach $\sim 70^\circ$ (Fig. \ref{traj}). 
This is consistent with the direction of the jet near the location 
of $C$ lying very close to the line of sight, thereby producing 
large differences in projected trajectories of moving components beyond $C$ even 
if the intrinsic trajectories differ only slightly. 
The proper motions that we measure cover the entire range from 0.12 to 0.53~mas~yr$^{-1}$
found previously in the jet of 3C~454.3 \citep[][and references therein]{GMA99,J01}.
Component $B6$ accelerates 
markedly (see Table \ref{Accel}). A similar acceleration is derived for
a bright feature detected in 1998 beyond $C$ if we associate it with $B3$ 
found by \citet{J01} based on its brightness, position, and time of ejection. 
This gives an increase in apparent speed of $B3$
from 6$c$ near the core to 13$c$ beyond $C$. Both of these components 
belong to the northwestern group (Fig. \ref{traj}). Northwestern components 
appear to accelerate significantly without a change in direction of motion
while southwestern knots (e.g., $B4$) show mostly a change in direction 
(Fig \ref{V_Change}). A diffuse feature (designated in Fig. \ref{Ident} 
by the symbol $+$) is located in the northern part of the jet and appears to
move after $B4$ passes $C$. Figure \ref{Map} displays 
component $S$ to the south of $C$ when $B4$ approaches $C$.
As observed in other objects discussed above, more slowly moving knots 
$b3$ and $b6$ trail some of the bright fast components ($B3$ and $B6$,
respectively). Figure \ref{3c454_D} shows knot $D$ $\sim$6~mas from the
core. This can be associated with a feature ($A_{\rm PT}$) imaged at longer wavelengths 
by \citet{PT98}. According to \citet{PT98} $A_{\rm PT}$ accelerated in 1984-1989 
from 16$c$ to 30$c$ and then decelerated to 20$c$ in 1992. Our observations
suggest a possible continuation of the deceleration such that the apparent 
speed of the knot has decreased to $6c$. 

\acknowledgements
This material is based on work supported by the National Science Foundation under grants
no. AST-0098579 and AST-0406865. J.L. G\'omez acknowledges support from
the Spanish Ministerio de Educaci\'on y Ciencia and the European Fund for
Regional Development through grant AYA2004-08067-C03-03.
The VLBA is a facility of the National Radio Astronomy
Observatory, operated by Associated Universities Inc. under cooperative agreement
with the National Science Foundation. This research has made use of data from the
University of Michigan Radio Astronomy Observatory, which is supported by the
National Science Foundation and by funds from the University of Michigan.
We are grateful to the referee for the thorough, thoughtful, and useful
criticism.

\clearpage

\begin{deluxetable}{ccrcccccc}
\singlespace
\tablecolumns{9}
\tablecaption{\small\bf Sample and Parameters of Maps shown in Figures \ref{Alla}-\ref{Allo} \label{Sample}}
\tabletypesize{\footnotesize}
\tablehead{
\colhead{Source}&\colhead{Type}&\colhead{z}&\colhead{Beam}&\colhead{P.A.}&\colhead{$I_{\rm peak}$}&\colhead{$I^{\rm p}_{\rm peak}$}&\colhead{$I_{\rm min}$}&\colhead{$I^{\rm p}_{\rm min}$} \\
\colhead{}&\colhead{}&\colhead{}&\colhead{mas$\times$mas}&\colhead{$^\circ$}&\colhead{mJy~Beam$^{-1}$}&\colhead{mJy~Beam$^{-1}$}&\colhead{mJy~Beam$^{-1}$}&\colhead{mJy~Beam$^{-1}$}
}
\startdata
3C~66A&BL&0.444&0.33$\times$0.20&$-$15&745&60&1.9&5 \\
3C~111&RG&0.048&0.30$\times$0.20&$-$10&2960&14&6.0&3 \\
0420$-$014&Q&0.915&0.38$\times$0.15&$-$8&6265&211&12.5&11 \\
3C~120&RG&0.033&0.40$\times$0.15&$-$6&1520&36&7.6&3 \\
0528+134&Q&2.06&0.30$\times$0.15&$-$7&3930&131&18.6&6 \\
OJ~287&BL&0.306&0.30$\times$0.14&$-$7&2330&96&8.1&5 \\
3C~273&Q&0.158&0.38$\times$0.14&$-$6&12800&943&44.8&45 \\
3C~279&Q&0.538&0.38$\times$0.14&$-$6&17240&1240&26.0&62 \\
1510$-$089&Q&0.361&0.40$\times$0.15&$-$6&2580&147&9.0&7 \\
3C~345&Q&0.595&0.22$\times$0.16&$-$9&5055&358&17.7&21 \\
1803+784&BL&0.68&0.16$\times$0.15&25&1490&94&5.2&5 \\
1823+568&BL&0.664&0.19$\times$0.16&$-$10&1580&131&7.9&5 \\
BL~Lac&BL&0.069&0.28$\times$0.15&$-$3&3810&184&9.5&9 \\
CTA~102&Q&1.037&0.25$\times$0.15&$-$3&4180&124&10.4&6 \\
3C~454.3&Q&0.859&0.31$\times$0.15&$-$5&5070&358&17.7&18 \\
\enddata
\end{deluxetable}

\begin{deluxetable}{ccrrrcccc}
\singlespace
\tablecolumns{9}
\tablecaption{\small\bf Log of VLBA Observations  \label{Antenna}}
\tabletypesize{\footnotesize}
\tablehead{
\colhead{Epoch}&\colhead{Antennas}&\colhead{Beam}&\colhead{P.A.}&\colhead{$f_{\rm amp}$}&\colhead{$\sigma$(EVPA)}&\colhead{BIMA}&\colhead{JCMT}&\colhead{Optic.} \\
\colhead{}&\colhead{}&\colhead{mas$\times$mas}&\colhead{$^\circ$}&\colhead{}&\colhead{$^\circ$}
&\colhead{obs.}&\colhead{obs.}&\colhead{obs.}
}
\startdata
03/25/98(1998.23)&ALL&0.29$\times$0.14&$-$5&1.26$\pm$0.10&7&NO&YES&NO \\
05/30/98(1998.41)&ALL&0.29$\times$0.14&$-$7&1.24$\pm$0.08&5&NO&YES&NO \\
08/01/98(1998.58)&ALL&0.29$\times$0.15&$-$11&1.30$\pm$0.07&5&NO&YES&NO \\
10/05/98(1998.76)&ALL&0.29$\times$0.14&$-$4&1.48$\pm$0.20&8&NO&YES&NO \\
12/10/98(1998.94)&ALL&0.29$\times$0.13&$-$9&1.57$\pm$0.11&8&NO&YES&NO \\
02/11/99(1999.12)&ALL&0.31$\times$0.14&$-$12&1.48$\pm$0.15&8&NO&YES&YES \\
04/29/99(1999.33)&ALL,bad weather&0.27$\times$0.14&$-$6&1.27$\pm$0.11&8&NO&YES&YES \\
07/18/99(1999.55)&ALL&0.31$\times$0.15&$-$12&0.7$\pm$0.2&10&NO&NO&NO \\
10/06/99(1999.76)&ALL&0.31$\times$0.16&$-$3&1.38$\pm$0.10&6&NO&YES&NO \\
12/05/99(1999.94)&ALL besides SC&0.38$\times$0.15&$-$20&1.38$\pm$0.12&8&NO&YES&YES \\
01/24/00(2000.07)&ALL&0.27$\times$0.15&$-$5&1.25$\pm$0.08&6&NO&NO&NO \\
04/05/00(2000.26)&ALL&0.28$\times$0.15&$-$8&1.24$\pm$0.08&6&YES&YES&YES \\
07/17/00(2000.54)&ALL&0.28$\times$0.14&$-$7&1.13$\pm$0.10&8&NO&NO&NO \\
10/01/00(2000.75)&ALL besides BR&0.32$\times$0.14&$-$8&1.23$\pm$0.14&8&NO&YES&NO \\
12/11/00(2000.95)&ALL besides MK&0.27$\times$0.14&$-$2&1.40$\pm$0.20&12&YES&YES&YES \\
01/28/01(2001.08)&ALL,bad weather&0.29$\times$0.14&$-$1&1.32$\pm$0.10&15&YES&YES&NO \\
04/14/01(2001.28)&ALL&0.28$\times$0.14&$-$8&1.19$\pm$0.08&5&YES&NO&NO \\
\enddata
\end{deluxetable}

\begin{deluxetable}{lllllllll}
\singlespace
\tablecolumns{9}
\tablecaption{\small\bf Best-Fit Polynomials for Component $C1$ of 3C 66A \label{Poly1}}
\tabletypesize{\footnotesize}
\tablehead{
\colhead{$l$}&\colhead{$a_o$}&\colhead{$a_1$}&\colhead{$a_2$}&\colhead{$a_3$}&\colhead{$a_4$}
&\colhead{$f$}&\colhead{$M$}&\colhead{$M_{\chi^2}$}
}
\startdata
0&$-$0.1566&&&&&15&8.687&7.261 \\
1&$-$0.1575&$-$0.0236&&&&14&6.344&6.571 \\
2&$-$0.1458&$-$0.0233&$-$0.0123&&&13&5.933&5.892 \\
3&$-$0.1448&$-$0.0060&$-$0.0129&$-$0.0109&&12&5.731&5.226 \\
4&$-$0.1327&$-$0.0065&$-$0.0577&$-$0.0106&0.0201&11&5.261&4.575 \\
\tableline
$l$&$b_o$&$b_1$&$b_2$&$b_3$&$b_4$&$f$&$M$&$M_{\chi^2}$ \\
\tableline
0&$-$0.6114&&&&&15&36.30&7.261 \\
1&$-$0.6164&$-$0.1288&&&&14&5.194&6.571 \\
2&$-$0.6410&$-$0.1293&$-$0.0258&&&13&4.395&5.892 \\
3&$-$0.6421&$-$0.1486& 0.0264&$-$0.0121&&12&4.284&5.226 \\
4&$-$0.6424&$-$0.1486& 0.0277&$-$0.0121&$-$0.0006&11&4.284&4.575 \\
\enddata 
\end{deluxetable}

\begin{deluxetable}{lllllllll}
\singlespace
\tablecolumns{9}
\tablecaption{\small\bf Best-fit Polynomials for Component $B1$ of 3C 273 \label{Poly2}}
\tabletypesize{\footnotesize}
\tablehead{
\colhead{$l$}&\colhead{$a_o$}&\colhead{$a_1$}&\colhead{$a_2$}&\colhead{$a_3$}&\colhead{$a_4$}
&\colhead{$f$}&\colhead{$M$}&\colhead{$M_{\chi^2}$}
 }
\startdata
0&$-$0.9252&&&&&15&1009.0&7.261 \\
1&$-$1.384 &$-$0.6840&&&&14&11.87&6.571 \\
2&$-$1.385 &$-$0.6361&0.0415&&&13&10.13&5.892 \\
3&$-$1.367 &$-$0.5981&$-$0.0114&$-$0.0545&&12&7.188&5.226 \\
4&$-$1.364 &$-$0.7077&$-$0.1515&0.0150&0.0791&11&4.242&4.575 \\
\tableline
$l$&$b_o$&$b_1$&$b_2$&$b_3$&$b_4$&$f$&$M$&$M_{\chi^2}$ \\
\tableline
0&$-$0.7363&&&&&15&215.4&7.261 \\
1&$-$1.018&$-$0.4385&&&&14&7.262&6.571 \\
2&$-$1.029&$-$0.3713&0.0683&&&13&4.002&5.892 \\
3&$-$1.027&$-$0.3657&0.0649&$-$0.0051&&12&3.988&5.226 \\
4&$-$1.026&$-$0.3740&0.0527&$-$0.0001&0.0065&11&3.975&4.575 \\
\enddata
\end{deluxetable}

\clearpage
\begin{deluxetable}{lrrrrrlrrrl}
\singlespace
\tablecolumns{11}
\tablecaption{\small\bf Jet Velocities \label{Speed}}
\tabletypesize{\footnotesize}
\tablehead{
\colhead{Source}&\colhead{Knot}&\colhead{$N$}&\colhead{$<S>$}&\colhead{$<R>$}&\colhead{$<\Theta>^\circ$}&
\colhead{$l$}&\colhead{$<\Phi^\circ>$}&\colhead{$<\mu>$}&\colhead{$<\beta_{app}>$}&\colhead{$T_\circ$}\\
\colhead{}&\colhead{}&\colhead{}&\colhead{Jy}&\colhead{mas}&\colhead{}&\colhead{}&\colhead{}
&\colhead{mas~yr$^{-1}$}&\colhead{c}&\colhead{}
 }
\startdata
3C 66A&$C4$&6&0.20&0.12&$-$160.5$\pm$5.6&1&$-$170.3$\pm$1.1&0.06$\pm$0.04&1.5$\pm$1.0&1998.1$\pm$1.2 \\
&$C3$&12&0.10&0.23&$-$158.1$\pm$6.2&1&$-$164.4$\pm$0.7&0.12$\pm$0.02&3.1$\pm$0.5&1998.04$\pm$0.34 \\
&$C2$&16&0.06&0.37&$-$160.1$\pm$4.6&1&$-$161.3$\pm$0.5&0.19$\pm$0.01&5.1$\pm$0.3&1997.94$\pm$0.02 \\
&$C1$&16&0.03&0.63&$-$165.3$\pm$4.2&1&$-$169.6$\pm$0.3&0.13$\pm$0.01&3.5$\pm$0.3&1994.5$\pm$1.1 \\
&$B7$&5&0.03&0.48&$-$163.5$\pm$9.0&1&$-$170.0$\pm$3.7&0.98$\pm$0.13&26.4$\pm$3.6&1998.03$\pm$0.12 \\
&$B6$&11&0.02&1.55&$-$165.6$\pm$3.3&1&$-$173.0$\pm$1.3&0.81$\pm$0.06&21.8$\pm$1.6&1997.27$\pm$0.12 \\
&$B4$&7&0.03&2.04&$-$169.5$\pm$2.9&1$^*$&173.3$\pm$2.5&1.00$\pm$0.12&26.9$\pm$3.3&1996.74$\pm$0.19 \\
&$A1$&16&0.02&1.19&$-$164.4$\pm$2.0&1$^*$&19.3$\pm$4.0&0.14$\pm$0.10&3.8$\pm$2.3&\nodata \\
&$A2$&17&0.05&2.50&$-$176.7$\pm$2.2&1$^*$&$-$42.3$\pm$6.2&0.09$\pm$0.06&2.5$\pm$1.5&\nodata \\ 
3C 111&$B7$&5&0.23&0.43&62.2$\pm$4.1&1&70.4$\pm$6.5&0.77$\pm$0.18&2.6$\pm$0.6&2000.36$\pm$0.12 \\
&$B6$&6&0.09&0.75&58.5$\pm$2.5&1&61.2$\pm$5.7&0.97$\pm$0.12&3.3$\pm$0.4&1999.84$\pm$0.06 \\
&$B5$&8&0.16&0.94&57.2$\pm$1.6&1&56.1$\pm$3.1&0.98$\pm$0.06&3.3$\pm$0.2&1999.23$\pm$0.03 \\
&$B4$&7&0.25&0.66&59.0$\pm$0.8&1&58.8$\pm$5.2&1.00$\pm$0.10&3.4$\pm$0.3&1999.05$\pm$0.03 \\
&$B3$&8&0.20&0.93&63.0$\pm$1.1&1&62.4$\pm$3.9&1.39$\pm$0.08&4.7$\pm$0.3&1998.76$\pm$0.03 \\
&$B2$&6&0.03&0.91&64.4$\pm$0.9&1&65.5$\pm$6.0&1.11$\pm$0.14&3.7$\pm$0.5&1997.85$\pm$0.05 \\
&$B1$&4&0.05&1.08&58.9$\pm$0.9&1&62.6$\pm$12.3&0.83$\pm$0.26&2.8$\pm$0.9&1997.24$\pm$0.11 \\
&$C1$&13&0.20&4.38&65.8$\pm$1.5&1&60.9$\pm$2.9&1.77$\pm$0.06&6.0$\pm$0.2&1996.93$\pm$0.24 \\
&$c1$&16&0.09&3.73&67.1$\pm$2.7&2&56.9$\pm$0.9&1.07$\pm$0.02&3.6$\pm$0.1&1996.31$\pm$0.62 \\
&$c2$&9&0.07&5.29&65.3$\pm$1.8&1&54.5$\pm$2.3&1.23$\pm$0.04&4.1$\pm$0.2&1996.1$\pm$0.9 \\
&$A1$&17&0.58&0.12&63.5$\pm$4.7&0&76.8$\pm$0.7&0.02$\pm$0.03&0.05$\pm$0.09&\nodata \\
&$A2$&12&0.37&0.27&60.2$\pm$3.5&0&31.8$\pm$1.8&0.02$\pm$0.03&0.07$\pm$0.10&\nodata \\
0420$-$014&$B2$&4&1.30&0.13&$-$79.2$\pm$8.1&1&$-$71.7$\pm$5.2&0.16$\pm$0.07&7.6$\pm
$3.5&2000.22$\pm$0.35 \\
&$B1$&14&0.27&0.32&$-$147$\pm$25&3&171.8$\pm$2.2&0.22$\pm$0.06&10.9$\pm$2.8&1998.59$\pm$0.16 \\
3C 120&$v$&4&0.09&1.03&$-$117.8$\pm$1.4&1&$-$116.1$\pm$8.9&1.96$\pm$0.20&4.3$\pm$0.4&2000.36$\pm$0.06 \\
&$u$&8&0.27&1.28&$-$123.1$\pm$1.9&1&$-$122.9$\pm$0.5&1.69$\pm$0.07&3.7$\pm$0.2&1999.70$\pm$0.02 \\
&$u1$&5&0.13&1.10&$-$117.7$\pm$4.6&1$^*$&$-$120.2$\pm$7.5&1.84$\pm$0.15&4.0$\pm$0.3&1999.92$\pm$0.02 \\
&$t$&10&0.17&1.95&$-$125.1$\pm$5.9&1&$-$128.6$\pm$5.2&2.04$\pm$0.08&4.5$\pm$0.2&1999.27$\pm$0.03 \\
&$o1$&16&0.27&3.28&$-$121.4$\pm$3.0&2&$-$118.6$\pm$2.8&2.03$\pm$0.06&4.5$\pm$0.1&1998.07$\pm$0.03 \\
&$o2$&14&0.42&1.85&$-$122.9$\pm$3.2&2&$-$119.1$\pm$1.4&1.52$\pm$0.03&3.4$\pm$0.1&1998.14$\pm$0.07 \\
&$l$&7&0.16&2.85&$-$124.9$\pm$1.9&1&$-$120.9$\pm$6.8&1.66$\pm$0.13&3.6$\pm$0.3&1997.03$\pm$0.14  \\
&$h$&16&0.16&5.49&$-$113.2$\pm$4.1&1$^*$&$-$103.8$\pm$1.7&1.82$\pm$0.06&4.0$\pm$0.1&1996.35$\pm$0.19 \\                     
&$d$&7&0.06&6.95&$-$113.0$\pm$2.0&1&$-$95.0$\pm$5.9&1.66$\pm$0.45&3.6$\pm$1.0&1994.88$\pm$0.28  \\
&$A1$&15&0.22&0.19&$-$122.6$\pm$10.9&1&34.3$\pm$1.7&0.07$\pm$0.03&0.15$\pm$0.06&\nodata \\
0528$+$134&$B8$&11&1.33&0.13&77$\pm$16&2&44.3$\pm$1.0&0.11$\pm$0.02&9.3$\pm$1.4&1999.12$\pm$0.50 \\
&$B8p$&9&0.04&0.19&60$\pm$22&1&33.6$\pm$2.5&0.22$\pm$0.04&18.6$\pm$3.7&1999.61$\pm$0.23 \\
&$B7$&6&0.43&0.16&108$\pm$22&1$^*$&66.9$\pm$3.5&0.25$\pm$0.08&20.8$\pm$7.1&1998.69$\pm$0.55  \\
&$B6$&13&0.32&0.34&86$\pm$20&3&52.9$\pm$3.2&0.26$\pm$0.04&21.6$\pm$3.1&1997.95$\pm$0.28 \\
&$B5$&10&0.08&0.43&75$\pm$12&1$^*$&53.1$\pm$2.0&0.25$\pm$0.04&20.7$\pm$3.1&1997.44$\pm$0.62  \\
&$B4$&9&0.71&0.39&69$\pm$14&3&36.6$\pm$5.6&0.21$\pm$0.08&17.8$\pm$8.2&1995.58$\pm$0.15 \\
&$B3$&6&0.65&0.53&68$\pm$17&1$^*$&31.9$\pm$2.5&0.31$\pm$0.04&26.1$\pm$3.7&1995.46$\pm$0.85 \\
OJ 287&$B6$&4&0.28&0.39&$-$113.9$\pm$11.0&1$^*$&$-$131.9$\pm$12.5&0.93$\pm$0.22&18.0$\pm$4.1&1999.57$\pm$0.11 \\
&$B5$&6&0.28&0.39&$-$103.6$\pm$22.1&1$^*$&$-$117.6$\pm$4.5&0.60$\pm$0.10&11.6$\pm$1.9&1998.61$\pm$0.09 \\
&$B4$&5&0.27&0.33&$-$99.2$\pm$20.5&1$^*$&$-$115.9$\pm$6.5&0.92$\pm$0.14&17.8$\pm$2.8&1998.23$\pm$0.04 \\
&$A1$&17&0.24&0.10&$-$90$\pm$27&1$^*$&$-$153.4$\pm$0.4&0.03$\pm$0.01&0.5$\pm$0.2&1992.67$\pm$5.0 \\
&$A2$&17&0.41&0.30&$-$111$\pm$4&1$^*$&$-$101.4$\pm$0.4&0.03$\pm$0.02&0.6$\pm$0.3&1989.7$\pm$5.0 \\
&$A2*$&24&0.36&0.28&$-$109$\pm$9&1$^*$&$-$178.2$\pm$6.5&0.02$\pm$0.01&0.4$\pm$0.2&$\sim$1970.5 \\
&$A3$&16&0.12&1.00&$-$108.9$\pm$2.2&1&69.3$\pm$1.2&0.06$\pm$0.03&1.3$\pm$0.6&\nodata \\
3C 273&$B2$&10&3.29&1.16&$-$129.6$\pm$4.5&2&$-$123.5$\pm$3.3&1.18$\pm$0.06&12.2$\pm$0.6&1999.25$\pm$0.02 \\
&$b2$&6&1.70&1.19&$-$121.1$\pm$1.0&1$^*$&$-$127.8$\pm$8.1&0.36$\pm$0.15&3.7$\pm$1.5&1997.5$\pm$0.4 \\
&$B1$&16&7.58&1.61&$-$127.7$\pm$4.0&4&$-$120.3$\pm$3.0&0.79$\pm$0.06&8.2$\pm$0.6&1997.89$\pm$0.16 \\
&$b1$&17&4.04&1.16&$-$122.6$\pm$5.0&1&$-$119.4$\pm$1.5&0.47$\pm$0.03&4.9$\pm$0.3&1997.27$\pm$0.16 \\
&$b_s$&8&0.55&2.67&$-$135.1$\pm$2.4&1&$-$139.6$\pm$2.8&1.12$\pm$0.05&11.6$\pm$0.5&1998.13$\pm$0.18 \\
&$A1$&17&1.14&0.16&$-$131.8$\pm$12.3&0&$-$11.8$\pm$0.6&0.00$\pm$0.02&0.0$\pm$0.2&\nodata \\
&$A2$&11&5.74&0.78&$-$123.3$\pm$3.8&1$^*$&$-$104.1$\pm$1.7&0.06$\pm$0.06&0.6$\pm$0.6&1984$\pm$12 \\
&$C1$&10&0.32&8.27&$-$114.5$\pm$1.2&1$^*$&$-$123.9$\pm$8.9&0.67$\pm$0.16&6.9$\pm$1.7&1988.2$\pm$2.2 \\
3C~279&$C16$&6&3.33&0.30&$-$144.9$\pm$7.3&1&$-$135.8$\pm$6.8&0.53$\pm$0.11&16.9$\pm$3.5&2000.27$\pm$0.05 \\
&$C15$&8&2.51&0.42&$-$137.4$\pm$4.9&1&$-$132.4$\pm$4.3&0.54$\pm$0.07&17.2$\pm$2.3&1999.85$\pm$0.04 \\
&$C14$&10&1.34&0.53&$-$134.5$\pm$5.5&1&$-$128.8$\pm$4.2&0.54$\pm$0.08&17.2$\pm$2.4&1999.50$\pm$0.09 \\
&$C13$&13&2.03&0.59&$-$129.8$\pm$4.2&1&$-$126.5$\pm$2.9&0.51$\pm$0.05&16.2$\pm$1.6&1998.99$\pm$0.07 \\
&$C12$&11&2.02&0.54&$-$129.5$\pm$3.1&1&$-$124.2$\pm$2.3&0.49$\pm$0.04&15.7$\pm$1.4&1998.50$\pm$0.08 \\
&$C11$&7&1.93&0.38&$-$135.0$\pm$3.6&1&$-$129.4$\pm$5.8&0.32$\pm$0.10&10.1$\pm$3.2&1997.59$\pm$0.12 \\
&$C10$&14&1.75&0.70&$-$131.1$\pm$5.9&1&$-$119.2$\pm$1.6&0.35$\pm$0.03&11.1$\pm$1.1&1997.23$\pm$0.47  \\
&$C9$&21&2.23&0.98&$-$129.8$\pm$4.9&1&$-$123.4$\pm$0.4&0.40$\pm$0.01&12.8$\pm$0.3&1996.89$\pm$0.22 \\
&$C8*$&27&3.13&0.81&$-$129.3$\pm$3.2&3&$-$126.6$\pm$0.7&0.33$\pm$0.01&10.5$\pm$0.4&1996.10$\pm$0.31 \\
&$C4$&46&1.28&3.54&$-$116.2$\pm$2.0&3&$-$127.9$\pm$0.8&0.31$\pm$0.01&9.9$\pm$0.5&1987.7$\pm$2.7 \\
&$C4a$&5&1.33&2.66&$-$115.0$\pm$1.0.3&1&$-$97.5$\pm$2.1&0.41$\pm$0.13&13.0$\pm$4.2&1988.2$\pm$9.8 \\
&$C4b$&11&1.56&3.19&$-$114.3$\pm$0.5&1&$-$123.8$\pm$3.7&0.18$\pm$0.07&5.7$\pm$2.2&1979.1$\pm$3.4 \\
&$C4c$&30&1.26&3.81&$-$117.0$\pm$1.8&1&$-$140.2$\pm$1.3&0.35$\pm$0.02&11.1$\pm$0.7&1987.5$\pm$4.5 \\
&$E$&4&1.12&0.28&$-$136.2$\pm$4.2&1$^*$&$-$147.8$\pm$1.0&0.04$\pm$0.02&1.3$\pm$0.5&1992.88$\pm$1.41 \\
1510$-$089&$B3$&5&0.24&0.55&$-$46.6$\pm$6.4&1&$-$41.9$\pm$6.0&1.25$\pm$0.10&28.0$\pm$2.4&1999.67$\pm$0.03 \\
&$B2$&4&0.10&0.64&$-$33.8$\pm$2.3&1&$-$29.5$\pm$7.0&1.56$\pm$0.13&35.0$\pm$3.0&1998.81$\pm$0.04 \\
&$B1$&4&0.04&0.78&$-$48.5$\pm$3.6&1&$-$43.2$\pm$9.3&2.04$\pm$0.16&45.9$\pm$3.6&1998.11$\pm$0.09 \\
&$A1$&15&0.28&0.14&$-$49.7$\pm$21.6&1$^*$&$-$72.7$\pm$2.6&0.02$\pm$0.02&0.4$\pm$0.4&\nodata \\
&$A2$&16&0.05&0.54&$-$33.9$\pm$7.6&1$^*$&105.8$\pm$0.9&0.03$\pm$0.02&0.5$\pm$0.4&\nodata \\
&$S$&9&0.18&0.13&127.1$\pm$32.8&1$^*$&17$\pm$1&0.07$\pm$0.02&1.6$\pm$0.5&\nodata \\
3C~345&$C13$&5&0.96&0.40&$-$87.7$\pm$2.3&1$^*$&$-$84.9$\pm$1.3&0.42$\pm$0.09&14.6$\pm$3.3&2000.00$\pm$0.34 \\
&$C12$&7&0.74&0.59&$-$86.5$\pm$2.3&1$^*$&$-$88.6$\pm$0.2&0.66$\pm$0.05&23.1$\pm$1.7&1999.80$\pm$0.59 \\
&$C11$&10&0.78&0.69&$-$95.2$\pm$2.7&2&$-$94.5$\pm$0.4&0.69$\pm$0.03&23.9$\pm$1.1&1999.24$\pm$0.29 \\
&$C10$&14&0.35&0.83&$-$95.0$\pm$4.6&1&$-$102.8$\pm$0.6&0.55$\pm$0.02&19.2$\pm$0.7&1998.60$\pm$0.41 \\
&$C9$&17&0.40&1.23&$-$106.2$\pm$7.5&4&$-$79.1$\pm$1.0&0.45$\pm$0.04&15.6$\pm$1.4&1996.71$\pm$0.26 \\
&$c9$&8&0.78&1.23&$-$110.4$\pm$2.9&1&$-$119.6$\pm$2.1&0.27$\pm$0.04&9.2$\pm$1.5&1996.11$\pm$0.57 \\
&$C8$&10&0.33&1.54&$-$97.2$\pm$5.6&3&$-$65.6$\pm$4.7&0.32$\pm$0.10&11.0$\pm$3.6&1994.0$\pm$0.8 \\
&$C7$&16&0.14&2.89&$-$96.8$\pm$1.4&2&$-$81.4$\pm$0.3&0.29$\pm$0.02&10.1$\pm$0.5&1993.77$\pm$0.36 \\
&$A1$&14&2.29&0.10&$-$84.6$\pm$5.2&0&$-$100.4$\pm$0.5&0.02$\pm$0.02&0.7$\pm$0.6&\nodata \\
1803$+$784&$B2$&13&0.12&0.38&$-$62.2$\pm$20.8&4&$-$75.4$\pm$1.6&0.42$\pm$0.05&15.9$\pm$1.9&1998.58$\pm$0.15 \\
&$B1$&14&0.14&0.70&$-$77.5$\pm$5.1&4&$-$60.0$\pm$2.4&0.42$\pm$0.05&15.7$\pm$1.8&1997.64$\pm$0.08 \\
&$A1$&17&0.21&0.28&$-$77.9$\pm$4.5&1$^*$&150.0$\pm$0.7&0.01$\pm$0.01&0.4$\pm$0.5&\nodata \\
&$A2$&17&0.11&1.44&$-$90.1$\pm$2.2&1$^*$&$-$36.1$\pm$0.8&0.04$\pm$0.01&1.5$\pm$0.5&\nodata \\
&$C1$&7&0.12&1.66&$-$96.0$\pm$1.1&1&$-$93.2$\pm$0.3&0.28$\pm$0.04&10.5$\pm$1.5&1992.6$\pm$3.6 \\
1823$+$568&$A1$&17&0.54&0.06&$-$163.2$\pm$10.4&0&$-$23.4$\pm$0.4&0.00$\pm$0.01&0.1$\pm$0.3&\nodata \\
&$A2$&17&0.07&0.21&$-$164.4$\pm$5.2&0&$-$40.9$\pm$0.5&0.02$\pm$0.01&0.6$\pm$0.3&\nodata \\
&$A3$&17&0.11&0.52&$-$162.5$\pm$2.2&0&$-$89.7$\pm$0.6&0.02$\pm$0.01&0.7$\pm$0.4&\nodata \\
&$C2$&11&0.04&1.61&$-$162.1$\pm$1.1&1&$-$174.9$\pm$0.3&0.20$\pm$0.02&7.7$\pm$0.6&1991.5$\pm$1.0 \\
&$C1$&16&0.04&1.91&$-$160.8$\pm$2.4&1&$-$153.1$\pm$0.6&0.10$\pm$0.01&3.8$\pm$0.5&1979.4$\pm$1.3 \\
BL Lac&$S13$&10&0.10&2.27&$-$158.3$\pm$5.1&3&$-$170.2$\pm$2.9&1.95$\pm$0.11&9.0$\pm$0.5&1999.29$\pm$0.04 \\
&$S12$&9&0.10&1.83&$-$158.3$\pm$1.2&1&$-$163.2$\pm$2.4&1.82$\pm$0.06&8.4$\pm$0.3&1998.93$\pm$0.12 \\
&$S11$&10&0.11&1.58&$-$159.3$\pm$4.4&3&$-$167.1$\pm$4.2&1.36$\pm$0.13&6.3$\pm$0.6&1998.02$\pm$0.12 \\
&$S10$&10&0.33&2.20&$-$167.6$\pm$6.1&2&$-$175.8$\pm$2.4&0.91$\pm$0.03&4.2$\pm$0.2&1996.93$\pm$0.08 \\
&$A1$&17&0.73&0.10&$-$160.4$\pm$9.7&0&$-$178.9$\pm$0.3&0.01$\pm$0.01&0.1$\pm$0.1&\nodata \\
&$A2$&17&0.34&0.29&$-$158.9$\pm$6.4&1$^*$&143.9$\pm$0.6&0.01$\pm$0.01&0.1$\pm$0.1&\nodata \\
CTA 102&$B6$&10&0.92&0.30&118.2$\pm$9.4&1&121.9$\pm$1.9&0.34$\pm$0.04&18.2$\pm$1.9&1999.54$\pm$0.04 \\
&$b6$&5&0.24&0.32&121.1$\pm$2.4&1&125.0$\pm$6.0&0.16$\pm$0.11&8.5$\pm$6.0&1998.9$\pm$1.5 \\
&$B5$&16&0.99&0.50&112.8$\pm$3.9&1&119.7$\pm$0.9&0.29$\pm$0.02&15.4$\pm$0.9&1997.9$\pm$0.2 \\
&$b5$&12&0.29&0.52&115.3$\pm$3.9&1$^*$&123.8$\pm$1.4&0.23$\pm$0.03&12.5$\pm$1.5&1998.0$\pm$0.3 \\
&$A1$&16&0.59&0.12&117.1$\pm$13.0&0&$-$129.5$\pm$0.7&0.01$\pm$0.01&0.6$\pm$0.7&\nodata \\
&$C$&15&0.13&1.99&137.6$\pm$1.6&0&96.2$\pm$0.7&0.01$\pm$0.05&0.6$\pm$2.6&\nodata \\
3C 454.3&$B6$&7&1.64&0.31&$-$101.0$\pm$21.8&2&$-$79.5$\pm$1.2&0.53$\pm$0.05&24.8$\pm$2.5&1999.80$\pm$0.37 \\
&$b6$&4&0.65&0.26&$-$92.1$\pm$2.8&0&$-$67.2$\pm$3.8&0.03$\pm$0.09&1.5$\pm$4.3&\nodata \\
&$B5$&5&0.31&0.19&$-$85.1$\pm$4.7&1&$-$88.7$\pm$0.3&0.22$\pm$0.09&10.0$\pm$4.3&1999.04$\pm$0.44 \\
&$B4$&15&0.68&0.61&$-$86.2$\pm$10.5&2&$-$98.3$\pm$0.4&0.41$\pm$0.02&19.0$\pm$1.1&1998.36$\pm$0.07 \\
&$B3$&7&0.85&1.37&$-$74.6$\pm$1.2&1&$-$78.1$\pm$2.7&0.28$\pm$0.11&13.2$\pm$5.3&1995.70$\pm$0.27 \\
&$b3$&6&0.79&0.73&$-$90.9$\pm$2.7&1$^*$&$-$102.1$\pm$3.5&0.23$\pm$0.15&10.7$\pm$6.8&1995.3$\pm$2.2 \\
&$C$&17&1.27&0.63&$-$79.5$\pm$4.1&1$^*$&2.7$\pm$0.2&0.01$\pm$0.01&0.4$\pm$0.6&\nodata \\
&$C^*$&29&1.44&0.63&$-$73.4$\pm$8.4&1$^*$&$-$165.1$\pm$0.3&0.03$\pm$0.01&1.6$\pm$0.3&\nodata \\
&$N$&14&0.37&0.86&$-$43.9$\pm$7.0&1$^*$&$-$14.6$\pm$0.9&0.16$\pm$0.02&7.7$\pm$1.0&1992$\pm$5 \\
&$D$&16&0.11&5.80&$-$83.1$\pm$1.3&1&$-$21.6$\pm$1.9&0.12$\pm$0.04&5.5$\pm$1.9&\nodata \\
\enddata
\end{deluxetable}     
\clearpage
\begin{deluxetable}{rrrrrrrrr}
\singlespace
\tablecolumns{9}
\tablecaption{\small\bf Acceleration in the Jets \label{Accel}}
\tabletypesize{\footnotesize}        
\tablehead{
\colhead{Source}&\colhead{Knot}&\colhead{$N$}&\colhead{$<\mu>$}&\colhead{$<\Phi^\circ>$}
&\colhead{$\dot{\mu}_\parallel$}&\colhead{$\dot{\mu}_\perp$}\\
\colhead{}&\colhead{}&\colhead{}&\colhead{mas~yr$^{-1}$}&\colhead{}&\colhead{mas~yr$^{-2}$}&\colhead{mas~yr$^{-2}$}
 }
\startdata 
0420$-$014&$B1$&14&0.22$\pm$0.06&171.8$\pm$2.2&0.07$\pm$0.02&0.02$\pm$0.01 \\
3C 111&$c1$&16&1.07$\pm$0.02&56.9$\pm$0.9&$-$0.22$\pm$0.02&$-$0.26$\pm$0.02 \\
3C 120&$o1$&16&2.04$\pm$0.06&$-$118.6$\pm$2.8&0.29$\pm$0.02&0.16$\pm$0.02 \\
&$o2$&14&1.52$\pm$0.03&$-$119.1$\pm$1.4&$-$0.66$\pm$0.03&$-$0.12$\pm$0.04 \\
0528$+$134&$B8$&11&0.11$\pm$0.02&44.3$\pm$1.0&0.10$\pm$0.01&$-$0.07$\pm$0.02 \\
&$B6$&13&0.26$\pm$0.04&52.9$\pm$2.2&0.06$\pm$0.03&$-$0.17$\pm$0.04 \\
&$B4$&9&0.21$\pm$0.08&36.6$\pm$5.6&0.11$\pm$0.02&$-$0.09$\pm$0.03 \\
3C 273&$B1$&16&0.79$\pm$0.06&$-$120.3$\pm$3.0&$-$0.14$\pm$0.03&0.10$\pm$0.02 \\
&$B2$&10&1.18$\pm$0.06&$-$123.5$\pm$3.3&0.04$\pm$0.03&0.26$\pm$0.04 \\
3C 279&$C8*$&27&0.33$\pm$0.01&$-$126.6$\pm$0.7&0.07$\pm$0.01&0.02$\pm$0.01 \\
&$C4$&46&0.31$\pm$0.01&$-$127.9$\pm$0.8&0.02$\pm$0.01&$-$0.05$\pm$0.01 \\
3C 345&$C11$&10&0.69$\pm$0.03&$-$94.5$\pm$0.4&0.22$\pm$0.03&0.19$\pm$0.04 \\
&$C9$&17&0.45$\pm$0.04&$-$79.1$\pm$1.0&0.11$\pm$0.02&$-$0.08$\pm$0.03 \\
&$C8$&10&0.32$\pm$0.10&$-$65.6$\pm$4.7&$-$0.20$\pm$0.03&$-$0.23$\pm$0.03 \\
&$C7$&16&0.29$\pm$0.02&$-$81.4$\pm$0.3&$-$0.07$\pm$0.01&$-$0.01$\pm$0.01 \\
1803$+$784&$B2$&13&0.42$\pm$0.05&$-$75.4$\pm$1.6&0.30$\pm$0.05&$-$0.10$\pm$0.04 \\
&$B1$&14&0.42$\pm$0.05&$-$60.0$\pm$2.4&0.52$\pm$0.04&$-$0.29$\pm$0.03 \\
BL Lac&$S13$&10&1.95$\pm$0.11&$-$170.2$\pm$2.9&$-$0.21$\pm$0.07&$-$0.76$\pm$0.06 \\
&$S11$&10&1.36$\pm$0.13&$-$167.1$\pm$4.2&0.07$\pm$0.06&$-$0.62$\pm$0.05 \\
&$S10$&10&0.91$\pm$0.03&$-$175.8$\pm$2.4&$-$0.32$\pm$0.03&$-$0.71$\pm$0.04 \\
3C 454.3&$B6$&7&0.53$\pm$0.05&$-$79.5$\pm$1.2&0.37$\pm$0.06&0.24$\pm$0.07 \\
&$B4$&15&0.41$\pm$0.02&$-$98.3$\pm$0.4&$-$0.04$\pm$0.01&$-$0.09$\pm$0.02 \\
\enddata
\end{deluxetable}     
\clearpage
\begin{deluxetable}{crrccccc}
\singlespace
\tablecolumns{8}
\tablecaption{\small\bf Parameters of Total and Polarized Intensity Maps shown in Figure \ref{Map} \label{Tmap}}
\tabletypesize{\footnotesize}
\tablehead{
\colhead{Source}&\colhead{Beam}&\colhead{P.A.}&\colhead{$I_{\rm peak}$}&\colhead{$I^p_{\rm peak}$}&\colhead{$I_{\rm min}$}&\colhead{rms(I)}&\colhead{rms$(I^{\rm p})$} \\
\colhead{}&\colhead{mas$\times$mas}&\colhead{$^\circ$}&\colhead{mJy~Beam$^{-1}$}
&\colhead{mJy~Beam$^{-1}$}&\colhead{mJy~Beam$^{-1}$}&\colhead{mJy~Beam$^{-1}$}&\colhead{mJy~Beam$^{-1}$}
}
\startdata
3C~66A&0.32$\times$0.19&$-$13&500&30&1.5&0.7&0.5 \\
3C~111&0.30$\times$0.18&$-$10&803&6&2.0&0.9&0.9 \\
0420$-$014&0.36$\times$0.14&$-$8&3504&112&10.5&4.0&2.7 \\
3C~120&0.38$\times$0.15&$-$6&650&11&2.9&0.7&1.4 \\
0528+134&0.30$\times$0.15&$-$7&1426&53&3.8&1.8&1.2 \\
OJ~287&0.30$\times$0.14&$-$8&1115&32&4.4&2.5&2.6 \\
3C~273&0.42$\times$0.17&$-$6&7178&655&35.9&7.6&14.1 \\
3C~279&0.38$\times$0.14&$-$7&9695&432&24.2&3.5&3.1 \\
1510$-$089&0.40$\times$0.15&$-$5&1517&36&7.6&1.5&3.4 \\
3C~345&0.22$\times$0.15&$-$10&1470&144&7.4&1.2&1.4 \\
1803+784&0.16$\times$0.15&30&1211&61&4.2&1.3&2.2 \\
1823+568&0.19$\times$0.15&$-$11&781&55&3.9&0.6&1.4 \\
BL~Lac&0.28$\times$0.15&$-$4&2709&128&6.8&2.2&2.7 \\
CTA~102&0.24$\times$0.16&$-$9&2456&109&6.1&3.1&4.6 \\
3C~454.3&0.30$\times$0.15&$-$6&3729&206&9.3&1.8&2.9 \\
\enddata
\end{deluxetable}
\clearpage
\begin{deluxetable}{lrrcrcc}
\singlespace
\tablecolumns{7}
\tablecaption{\small\bf Jet Components in Quasars \label{Q_Param}}
\tabletypesize{\footnotesize}
\tablehead{
\colhead{Source}&\colhead{Knot}&\colhead{$\Gamma^{\rm var}$}&\colhead{$\Theta_\circ^{\rm var}(^\circ)$}&\colhead{$\delta^{\rm var}$}&\colhead{$T_{\rm b,obs}(10^{10}~K)$}&\colhead{$T_{\rm b,int}(10^9~K)$}
 }
\startdata
0420$-$014&$B2$&10.4&2.4&$<$17.5&4.69&1.09\\
&$B1$&11.5&3.6&$<$15.0&1.64&0.49 \\
0528$+$134&$B8$&27.4&0.4&$<$53.2&7.94&0.62 \\
&$B7$&35.2&0.5&$<$63.6&2.17&0.125 \\
&$B6$&25.5&1.2&39.2&2.58&0.34 \\
&$B5$&21.1&3.3&17.3&0.38&0.20 \\  
&$B4$&27.5&0.8&48.5&3.72&0.34 \\
&$B3$&39.5&0.6&69.2&1.07&0.05  \\
3C 273&$B2$&13.8&6.9&7.4&15.4&6.62 \\
&$B1$&8.3&6.1&9.4&16.4&4.66 \\
&$C1$&8.2&3.8&12.6&0.23&0.04  \\
3C~279&$C16$&17.1&2.9&19.7&20.2&2.65 \\
&$C15$&18.2&2.3&23.9&25.3&2.38 \\
&$C14$&17.4&2.8&20.0&9.37&1.19 \\
&$C13$&16.3&3.3&17.6&16.1&2.57 \\
&$C12$&16.6&2.5&22.0&23.3&2.52 \\
&$C11$&12.9&2.2&$<$20.9&12.6&1.49 \\
&$C10$&14.9&1.7&24.7&4.96&0.44 \\
&$C9$&16.7&1.6&27.3&3.74&0.28 \\
&$C8$&17.2&0.6&33.4&6.47&0.35 \\
&$C4$&10.5&3.9&13.8&1.05&0.25 \\
1510$-$089&$B3$&30.1&1.3&41.3&1.10&0.03 \\
&$B2$&35.2&1.5&39.0&0.73&0.02 \\
&$B1$&47.9&1.6&34.2&0.21&0.01 \\
3C~345&$C13$&16.2&2.3&23.1&7.41&0.79 \\
&$C12$&24.6&1.6&32.9&3.25&0.19  \\
&$C11$&29.8&3.8&12.0&7.37&2.39 \\
&$C10$&19.3&3.3&17.6&12.1&2.05 \\
&$C9$&16.2&4.6&11.9&1.51&0.49 \\
&$C8$&19.9&0.9&36.5&0.50&0.025 \\
&$C7$&13.3&2.0&21.9&0.02&0.003 \\
CTA 102&$B6$&20.5&1.7&$<$29.8&3.54&0.37 \\
&$B5$&15.7&3.0&18.8&14.4&3.29 \\
3C 454.3&$B6$&25.2&1.9&29.7&17.6&1.58 \\
&$B5$&14.8&1.5&$<$25.8&1.58&0.18 \\
&$B4$&20.0&3.9&13.9&4.54&1.49 \\
&$B3$&13.4&3.7&15.4&24.7&6.76 \\
&$C$&14.2&0.2&28.2&6.76&0.66 \\
&$D$&10.2&1.7&18.7&0.03&0.007 \\
\enddata
\end{deluxetable} 

\begin{deluxetable}{lrrcrcc}
\singlespace
\tablecolumns{7}
\tablecaption{\small\bf Jet Components in BL Lac Objects\label{B_Param}}
\tabletypesize{\footnotesize}
\tablehead{
\colhead{Source}&\colhead{Knot}&\colhead{$\Gamma^{\rm var}$}&\colhead{$\Theta_\circ^{\rm var}(^\circ)$}&\colhead{$\delta^{\rm var}$}&\colhead{$T_{\rm b,obs}(10^{10}~K)$}&\colhead{$T_{\rm b,int}(10^9~K)$}
 }
\startdata
3C 66A&$C4$&5.0&1.8&$<$9.7&0.73&0.285 \\
&$C3$&5.9&2.8&$<$10.9&0.44&0.141 \\
&$C2$&7.7&2.8&$<$13.5&0.35&0.079 \\
&$C1$&11.4&0.8&22.2&0.15&0.014 \\
&$B7$&29.7&3.3&$<$15.4&0.21&0.037  \\
&$B6$&26.1&4.0&12.1&0.12&0.032 \\
&$B4$&27.6&2.5&22.1&0.02&0.002 \\
OJ 287&$B6$&20.7&4.7&10.6&0.41&0.117 \\
&$B5$&12.3&3.4&16.1&1.31&0.183 \\
&$B4$&19.9&1.8&$<$28.8&1.70&0.088 \\
1803$+$784&$B2$&17.2&2.2&23.7&0.34&0.038 \\
&$B1$&15.7&3.6&16.1&0.25&0.054 \\
&$C1$&20.8&0.8&38.7&0.02&0.001 \\
1823$+$568&$C2$&9.8&2.9&15.7&0.20&0.005 \\
&$C1$&13.8&0.6&27.0&0.14&0.012 \\
BL Lac&$S13$&9.1&7.0&8.1&0.03&0.009 \\
&$S12$&8.6&7.9&7.2&0.16&0.061 \\
&$S11$&7.2&13.2&3.9&0.07&0.082 \\
&$S10$&5.1&6.0&8.0&0.07&0.025 \\
\enddata
\end{deluxetable} 
  
\begin{deluxetable}{lrrcrcc}
\singlespace
\tablecolumns{7}
\tablecaption{\small\bf Jet Components in Radio Galaxies\label{G_Param}}
\tabletypesize{\footnotesize}
\tablehead{
\colhead{Source}&\colhead{Knot}&\colhead{$\Gamma^{\rm var}$}&\colhead{$\Theta_\circ^{\rm var}(^\circ)$}&\colhead{$\delta^{\rm var}$}&\colhead{$T_{\rm b,obs}(10^{10}~K)$}&\colhead{$T_{\rm b,int}(10^9~K)$}
}
\startdata
3C 111&$B7$&2.8&24.8&2.3&1.49&3.82 \\
&$B6$&3.7&22.9&2.4&8.00&0.20 \\
&$B5$&3.5&19.2&$<$3.0&1.44&2.40 \\
&$B4$&4.0&24.4&2.1&1.61&4.94 \\
&$B3$&4.8&10.4&$<$5.5&0.06&0.04 \\
&$B2$&4.2&21.6&2.5&0.29&0.69 \\
&$B1$&3.2&12.5&$<$4.3&0.23&0.21 \\
&$C1$&6.6&13.0&4.1&0.15&0.15 \\
3C 120&$v$&6.6&23.0&1.7&0.19&0.85 \\
&$u$&4.3&22.3&2.3&0.75&1.89 \\
&$t$&4.9&16.9&3.2&0.11&0.16 \\
&$o1$&7.0&22.0&1.7&3.69&15.2 \\
&$l$&4.0&21.4&2.5&0.88&1.91 \\
&$h$&4.5&19.8&2.7&0.31&0.61  \\
&$d$&3.9&11.3&4.9&0.03&0.02 \\
\enddata
\end{deluxetable}
\clearpage
\begin{deluxetable}{lrrrrrrrl}
\singlespace
\tablecolumns{9}
\tablecaption{\small\bf Global Parameters of the Jets\label{Source_par}}
\tabletypesize{\footnotesize}
\tablehead{
\colhead{Source}&\colhead{$D$(Gpc)}&\colhead{$\Theta_{jet}^\circ$}&\colhead{$\theta_p^\circ$}
&\colhead{$<\Theta_\circ>^\circ$}&\colhead{$\theta^\circ$}&
\colhead{$<\Gamma>$}&\colhead{$<\delta>$}&\colhead{$f_{\rm ej}$}
 }
\startdata
3C 66A&2.46&$-$162.0$\pm$25&9.5$\pm$4.6&2.1$\pm$1.0&0.4$\pm$0.2&10.4$\pm$3.5&16.0$\pm$4.9&1.9 \\
3C 111&0.222&63.5$\pm$4.5&9.1$\pm$2.7&18.1$\pm$5.0&2.8$\pm$0.8&4.4$\pm$1.3&3.4$\pm$1.1&2.4 \\
0420$-$014&5.925&$-$139.2$\pm$21.9&21$\pm$10&3.0$\pm$0.6&1.1$\pm$0.5&11.0$\pm$0.5&16.2$\pm$1.2&1.1 \\
3C 120&0.145&$-$121.7$\pm$6.1&10.9$\pm$3.8&20.5$\pm$1.8&3.8$\pm$1.3&5.3$\pm$1.2&2.4$\pm$0.6&2.4 \\
0528+134&16.12&74.7$\pm$22.4&35.1$\pm$9.3&1.1$\pm$1.0&0.7$\pm$0.2&28.3$\pm$4.5&47.5$\pm$13.8&7.0 \\
OJ 287&1.59&$-$105.8$\pm$16.9&11.1$\pm$5.4&3.2$\pm$0.9&0.8$\pm$0.4&16.5$\pm$4.0&18.9$\pm$6.4&1.3 \\
3C 273&0.76&$-$126.4$\pm$8.6&13.4$\pm$2.5&6.1$\pm$0.8&1.4$\pm$0.3&10.6$\pm$2.8&9.0$\pm$1.4&0.7 \\
3C 279&3.10&$-$131.5$\pm$10.8&8.3$\pm$2.7&2.1$\pm$1.1&0.4$\pm$0.2&15.5$\pm$2.5&24.1$\pm$6.5&3.5 \\
1510$-$089&1.93&$-$42.6$\pm$16.4&4.8$\pm$6.5&1.4$\pm$0.4&0.2$\pm$0.2&36.6$\pm$7.0&38.6$\pm$2.8&2.6 \\
3C 345&3.50&$-$94.9$\pm$8.6&10.7$\pm$4.4&2.7$\pm$0.9&0.5$\pm$0.2&18.7$\pm$5.5&20.2$\pm$5.0&3.7 \\
1803+784&4.11&$-$78.6$\pm$31.1&7.5$\pm$3.6&2.1$\pm$1.2&0.3$\pm$0.2&18.1$\pm$2.3&27.0$\pm$9.9&1.0 \\
1823+568&4.00&$-$162.6$\pm$5.9&2.4$\pm$2.0&1.6$\pm$1.1&0.1$\pm$0.2&12.0$\pm$2.0&21.9$\pm$5.6&0.5 \\
BL Lac&0.31&$-$161.1$\pm$7.3&13.9$\pm$4.3&7.7$\pm$1.9&1.9$\pm$0.6&7.0$\pm$1.8&7.2$\pm$1.1&1.1 \\
CTA 102&6.92&123.1$\pm$19.9&4.8$\pm$1.9&2.6$\pm$0.5&0.2$\pm$0.1&17.2$\pm$2.0&22.3$\pm$4.5&1.2 \\
3C 454.3&5.48&$-$76.8$\pm$27.8&36.8$\pm$9.8&1.3$\pm$1.2&0.8$\pm$0.2&15.6$\pm$2.2&24.6$\pm$4.5&1.9 \\
\enddata
\end{deluxetable}
\clearpage
\begin{figure}
\epsscale{0.85}
\vspace{10cm}
\caption{Total intensity images of 3C~66A at 43~GHz. North is toward the top
and east is toward the left. Line segments 
within each image indicate direction of electic vectors, with 
length proportional to the polarized intensity. Line segments outside the images give 
direction of polarization at 3~mm (dark gray), 0.85/1.3~mm (bold black), and at optical 
wavelengths (light gray) with length reflecting the percent polarization.
Percent polarization is printed near corresponding line segments along with
the epoch of observation. Darkness of the print indicates the wavelength
of observation. If epochs of the JCMT and optical observations coincide, only the JCMT
epoch is indicated. Parameters of the maps are listed in Table \ref{Sample}. 
The resolution beam is shown by the cross-hatched ellipse in the lower left
corner. Designation of knots follows Table \ref{Speed}. The figure 
can be found at the web-site {\bf www.bu.edu/blazars/multi.html}. \label{Alla} }
\end{figure}
\clearpage
\begin{figure}
\vspace{10.0cm}
\caption{{\it Left panel:} Total intensity images of 3C~111 at 43~GHz along with polarization measurements
at 7~mm, 3~mm, 1.3/0.85~mm, and at optical wavelenths. The epochs of BIMA observations
are marked by the letter B. See caption for Figure \ref{Alla} for details.
{\it Right panel:}
Total intensity images of the inner jet of 3C~111 at 43~GHz. The figure 
can be found at the web-site {\bf www.bu.edu/blazars/multi.html}.\label{Allb}}
\end{figure} 
\clearpage
\begin{figure}
\epsscale{1.1}
\vspace{10cm}
\caption{Total intensity images of 0420$-$014 at 43~GHz along with polarization measurements
at 7~mm, 3~mm, 1.3/0.85~mm, and at optical wavelenths. The epochs of BIMA observations
coincide with the epochs of corresponding VLBA images. See caption for Figure \ref{Alla}.
\label{Allc}}
\end{figure}
\clearpage 
\begin{figure}
\epsscale{1.1}
\vspace{10cm}
\caption{Total intensity images images of 3C 120 at 43~GHz along with polarization measurements
at 7~mm, 3~mm, and 1.3/0.85~mm. Measurements of the optical polarization are consistent with 
Galactic interstellar polarization for this sight line and are not included. See caption for Figure 
\ref{Alla}. \label{Alld}}
\end{figure} 
\clearpage
\begin{figure}
\epsscale{0.85}
\vspace{10cm}
\caption{Total intensity images of 0528+134 at 43~GHz along with polarization measurements
at 7~mm, 3~mm, 1.3/0.85~mm, and at optical wavelenths. The epochs of BIMA
observations coincide with the epochs of corresponding VLBA images. See caption for Figure 
\ref{Alla}. \label{Alle}}
\end{figure}
\clearpage
\begin{figure}
\epsscale{1.1}
\vspace{10cm}
\caption{Total intensity images of OJ 287 at 43~GHz along with polarization measurements
at 7~mm, 3~mm, 1.3/0.85~mm, and at optical wavelenths. The epochs of BIMA
observations coincide with the epochs of corresponding VLBA images. See caption for Figure 
\ref{Alla}. \label{Allf}}
\end{figure}
\clearpage
\begin{figure}
\epsscale{1.0}
\vspace{10cm}
\caption{Total intensity images of 3C 273 at 43~GHz along with polarization measurements
at 7~mm, 3~mm, and 1.3/0.85~mm. The epochs of BIMA
observations coincide with the epochs of corresponding VLBA images. See caption for Figure 
\ref{Alla}. \label{Allg}}
\end{figure}
\clearpage
\begin{figure}
\epsscale{1.1}
\vspace{10cm}
\caption{Total intensity images of 3C 279 at 43~GHz along with polarization measurements
at 7~mm, 3~mm, 1.3/0.85~mm, and at optical wavelenths. The epochs of BIMA
observations coincide with the epochs of corresponding VLBA images. See caption for Figure 
\ref{Alla}. \label{Allh}}
\end{figure}
\clearpage
\begin{figure}
\epsscale{1.0}
\vspace{10cm}
\caption{Total intensity images of PKS 1510$-$089 at 43~GHz along with polarization measurements
at 7~mm, 3~mm, 1.3/0.85~mm, and at optical wavelengths. The epochs of BIMA
observations coincide with the epochs of corresponding VLBA images. See caption for Figure 
\ref{Alla}. \label{Alli}}
\end{figure}
\clearpage
\begin{figure}
\epsscale{1.1}
\vspace{10cm}
\caption{Total intensity images of 3C 345 at 43~GHz along with polarization measurements
at 7~mm, 3~mm, 1.3/0.85~mm, and at optical wavelengths. The epochs of BIMA
observations coincide with the epochs of corresponding VLBA images. See caption for Figure 
\ref{Alla}. \label{Allj}}
\end{figure}
\clearpage
\begin{figure}
\epsscale{1.1}
\vspace{10cm}
\caption{Total intensity images of 1803+784 at 43~GHz  along with polarization measurements
at 7~mm, 3~mm, and 1.3/0.85~mm. The epochs of BIMA
observations coincide with the epochs of corresponding VLBA images. See caption for Figure 
\ref{Alla}. \label{Allk}}
\end{figure}
\clearpage
\begin{figure}
\epsscale{1.0}
\vspace{10cm}
\caption{Total intensity images of 1823+568 at 43~GHz along with polarization measurements
at 7~mm, 3~mm, 1.3/0.85~mm, and at optical wavelengths. The epochs of BIMA
observations coincide with the epochs of corresponding VLBA images. See caption for Figure 
\ref{Alla}. \label{Alll}}
\end{figure}
\clearpage
\begin{figure}
\epsscale{1.0}
\vspace{10cm}
\vspace{-3cm}
\caption{Total intensity images of BL Lac at 43~GHz along with polarization measurements
at 7~mm, 3~mm, 1.3/0.85~mm, and at optical wavelengths. The epochs of BIMA
observations coincide with the epochs of corresponding VLBA images. See caption for Figure 
\ref{Alla}. \label{Allm}}
\end{figure}
\clearpage
\begin{figure}
\epsscale{0.85}
\vspace{10cm}
\caption{Total intensity images of CTA~102 at 43~GHz along with polarization measurements
at 7~mm, 3~mm, 1.3/0.85~mm, and at optical wavelengths. The epochs of BIMA
observations coincide with the epochs of corresponding VLBA images. See caption for Figure 
\ref{Alla}. \label{Alln}}
\end{figure}
\clearpage
\begin{figure}
\epsscale{0.75}
\vspace{10cm}
\caption{Total intensity images of 3C 454.3 at 43~GHz along with polarization measurements
at 7~mm, 3~mm, 1.3/0.85~mm, and at optical wavelengths. The epochs of BIMA
observations coincide with the epochs of corresponding VLBA images. See caption for Figure 
\ref{Alla}.  \label{Allo}}
\end{figure}
\clearpage
\begin{figure}
\epsscale{1.0}
\vspace{-2.0cm}
\plotone{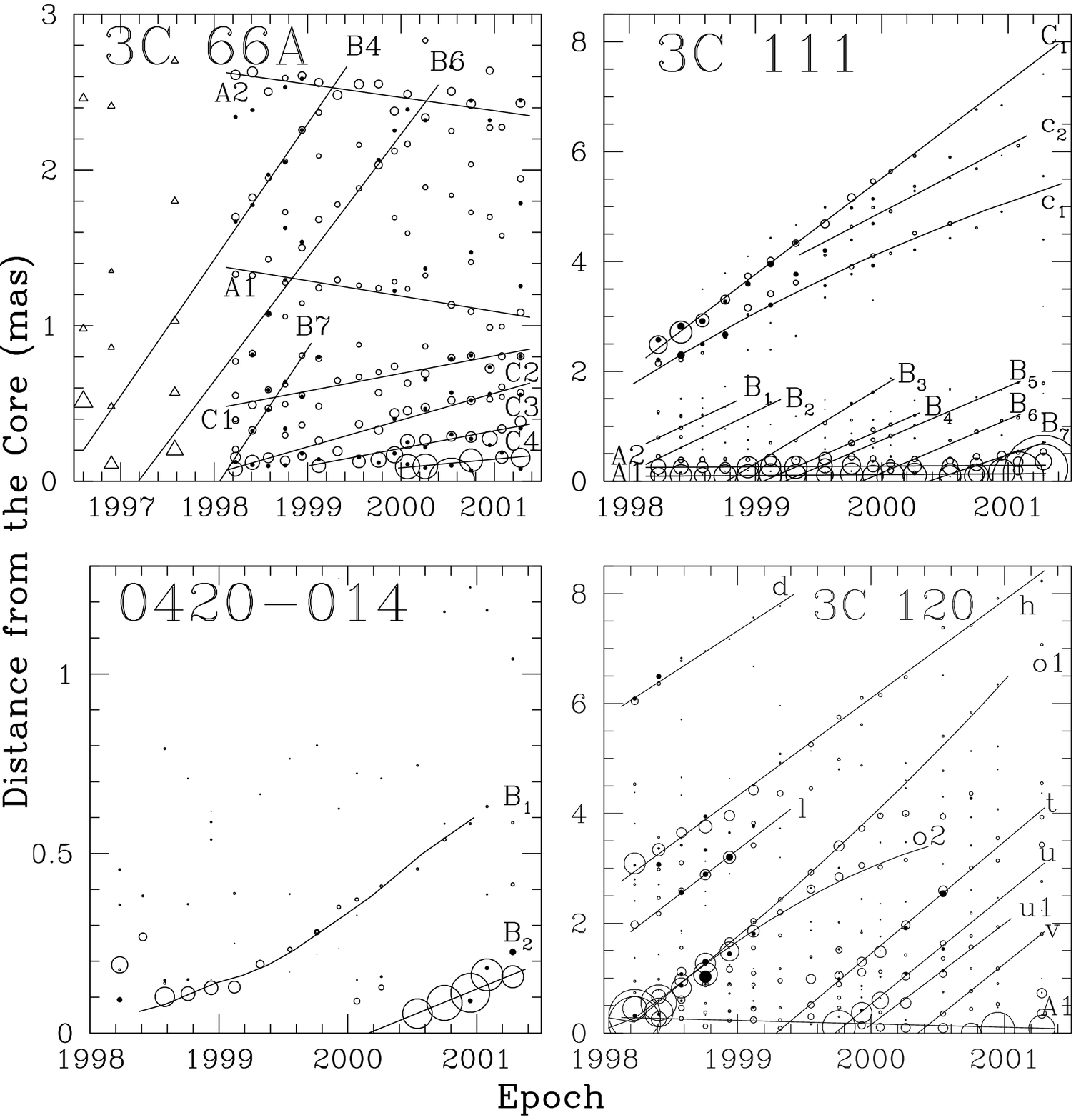}
\vspace{-5.0cm}
\caption{Separation of components from the VLBA core vs. epoch of 
observation.
Total intensity components are indicated by open circles, polarized intensity
components by filled circles. In a few cases (3C~66A, 0528+134, 3C~279, and 3C~345)
data prior to 1998 are from \citet{J01} (triangles), \citet{W01} (squares), and \citet{RZL00}
(asterisks) (see \S  A for details). For 3C~279 and 3C~454.3 different 
symbols are used to separate closely located components. The size of a symbol is 
related to the flux density. The solid lines(curves) represent the best polynomial
approximation to the data according to Table \ref{Speed}. Designation of knots follows 
Table \ref{Speed}. \label{Ident}}
\end{figure}
\begin{figure}
\figurenum{16}
\epsscale{1.0}
\vspace{-2.0cm}
\plotone{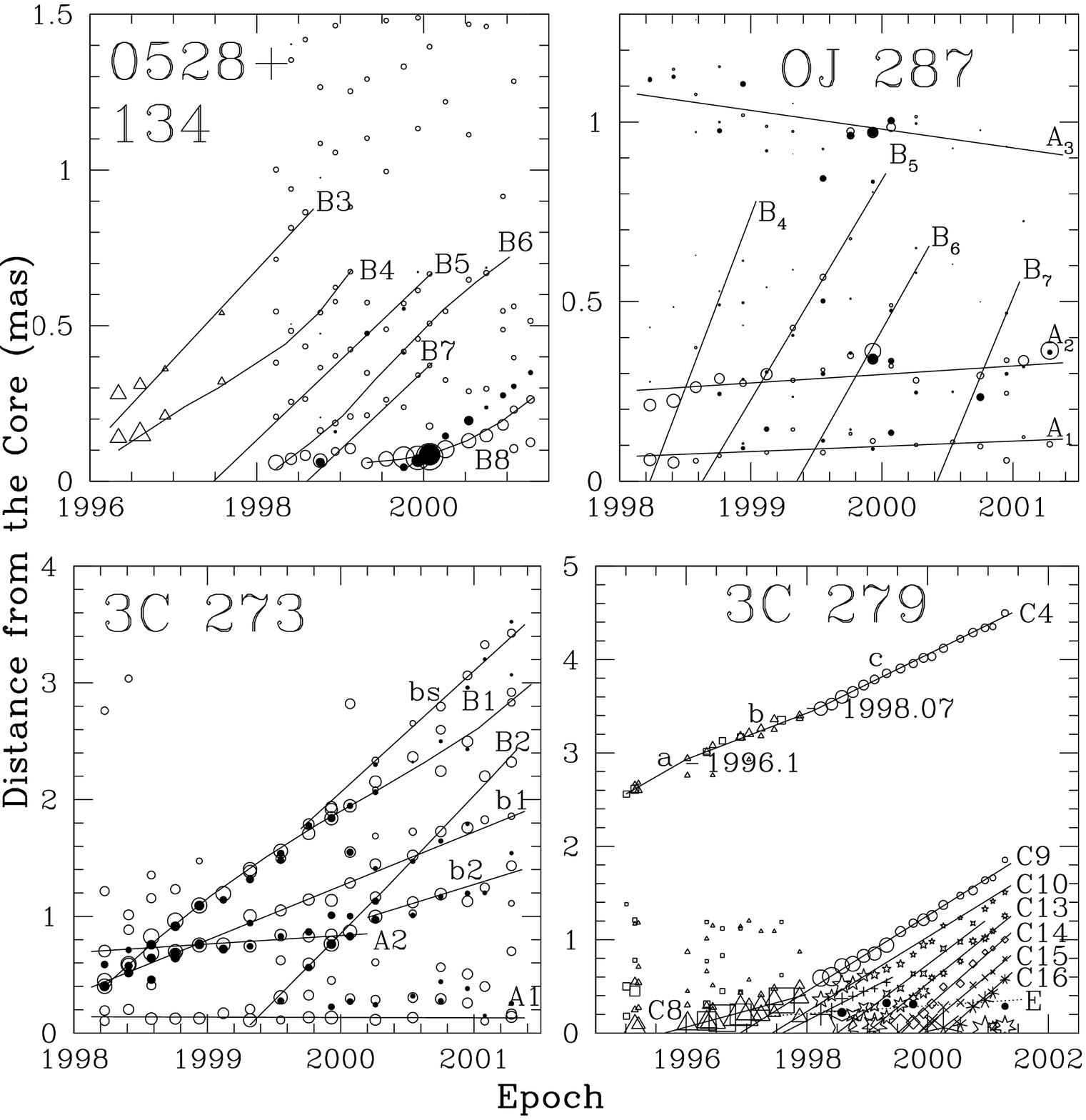}
\vspace{-5.0cm}
\caption{Continued}
\end{figure}
\begin{figure}
\figurenum{16}
\epsscale{1.0}
\vspace{-2.0cm}
\plotone{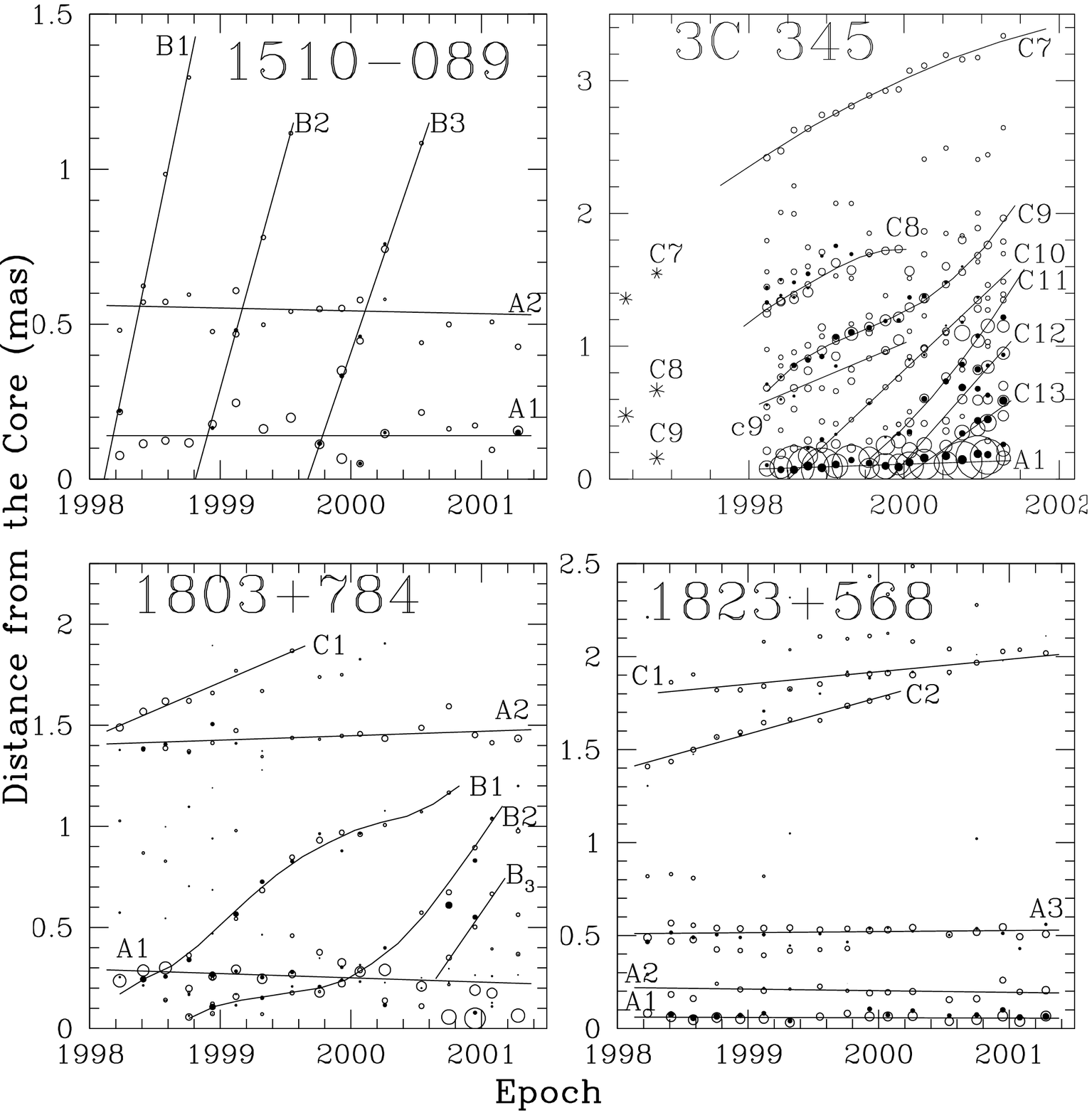}
\vspace{-5.0cm}
\caption{Continued}
\end{figure}
\begin{figure}
\figurenum{16}
\epsscale{1.0}
\vspace{-2.0cm}
\plotone{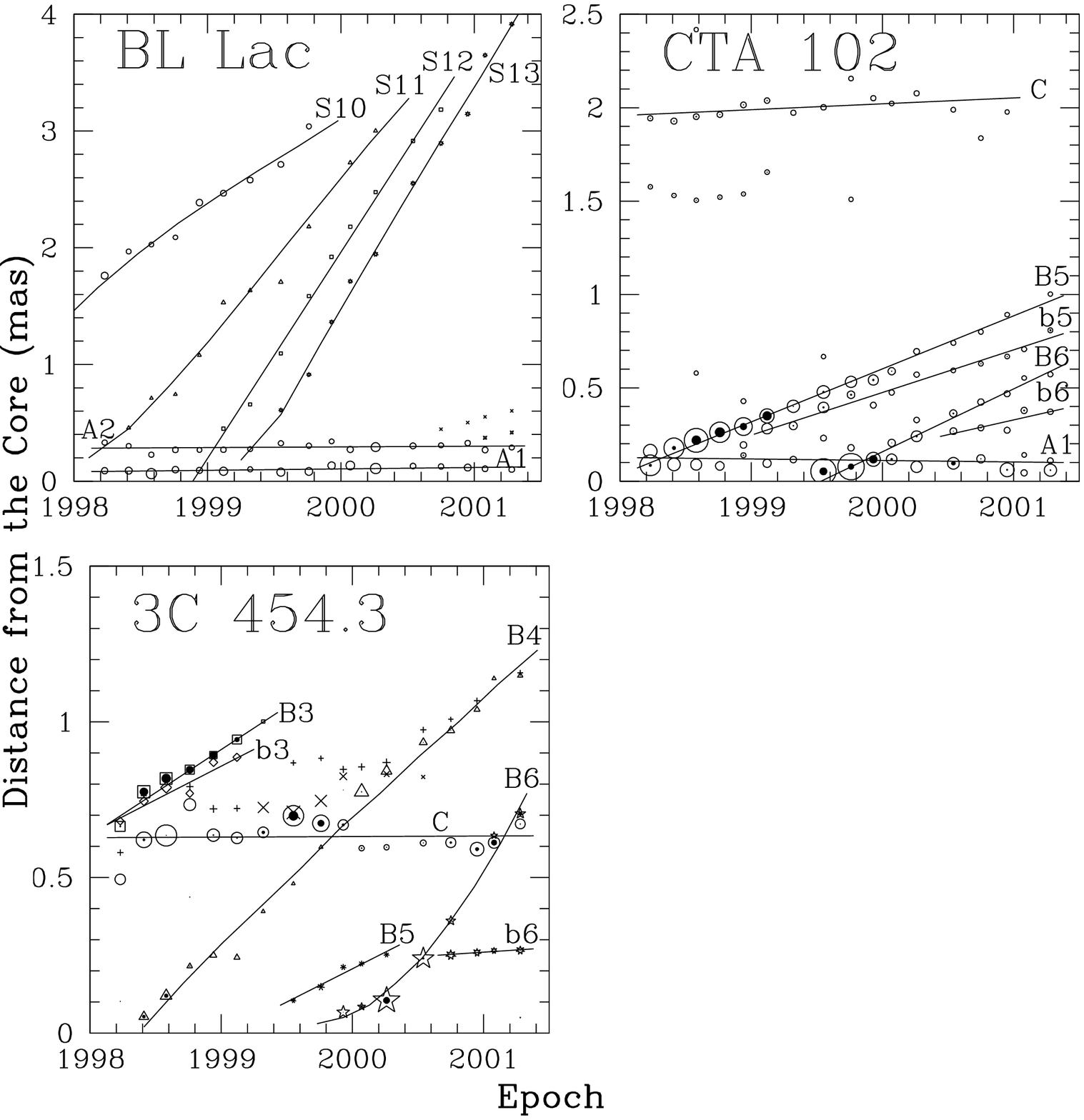}
\vspace{-5.0cm}
\caption{Continued}
\end{figure}
\clearpage
\begin{figure}
\epsscale{1.1}
\vspace{10cm}
\caption{Total (contours) and polarized (color scale) intensity images at 43~GHz. 
The yellow line segments show the direction of the electric vectors. Parameters of the maps
are given in Table \ref{Tmap}. Designation of components follows Table \ref{Speed}
and Figure \ref{Ident}. The figure 
can be found at the web-site {\bf www.bu.edu/blazars/multi.html}. \label{Map}}
\end{figure}
\begin{figure}
\figurenum{17}
\epsscale{1.1}
\vspace{10cm}
\caption{Continued}
\end{figure}
\begin{figure}
\epsscale{1.0}
\plotone{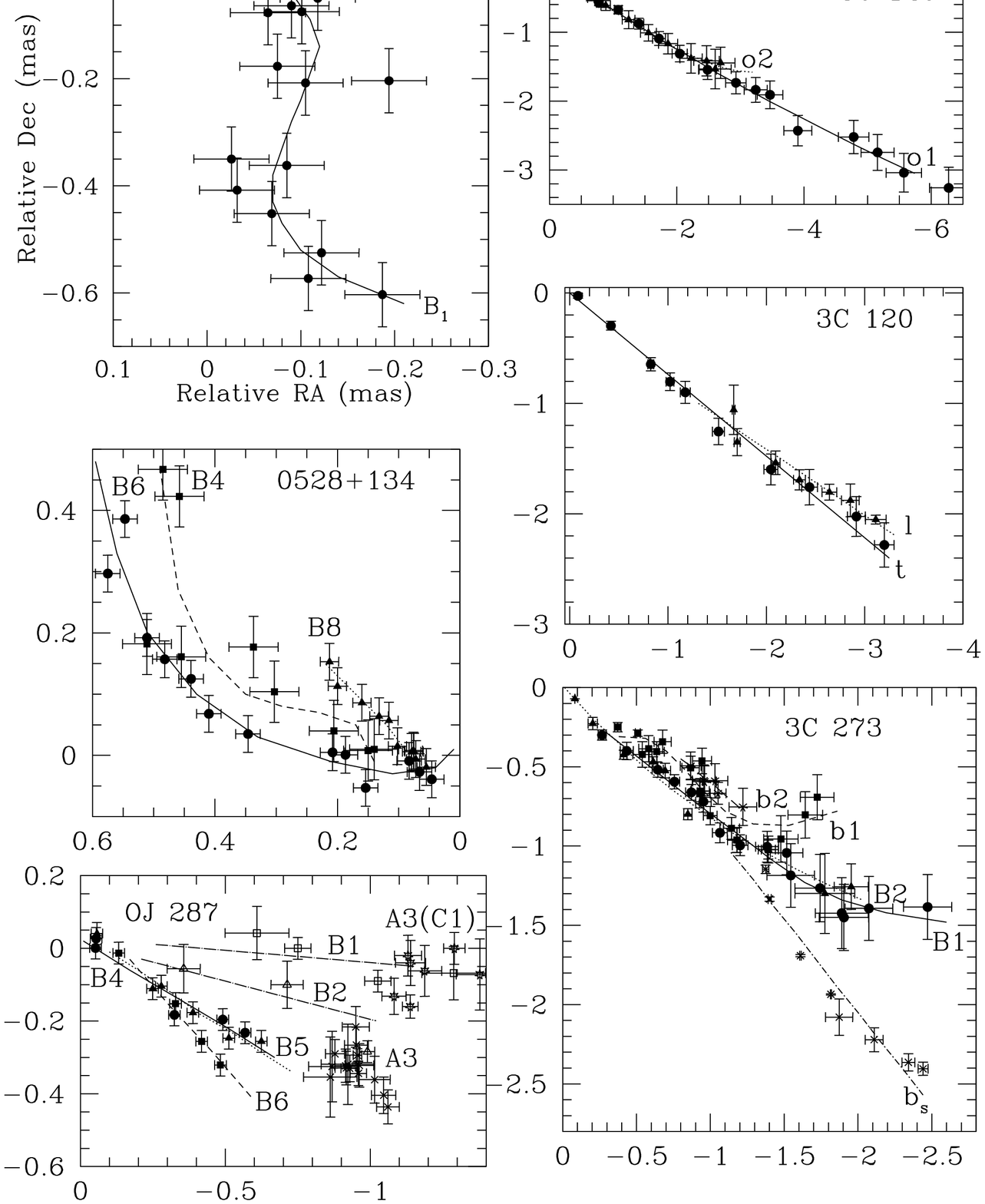}
\vspace{-3.5cm}
\caption{Trajectories of components that exhibit a statistically significant
change of apparent speed (Table \ref{Accel}). 
The symbols show measured positions of components while the curves represent
polynomial approximations to the trajectories. Ballistic trajectories 
are shown for 3C~120 (components $l$ and $t$), OJ 287, and CTA102 (see \S A
for explanation.)\label{traj}}
\end{figure}
\begin{figure}
\figurenum{18}
\epsscale{1.0}
\plotone{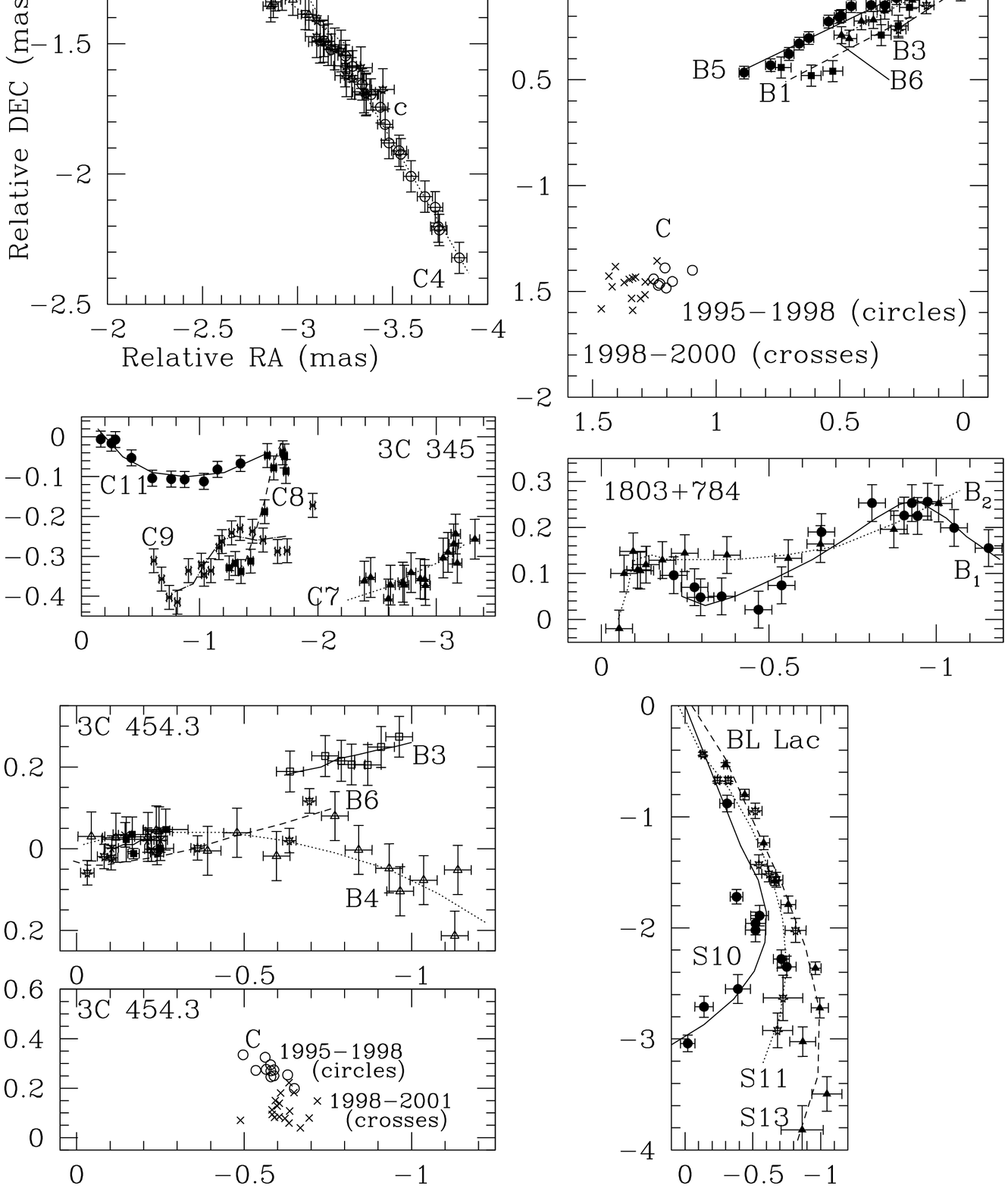}
\vspace{-3.5cm}
\caption{Continued}
\end{figure}
\clearpage
\begin{figure}
\epsscale{1.0}
\vspace{-1cm}
\plotone{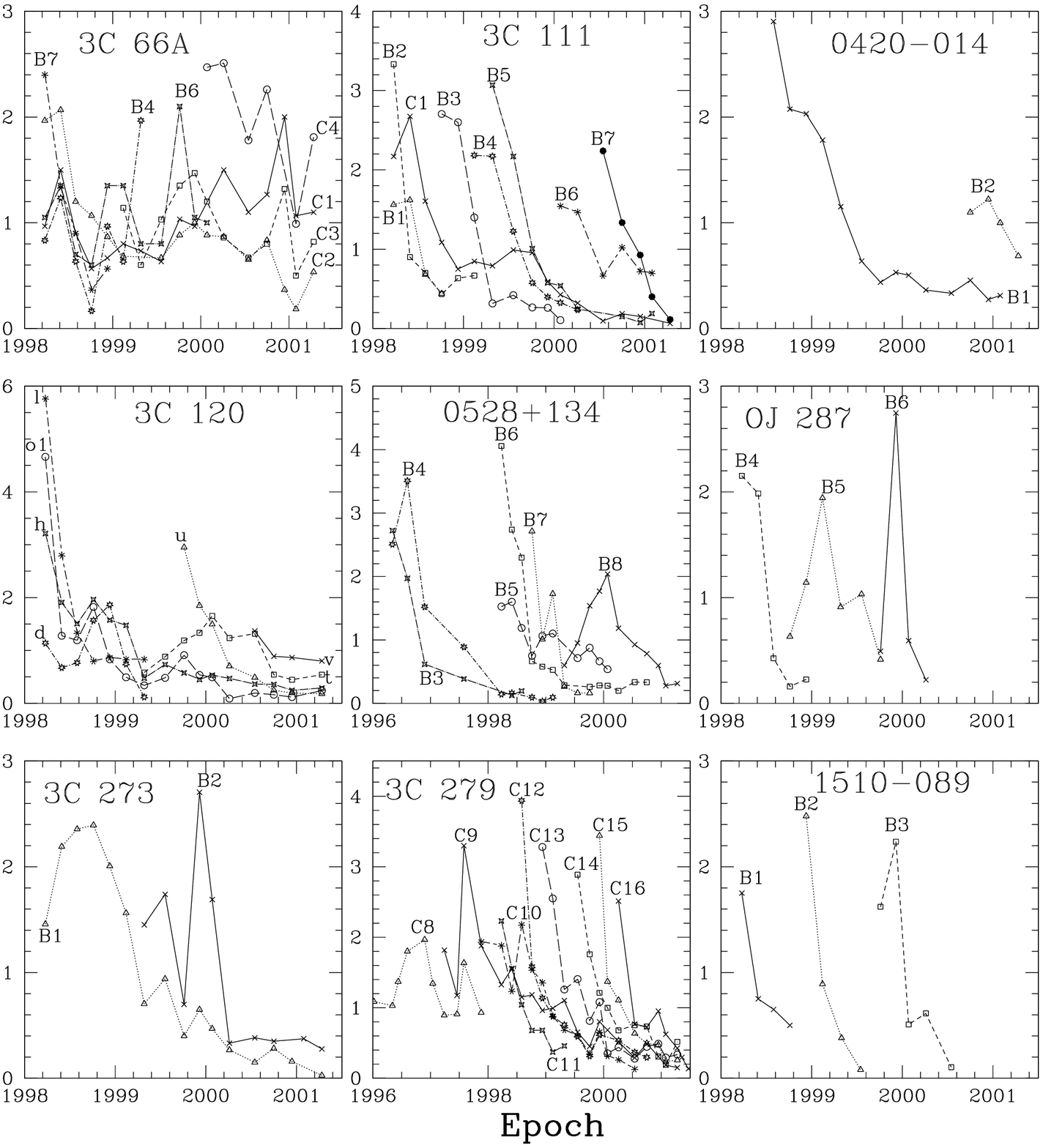}
\vspace{-3cm}
\caption{Light curves of superluminal jet components. 
Flux densities are 
normalized by the average flux density listed in Table \ref{Speed}. 
For each source different components are denoted by different symbols.  
\label{flux}}
\end{figure}
\begin{figure}
\epsscale{1.0}
\figurenum{19}
\vspace{-1cm}
\plotone{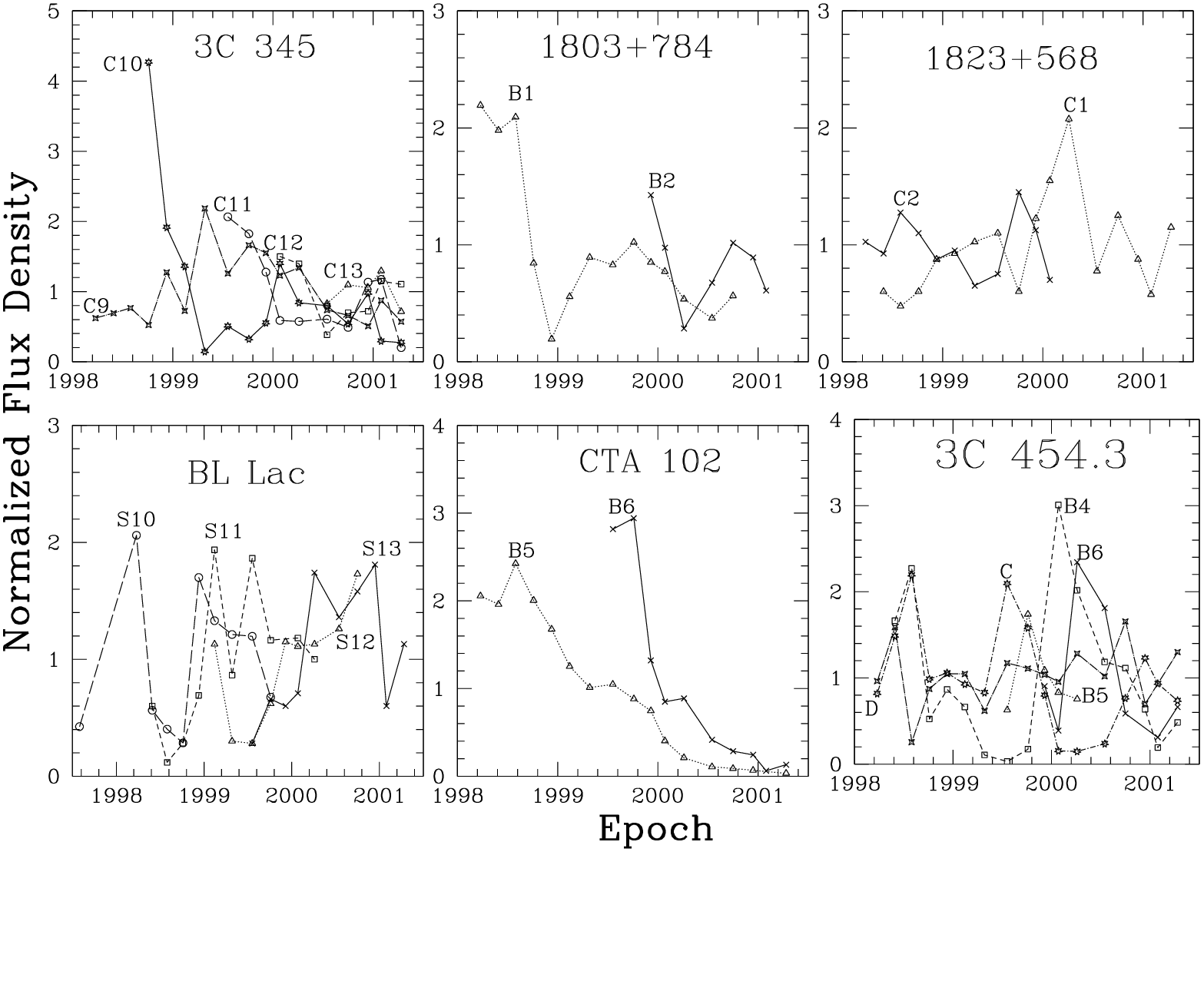}
\vspace{-3cm}
\caption{Continued}
\end{figure}
\clearpage
\clearpage
\clearpage
\clearpage
\begin{figure}
\epsscale{1.0}
\vspace{6cm}
\caption{Total (contours) and polarized (gray scale) intensity images of 3C~273. 
The total intensity peak is 1710 mJy~beam$^{-1}$, the polarized intensity peak is 
90 mJy~beam$^{-1}$, the beam is 0.39$\times$0.14~mas$^2$ at P.A.=$-$6.5$^\circ$,
and contours increase by a factor of $\sqrt{2}$, with the lowest contour 
corresponding to 0.25\% of the peak. The line segments indicate the electric vector direction. The figure 
can be found at the web-site {\bf www.bu.edu/blazars/multi.html}.\label{3c273_C1}}
\end{figure}
\clearpage
\begin{figure}
\epsscale{0.95}
\vspace{10cm}
\caption{Total intensity images with polarization vectors of the 
quasar PKS 1510$-$089 from 1998.94 to 1999.55 ({\it left panel}: $S_{\rm peak}$=1520~mJy~beam$^{-1}$,  
$S^{\rm p}_{\rm peak}$=107~mJy~beam$^{-1}$) and from 1999.76 to 2000.26 ({\it right panel:} 
$S_{\rm peak}$=1700~mJy~beam$^{-1}$, $S^{\rm p}_{\rm peak}$=147~mJy~beam$^{-1}$). 
The beam is 
0.40$\times$0.15~mas$^2$ at P.A.=$-5^\circ$, contour levels are 0.3, 0.6,...76.8, 90\% 
of the total intensity peak. The figure 
can be found at the web-site {\bf www.bu.edu/blazars/multi.html}.
\label{1510_bw4}}
\end{figure}
\clearpage
\begin{figure}
\epsscale{1.0}
\vspace{6cm}
\caption{Total (contours) and polarized (gray scale) intensity images of 3C~454.3.
The total intensity peak is 1860~mJy~beam$^{-1}$, and the polarized intensity peak is
71~mJy~beam$^{-1}$. The beam is 0.31$\times$0.15~mas$^2$ at P.A.=$-$6$^\circ$;
contour levels increase by a factor of $\sqrt{2}$ with the lowest contour corresponding 
to 0.25\% of the peak. The line segments indicate the polarization direction. The figure 
can be found at the web-site {\bf www.bu.edu/blazars/multi.html}.
\label{3c454_D}}
\end{figure}
\end{document}